\documentclass[iop]{emulateapj}
\pdfoutput=1
\usepackage{graphicx}
\usepackage{amsmath}
\usepackage{amsfonts}
\usepackage{float}
\usepackage{upgreek}
\usepackage{color}
\usepackage{changepage}

\usepackage{natbib}
\newcommand{\hst}{\textit{HST}}
\newcommand{\Mbh}{\ensuremath{M_\mathrm{BH}}}
\newcommand\sersic{S\'ersic}
\newcommand\csersic{core-S\'ersic}

\shorttitle{WFC3/IR Photometry of MASSIVE Galaxies}
\shortauthors{Goullaud et al.}

\begin{document}

\title{The MASSIVE Survey IX: \\
  Photometric Analysis of 35 High Mass Early-Type Galaxies with \emph{HST} WFC3/IR\altaffilmark{1}}

\altaffiltext{1}{Based on observations with the NASA/ESA
  \textit{Hubble Space Telescope}, obtained at the Space Telescope
  Science Institute, which is operated by the Association of
  Universities for Research in Astronomy, Inc., under NASA contract
  NAS 5-26555. These observations are associated with Program
  GO-14219.}

\author{Charles F. Goullaud}
\affil{Department of Physics, University of California, Berkeley, CA, USA;
{\tt goullaud@berkeley.edu}}

\author{Joseph B. Jensen}
\affil{Utah Valley University, Orem, UT, USA}

\author{John P. Blakeslee}
\affil{Herzberg Astrophysics, Victoria, BC, Canada}

\author{Chung-Pei Ma}
\affil{Department of Astronomy, University of California, Berkeley, CA, USA}

\author{Jenny E. Greene}
\affil{Princeton University, Princeton, NJ, USA}

\author{Jens Thomas}
\affil{Max Planck-Institute for Extraterrestrial Physics, Garching, Germany.}

\begin{abstract}
We present near-infrared observations of 35 of the most massive
early-type galaxies in the local universe.  
The observations were made using the infrared channel of the
\emph{Hubble Space Telescope} (\hst) \emph{Wide Field Camera 3}
(WFC3) in the F110W (1.1~$\mu$m) filter.  
We measured surface brightness profiles and elliptical isophotal fit parameters
from the nuclear regions 
out to a radius of ${\sim}10$ kpc in most cases. 
We find that $37\%$ ($13$) of the galaxies in our sample have isophotal
position angle rotations greater than $20^{\circ}$ over the radial range 
imaged by WFC3/IR,
which is often due to the presence of neighbors or multiple nuclei.
Most galaxies in our sample are significantly rounder near the center
than in the outer regions.
This sample contains six fast rotators and 28 slow rotators.
We find that all fast rotators are either disky or show no measurable
deviation from purely elliptical isophotes.
Among slow rotators, significantly disky and boxy galaxies
occur with nearly equal frequency.
The galaxies in our sample often exhibit changing isophotal shapes, 
sometimes showing both significantly disky and boxy isophotes at different radii.
The fact that parameters vary widely between galaxies and within
individual galaxies is evidence that these massive galaxies have
complicated formation histories, and some of them have experienced recent
mergers and have not fully relaxed.
These data demonstrate the value of high spatial resolution IR imaging
of galaxies and provide measurements necessary for determining stellar
masses, dynamics, and black hole masses in high mass galaxies.
\end{abstract}


\section{Introduction}
\label{sec:intro}

Massive early-type galaxies are the modern-day remnants of the earliest major
star-formation events in the history of the universe, and they comprise a substantial
fraction of the stellar mass in the local universe \citep[e.g.,][]{renzini2006}.  Thus,
understanding the formation histories of such galaxies is central to the problem of
galaxy formation in general.  The MASSIVE survey, described by
\citet[][Paper~I]{Ma2014}, is an ongoing systematic effort using integral field
spectroscopy and other data to study the detailed mass structure and abundance
profiles of the highest mass galaxies in the local universe. The complete MASSIVE
sample includes all early-type galaxies in the northern sky out to an estimated
distance of $108$ Mpc and absolute $K$-band magnitude $M_{K}\,{<}\,-25.3$~mag, corresponding
to stellar masses $M_{*} \gtrsim 10^{11.5} M_{\odot}$.

In Paper~II of the MASSIVE survey, \citet{Greene2015} presented the first
results on stellar population gradients from spatially resolved absorption line
measurements and found evidence that the
centers of massive galaxies formed rapidly at high redshift, while the outskirts were
accreted during later mergers.  \citet[][Paper~III]{Davis2016} presented a pilot
study of CO observations in 15 MASSIVE galaxies and found a significant amount of cold
gas in ten of the galaxies.
Paper~IV \citep{Goulding2016} used archival Chandra X-ray images
to analyze the hot X-ray gas properties in 33 MASSIVE galaxies and additional
lower mass galaxies from the ATLAS$^{3D}$ survey \citep{atlas3d}.
Paper~V \citep{Veale2017-v} presented stellar kinematics
for the subset of 41 MASSIVE galaxies with stellar masses $M_{*} \gtrsim 10^{11.8}
M_{\odot}$, and Paper~VI \citep{Pandya2017} studied the spatial distribution and kinematics of the
warm ionized gas.  
Paper VII \citep{Veale2017-vii} presented integral field spectroscopic measurements of the
specific spin parameter $\lambda_e$ in a complete mass-limited subsample of $75$ MASSIVE
galaxies, and examined the interrelationships between $\lambda_e$, stellar mass, and
environment in a combined sample of 370 MASSIVE and ATLAS$^{3D}$ survey galaxies.
Velocity dispersion profiles for 90~MASSIVE galaxies out to projected radii as large as
30~kpc were presented in Paper~VIII \citep{Veale2018-viii}, which found strong evidence
for a trend in the outer slope of the dispersion profile with environment.

All of the MASSIVE studies to date have
been based primarily on optical or radio-wavelength spectroscopic observations or
archival X-ray data.  However, photometric imaging data are also an essential element of
the MASSIVE project for determining the stellar mass profiles of the sample galaxies.
An important goal of the project is to constrain the central
supermassive black hole (SMBH) masses \Mbh\ in MASSIVE galaxies using Schwarzchild orbit
modeling \citep{Thomas2004}.
Multiple mass components contribute to the gravitational potential to varying degrees
as a function of galactocentric radius, and an individual star may travel from the
sphere of influence of the SMBH to large radii where dark matter
dominates, traversing the intermediate region dominated by the stellar mass.  
Therefore, in order to disentangle the contributions of the various components and
construct stellar orbits in the combined gravitational potential, the dynamical models
require accurate characterization of the stellar luminosity profile from the galaxy
center to its outskirts.
In particular, the derived SMBH mass is partly degenerate with the central
stellar mass profile; the inner luminosity profile and stellar mass-to-light ratio
must be well constrained in order to obtain a robust measurement of \Mbh.
For example, \citet{Thomas2016} used archival \textit{Hubble Space Telescope} (\hst)
NICMOS imaging to 
constrain the very flat core in NGC\,1600 and dynamically determined
$\Mbh=(1.7{\pm}0.15){\,\times\,}10^{10} M_\odot$, the largest SMBH known in the local
universe outside a rich galaxy cluster.  
We have been amassing new and archival high-resolution \hst\ and wide-field
ground-based photometry for the full sample of MASSIVE galaxies
to better understand their stellar and central black hole masses.

\hst\ has revolutionized the study of early-type galaxies, especially
with regard to the complex properties of their central regions.
\citet{Ferrarese1994} studied a sample of 14 early-type galaxies in
the Virgo cluster and found that they fell into two groups: those
whose surface brightness profiles became less steep near their centers
and were best fit by a broken power-law (``core'' galaxies), and those
whose interiors were well described by a single logarithmic slope
(``power-law'' galaxies).  Based on a larger, but more heterogeneous,
sample of 45 galaxies, \citet{Lauer1995} came to a similar conclusion
and fitted all the profiles with a double power-law model (dubbed the
``Nuker~law'') that included a transition sharpness parameter.  Their
results showed a clear bimodality in the value of the logarithmic slope
$\gamma$ inside the break (or transition) radius, with core galaxies
having $\gamma{\,<\,}0.3$ and power-law galaxies having
$\gamma\,{\gtrsim}\,0.6$; very few galaxies in their sample had
$\gamma$ between these two ranges 
\citep{Laine2003,Lauer2005,Lauer2007cen}.

\citet{Faber1997} studied a somewhat larger sample and demonstrated
that the central slope correlates strongly with luminosity: luminous
galaxies (those with $M_V{<}{-}22$) possessed central cores, while faint galaxies
with $M_V{>}{-}20$ had power-law profiles over the range of radii explored.
Intermediate luminosity galaxies occupied a transition region with a
mix of power-law and core profiles.  
\citet{Laine2003} observed 81 brightest cluster galaxies with \hst\
and came to the conclusion that the distribution of core and power-law
profiles is bimodal, a result supported by \citet{Lauer2007cen}. 
On the other hand, other studies have shown that the transition in 
central slope from power-law to core occurs smoothly with galaxy luminosity, 
or stellar mass \citep{Graham2003,Ferrarese2006,Cote2007,Glass2011}.  
The more
recent studies tend to parameterize early-type galaxy profiles using
\sersic\ \citep{Sersic1968} or \csersic\ \citep{Graham2003} profiles,
and associate the difference between these two profiles for a given
galaxy as representing the amount of ``missing,'' or depleted, stellar
mass at their centers
\citep[e.g.,][]{Trujillo2004,Ferrarese2006,Kormendy2009,Rusli2013b}.
\citet{Thomas2016} demonstrated that the break radius in the \csersic\
fit to a galaxy profile corresponds closely to the radius of the
sphere of influence of the central SMBH.  Thus, a single
high-resolution photometric measurement can provide a reasonable
prediction of \Mbh\ for massive galaxies. 
The aim of this paper is not to address the core/cusp issue,
but to explore the photometric properties of the MASSIVE galaxies
outside the nuclear regions to constrain the stellar mass distributions
in prepration for accurate black hole mass and SBF distance 
measurements.

While the studies above focused on the surface brightness profiles of
the nuclear regions of the galaxies, others have used the shapes of
the isophotes to study the structure and formation histories of
early-type galaxies. 
These measurements include ellipticity, position
angle (PA) rotation, isophote center shifts, and deviations from
purely elliptical isophotes 
\citep[i.e., diskiness or boxiness;][]{Franx1989,Kormendy1996}.
For example, attempts to determine the intrinsic shapes of galaxies
from the statistical frequency distribution of observed projected 
ellipticities have suggested that the brightest elliptical
galaxies must be intrinsically triaxial, compared to lower-luminosity
early-type galaxies, which appear to be axisymmetric
\citep[e.g.,][]{Tremblay1996,Vincent2005}. 
Likewise, kinematic misalignments, which cannot occur in
axisymmetric systems, are more frequent in more massive galaxies
\citep[e.g.,][]{Emsellem2007}.

Indeed, to explain the global photometric and kinematic structure of
elliptical galaxies, a rough paradigm has emerged where high-mass
galaxies primarily form in gas poor merging situations which produce
boxy, anisotropic, cored, triaxial giant ellipticals; lower-mass
elliptical galaxies preferentially form from the gas rich mergers of
disk galaxies that produce rotating, disky ellipticals
\citep{Ferrarese1994, vandenBosch1994, Lauer1995, Kormendy1996,
  Rest2001, Lauer2005, Ferrarese2006}.
More specifically, \citet{Weijmans2014}, for example, suggested that
angular momentum is the crucial parameter and that the fast rotators
make the flatter and nearly axisymmetric elliptical galaxy
subpopulation, while the statistically less flattened and triaxial
ones are slow rotators.
This was based on the ATLAS$^{3D}$ sample, which lacks the most
massive (predominantly slow rotator) galaxies targeted by MASSIVE, for
which we present new data here.

For individual galaxies, photometric data alone are not sufficient to
constrain the intrinsic axis ratios.
They must be complemented by stellar kinematics and detailed
dynamical modeling 
\citep[e.g.,][]{Statler1994,Magorrian1999,Gerhard1996}.
However, some clues on the intrinsic shape of a galaxy are directly
encoded in the isophotes.
PA rotation measurements, for example, provide a lower limit for the
statistical occurrence of non-axisymmetry and, hence, are an important
piece of information in the context of elliptical galaxy evolution
\citep[e.g.,][]{Stark1977,Williams1979}.

High-spatial resolution \hst\ imaging has also revealed the presence of dust 
in the centers of many early-type galaxies. 
\citet{Rest2001} analyzed an unbiased, though incomplete, magnitude- and
space-limited sample of 67 early-type galaxies within 54 Mpc, while
\citet{Lauer2005} studied a sample of 77 galaxies within 100 Mpc using archival observations. 
Both studies found nuclear dust in roughly half of the galaxies.
\citet{Ferrarese2006} conducted the first survey of early-type galaxies using 
the \emph{Advanced Camera for Surveys} (ACS), enabling surface photometry measurements
out to ${\gtrsim}\,1.5$ arcmin, as compared to the ${\sim}20$ arcsec
limit of previous studies that only utilized the PC chip of \textit{WFPC2}.
Their study examined 100 early-type galaxies from the ACS Virgo Cluster Survey
\citep{Cote2004}, with
the brightest 26 galaxies forming a magnitude-limited sample, and confirmed 
the presence of dust in roughly half of the most massive early-type galaxies.
Although this sample was much more complete and homogeneous than other
studies of early-type galaxies with \hst, the very limited volume meant
that it contained only a few galaxies at masses $\gtrsim 10^{11.5} M_{\odot}$.
Thus, no survey has yet included a complete sample of the most
massive early-type galaxies in the local universe.

This paper presents new IR imaging observations with the Infrared Channel of 
the \hst\ Wide Field Camera~3 (WFC3/IR) of a complete subsample of 35 
galaxies selected from the MASSIVE survey (GO-14219, P.I. J.~Blakeslee).
The purpose of this study is to: (1) measure accurate surface brightness
profiles to establish the stellar mass profiles in order to 
better constrain the black hole masses; and (2) to measure the radial 
variations in isophotal parameters to test the idea that the most massive 
galaxies are preferentially boxy and slowly rotating. 
The high-resolution IR imaging data of this sample of MASSIVE galaxies
provides detailed insight into the structures and formation histories of
these galaxies that goes beyond a simplistic characterization of each galaxy.
We analyze the radial variation of isophote center, position angle,  
ellipticity, and boxiness or diskiness for each galaxy, and we examine 
possible trends among various mean isophotal parameters as a function of 
radius out to ${\sim}10$ kpc.
We also discuss the prevalence and distribution of central dust in our sample 
galaxies and highlight the peculiar double nucleus in the
galaxy NGC\,1129.

The following section describes the observations and data reduction 
procedures, 
background sky estimation, mask construction, and the isophotal fitting 
process we used to measure the surface brightness profiles.
Sec.~3 presents our surface brightness profile measurements, which are shown
in detail for each galaxy in the Appendix along with IR images of each
galaxy.
Sec.~4 summarizes our results that provide a unique look into the photometric
properties of the most massive early-type galaxies, a population that has 
not been systematically studied prior to the MASSIVE survey.
Distances for our sample galaxies, measured using the surface brightness
fluctuations method, will be reported in a subsequent paper.

\section{Infrared Imaging}

\subsection{Observations and Data Reduction}
\label{sec:obs}

The current sample of 35 galaxies was selected from among the nearest of the
72 most luminous
MASSIVE galaxies, and includes all the early-type MASSIVE galaxies
within 80 Mpc for which surface brightness fluctuation (SBF) distances
can be measured in a single orbit \citep{Jensen2015}.
Accurate distances are key to measuring the black hole--galaxy mass
relationship, as well as determining the dynamical properties of the galaxy.
The current sample is limited to those MASSIVE galaxies for which accurate
SBF distances are feasible and currently unavailable in the literature.
Each galaxy was observed for one orbit using the F110W filter 
(Table~\ref{tbl:obs}). 
This paper reports the results of our surface brightness profile measurements
and the galaxy structural properties derived from them. 

Each orbit was divided into five dithered 
exposures (four galaxies received an additional shorter exposure when
it could be scheduled during the orbit). 
Two galaxies (NGC~545 and NGC~547) were observed simultaneously.
The total exposure time for each galaxy is listed in Table~\ref{tbl:obs}. 
The dithered exposures used a five point sub-pixel dither pattern
to improve sampling of the point-spread function (PSF); 
the pixel scale at F110W is 0.128 
arcsec~pix$^{-1}$, which is slightly undersampled for this wavelength.
Properties for each galaxy can be found in Table \ref{tbl:obs}.

\begin{deluxetable*}{crrcccccr}
\centering
\tabletypesize{\scriptsize}
\tablecaption{Properties for Observed Galaxies}
\tablewidth{0pt}
\tablehead{
\colhead{Galaxy} & \colhead{R.A.} & 
\colhead{Dec.} & \colhead{$D$} & 
\colhead{$M_{K}$} & \colhead{$A_{V}$} & 
\colhead{Exposure} & \colhead{Background} & 
\colhead{Background}\\
\colhead{} & \colhead{(deg)} & 
\colhead{(deg)} & \colhead{(Mpc)} & 
\colhead{(mag)} & \colhead{(mag)} & 
\colhead{(sec)} & \colhead{(mag arcsec$^{-2}$)} & 
\colhead{(e$^{-}$s$^{-1}$pix$^{-1}$)}\\
\colhead{(1)} & \colhead{(2)} & 
\colhead{(3)} & \colhead{(4)} & 
\colhead{(5)} & \colhead{(6)} & 
\colhead{(7)} & \colhead{(8)} & 
\colhead{(9)}
}
\startdata
NGC~0057 & $003.8787$ & $17.3284$ & $76.3$ & $-25.75$ & $0.212$ & $2496$ & $22.00$ & $1.39\pm 0.08$\phn \\
NGC~0315 & $014.4538$ & $30.3524$ & $70.3$ & $-26.30$ & $0.177$ & $2496$ & $21.77$ & $1.72\pm 0.23$\phn \\
NGC~0383 & $016.8540$ & $32.4126$ & $71.3$ & $-25.81$ & $0.194$ & $2496$ & $21.94$ & $1.47\pm 0.11$\phn \\
NGC~0410 & $017.7453$ & $33.152$\phn & $71.3$ & $-25.90$ & $0.161$ & $2496$ & $21.62$ & $1.98\pm 0.07^{*}$ \\
NGC~0507 & $020.9164$ & $33.2561$ & $69.8$ & $-25.93$ & $0.170$ & $2496$ & $21.56$ & $2.08\pm 0.16^{*}$ \\
NGC~0533 & $021.3808$ & $1.759$\phn & $77.9$ & $-26.05$ & $0.084$ & $2496$ & $21.50$ & $2.21\pm 0.08$\phn \\
NGC~0545 & $021.4963$ & $-1.3402$ & $74.0$ & $-25.83$ & $0.114$ & $2496$ & $21.85$ & $1.59\pm 0.06$\phn \\
NGC~0547 & $021.5024$ & $-1.3451$ & $74.0$ & $-25.83$ & $0.113$ & $2496$ & $21.85$ & $1.59\pm 0.06$\phn \\
NGC~0665 & $026.2338$ & $10.423$\phn & $74.6$ & $-25.51$ & $0.242$ & $2496$ & $21.79$ & $1.69\pm 0.08$\phn \\
NGC~0708 & $028.1937$ & $36.1518$ & $69.0$ & $-25.65$ & $0.247$ & $2496$ & $21.84$ & $1.61\pm 0.08$\phn \\
NGC~0741 & $029.0874$ & $5.6289$ & $73.9$ & $-26.06$ & $0.144$ & $2496$ & $21.81$ & $1.65\pm 0.12$\phn \\
NGC~0777 & $030.0622$ & $31.4294$ & $72.2$ & $-25.94$ & $0.128$ & $2496$ & $21.79$ & $1.69\pm 0.08$\phn \\
NGC~0890 & $035.5042$ & $33.2661$ & $55.6$ & $-25.50$ & $0.212$ & $2496$ & $21.94$ & $1.47\pm 0.20^{*}$ \\
NGC~1016 & $039.5815$ & $2.1193$ & $95.2$ & $-26.33$ & $0.085$ & $2496$ & $21.72$ & $1.80\pm 0.14$\phn \\
NGC~1060 & $040.8127$ & $32.425$\phn & $67.4$ & $-26.00$ & $0.532$ & $2496$ & $21.74$ & $1.77\pm 0.15$\phn \\
NGC~1129 & $043.6141$ & $41.5796$ & $73.9$ & $-26.14$ & $0.309$ & $2496$ & $21.53$ & $2.15\pm 0.09$\phn \\
NGC~1167 & $045.4265$ & $35.2056$ & $70.2$ & $-25.64$ & $0.496$ & $2496$ & $21.71$ & $1.82\pm 0.08^{*}$ \\
NGC~1272 & $049.8387$ & $41.4906$ & $77.5$ & $-25.80$ & $0.441$ & $2496$ & $21.87$ & $1.57\pm 0.04$\phn \\
NGC~1453 & $056.6136$ & $-3.9688$ & $56.4$ & $-25.67$ & $0.289$ & $2496$ & $21.65$ & $1.92\pm 0.16$\phn \\
NGC~1573 & $068.7666$ & $73.2624$ & $65.0$ & $-25.55$ & $0.377$ & $2895$ & $21.06$ & $3.29\pm 0.14^{*}$ \\
NGC~1600 & $067.9161$ & $-5.0861$ & $63.8$ & $-25.99$ & $0.118$ & $2496$ & $21.78$ & $1.70\pm 0.19^{*}$ \\
NGC~1684 & $073.1298$ & $-3.1061$ & $63.5$ & $-25.34$ & $0.159$ & $2496$ & $21.83$ & $1.62\pm 0.08$\phn \\
NGC~1700 & $074.2347$ & $-4.8658$ & $54.4$ & $-25.60$ & $0.119$ & $2496$ & $21.91$ & $1.51\pm 0.18$\phn \\
NGC~2258 & $101.9425$ & $74.4818$ & $59.0$ & $-25.66$ & $0.351$ & $2895$ & $21.32$ & $2.60\pm 0.12^{*}$ \\
NGC~2274 & $101.8224$ & $33.5672$ & $73.8$ & $-25.69$ & $0.286$ & $2496$ & $21.83$ & $1.62\pm 0.19^{*}$ \\
NGC~2513 & $120.6028$ & $9.4136$ & $70.8$ & $-25.52$ & $0.063$ & $2496$ & $21.50$ & $2.21\pm 0.06^{*}$ \\
NGC~2672 & $132.3412$ & $19.075$\phn & $61.5$ & $-25.60$ & $0.058$ & $2496$ & $21.66$ & $1.90\pm 0.19$\phn \\
NGC~2693 & $134.2469$ & $51.3474$ & $74.4$ & $-25.76$ & $0.054$ & $2695$ & $21.92$ & $1.50\pm 0.26^{*}$ \\
NGC~4914 & $195.1789$ & $37.3153$ & $74.5$ & $-25.72$ & $0.037$ & $2496$ & $22.30$ & $1.05\pm 0.08$\phn \\
NGC~5322 & $207.3133$ & $60.1904$ & $34.2$ & $-25.51$ & $0.038$ & $2745$ & $21.32$ & $2.60\pm 0.20$\phn \\
NGC~5353 & $208.3613$ & $40.2831$ & $41.1$ & $-25.45$ & $0.035$ & $2496$ & $21.72$ & $1.80\pm 0.21$\phn \\
NGC~5557 & $214.6071$ & $36.4936$ & $51.0$ & $-25.46$ & $0.016$ & $2496$ & $21.92$ & $1.49\pm 0.08^{*}$ \\
NGC~6482 & $267.9534$ & $23.0719$ & $61.4$ & $-25.60$ & $0.277$ & $2496$ & $22.41$ & $0.95\pm 0.04$\phn \\
NGC~7052 & $319.6377$ & $26.4469$ & $69.3$ & $-25.67$ & $0.337$ & $2496$ & $22.36$ & $1.00\pm 0.28$\phn \\
NGC~7619 & $350.0605$ & $8.2063$ & $54.0$ & $-25.65$ & $0.224$ & $2496$ & $21.58$ & $2.05\pm 0.06$\phn \\
\enddata

\tablecomments{
  Column (1) galaxy name. 
  Column (2) right ascension in degrees (J2000.0).
  Column (3) declination in degrees (J2000.0).
  Column (4) redshift distances from \citet{Ma2014}. NGC~1016 is included in this sample because it has Tully-Fisher and $D_n$-$\sigma$ distances near ${\sim}80$ Mpc \citep{theureau2007,Willick1997}, which is significantly lower than the group redshift distance tabulated here.
  Column (5) extinction-corrected total absolute $K$-band magnitude from \citet{Ma2014}.
  Column (6) foreground galactic extinction in Landolt $V$-band \citep{sf11} 
with reddening relation of \citet{Fitzpatrick1999}.
Column (7) total exposure time in seconds. 
Column (8) background value, determined in Section \ref{sec:background}. 
Column (9) background value in the native WFC3/IR resolution. 
Starred galaxies were affected by the variable He emission background.}
\label{tbl:obs}
\end{deluxetable*}
The WFC3/IR images were first reduced using the standard data
reduction pipeline, as outlined in the STScI data handbook for 
WFC3/IR.\footnote{\url{http://www.stsci.edu/hst/wfc3/pipeline/wfc3\_pipeline}} 
Approximately $40\%$ of the individual F110W exposures were affected
by the diffuse upper-atmosphere He emission line at 1.083~$\mu$m
\citep{Brammer2014}, resulting in an elevated background.
During each MULTIACCUM exposure, multiple nondestructive reads of the
detector were saved, allowing us to determine the rate at which signal
accumulated in each pixel and therefore identify pixels that were
affected by saturation or cosmic rays, which could be identified as a
change in the count rate of electrons being detected in a pixel.
The He emission, which appears as a variable background,
also causes the signal rate to change nonlinearly during an individual 
MULTIACCUM exposure.  

Each raw image was run through a specialized Python program\footnote{\url{https://github.com/gbrammer/wfc3}}
written by Gabriel Brammer at STScI to correct for variable
background levels and other events in MULTIACCUM sequences (like satellite passages)
that affected the flux in a large number of pixels at
once.
The correction algorithm starts by comparing the count rate over a
large region of the detector in the first and last reads of a
MULTIACCUM sequence, and then flattens the sequence of reads so that
each individual subframe has the same average background count rate.
We used Brammer's program to identify individual subframes that had varying
background and re-processed all images to have flattened (linear)
background count rates. 
This process does not adjust the median sky level, so some background
values are still higher than would otherwise be expected 
in the absence of the elevated He background (see Table~\ref{tbl:obs}). 
The individual exposures were then combined using the STSDAS PyRAF
task \emph{Astrodrizzle}. 
During stacking, we rescaled our images from the native 
$0.128$ arcsec~pix$^{-1}$ to $0.10$ arcsec~pix$^{-1}$ to improve spatial 
resolution, which is possible due to the fractional-pixel offsets of the
dither pattern.
The following parameters were changed from their default
values in order to rescale the final images: 
\begin{alignat*}{4}
  & \mathtt{skywidth} &&= 0.1\:\:&&\:\: \mathtt{skystat} &&= \mathtt{mode}\\
  & \mathtt{dirz\_sep\_kernel} &&= \mathtt{square}\:\:&&\:\: \mathtt{dirz\_sep\_bits} &&= 4928\\
  & \mathtt{combine\_maskpt} &&= 0.7\:\:&&\:\: \mathtt{combine\_type} &&= \mathtt{median}\\
  & \mathtt{driz\_cr\_corr} &&= \mathtt{True}\:\:&&\:\: \mathtt{driz\_cr\_snr} &&= 10.0\:8.0\\
  & \mathtt{final\_kernel} &&= \mathtt{gaussian}\:\:&&\:\: \mathtt{final\_pixfrac} &&= 0.6\\
  & \mathtt{final\_bits} &&= 4928\:\:&&\:\: \mathtt{final\_wcs} &&= \mathtt{True}\\
  & \mathtt{final\_scale} &&= 0.10 \:\:&&\:\: \\
\end{alignat*}

\subsection{Background Measurements}
\label{sec:background}

The MASSIVE galaxies in our sample are extended enough that, in most cases,
the entire field of view contained light from the galaxy that was a significant
fraction of the instrumental background. 
As a consequence, the standard background subtraction procedure
performed by \emph{Astrodrizzle}, which is based on the median value in the 
field of view, oversubtracted the sky. 
In order to accurately measure the surface brightness profiles of each
galaxy, we iteratively compared our measured profiles using different 
background offsets to $r^{1/4}$ models, Sersic profiles from the NASA-Sloan
Atlas\footnote{\url{http://www.nsatlas.org}} (NSA), and empirical
Two Micron All-Sky Survey (2MASS) $J$-band profiles.  

A first background estimate was made by measuring the median value in
the darkest corners of each image. 
This value provided an absolute upper limit on the background
value and provided an initial background estimate.  
We then created our initial elliptical model of each galaxy using two
different programs, ELLIPSE and ELLIPROF.
Each program determined the surface brightness profile in annuli that
were allowed to vary in ellipticity, ellipse center, and PA. 
ELLIPSE utilizes the fitting framework outlined by \citet{Carter1978}, and
later modified for full isophote fitting by \citet{Jedrzejewski1987}.
The intensity profile model $I(\phi)$ is constructed from the mean
intensity $\left< I_{ell} \right>$ along an elliptical path with Fourier
series azimuthal perturbations along the elliptical path:
\begin{equation}
  \label{eq:isophote}
  I(\phi) = \left< I_{ell} \right> + \sum_{n}\left( a_{n}\sin(n\phi) + b_{n}\cos(n\phi)\right)
\end{equation}
where $\phi$ is the azimuthal angle.
ELLIPSE fits the purely elliptical patterns first by minimizing the
residual between the model and data.
Higher order deviations are fitted sequentially.
We also utilized the galaxy isophote modeling task ELLIPROF
\citep{Tonry1997, Jordan2004}. ELLIPROF is similar to ELLIPSE
but provides an independent comparison for testing the robustness of
the fit parameters.
It also provides an extrapolation beyond the outer annulus, 
which is convenient for situations where a sharp cut-off in the model is
undesirable, such as in the iterative fitting of multiple galaxies in the
field of view.
Many galaxies have nearby neighbors in the frame, which
frequently skewed the isophotal fits.
To correct for bright companions, we used ELLIPROF to iteratively fit
each galaxy until the residuals after galaxy subtraction were minimized
within the field of view.
Stars and other bright objects in the frame were masked prior to
fitting with ELLIPROF.

The surface brightness profiles were then compared to Sersic and
$r^{1/4}$ models to determine the residual sky background, assuming that
the galaxies follow a Sersic or $r^{1/4}$ distribution.
The Sersic profiles were fitted to the outer ten isophotes of the surface
brightness profile with the intensity as the only free variable.
The $r^{1/4}$ profiles were fitted with both the effective radius $r_{e}$
and intensity as free variables.
The background estimate adopted for each iteration and for each profile 
was that for which the residuals were minimized.
This procedure was performed independently by two members of our
collaboration, one using ELLIPSE and the other using ELLIPROF, 
to determine the background levels.

As a third independent measurement of the background, we compared our F110W
surface brightness profiles to 2MASS $J$-band images 
\citep{Jarrett2000}.\footnote{2MASS profiles were kindly provided 
by John Lucey.} 
The 2MASS profiles were measured in a comparable way to our F110W
profiles in elliptical annuli after subtracting the sky value
determined much farther from the galaxies than is possible in the WFC3/IR
images. 
Generally speaking, the F110W and 2MASS $J$-band profiles agree very
well. 
In some cases the profiles were offset from one another, usually when
bright companion galaxies affected the surface brightness profiles. 
The procedure for removing companions from the 2MASS images was to
automatically identify and mask objects using SExtractor
\citep{Bertin1996}, which was less effective when companions were
large and close.

The final background measurements and their uncertainties were adopted after 
carefully considering both the comparison with 2MASS surface brightness 
profiles and the background levels that we found from Sersic and $r^{1/4}$
extrapolations. 
The Sersic, $r^{1/4}$, and 2MASS profiles all had their own 
uncertainties associated with their best fits.
To determine the best background value, we adopted the preferred value from the Sersic or $r^{1/4}$ profile,
depending on which was the closest fit to the surface brightness profile. 
In 17 cases, the $r^{1/4}$ profile was either the better or only fit
available.
The process of determining the best background level involved the following
criteria:
\begin{enumerate}
\item If the Sersic or $r^{1/4}$ profile background measurement was within the
  2MASS profile's uncertainty range, the Sersic or $r^{1/4}$ fit value
  for the background was adopted (17 galaxies).
\item If the Sersic or $r^{1/4}$ profile background measurement was outside 
  the 2MASS profile's uncertainty range, the background value was increased
  or decreased until it fell within the uncertainty range of both
  fits (12 galaxies).
\item In cases where a bright companion was present, the 2MASS profile was
  disregarded when determining the background value. Instead, the
  value was solely determined by the profile fits to the WFC3/IR
  images (four galaxies).
\item If the background measurement found in the corner of the image
  was lower than the value determined through the system above, the
  corner value was adopted, as this value functioned as a boundary
  condition on our background measurements (two galaxies). 
\end{enumerate}

The scatter in the different estimates of the instrumental background
served as a measurement of the uncertainty (see Table~\ref{tbl:obs}). 
The adopted sky levels were then applied to the final stacked images
from \emph{Astrodrizzle} before measuring the final surface brightness 
profiles.
The mean background level we measured for this sample was $1.78\,{\pm}\,0.18$
e$^-$s$^{-1}$pix$^{-1}$ for the native WFC3/IR average pixel scale 
(0.128 arcsec~pix$^{-1}$), or
$21.77\,{\pm}\,0.11$ AB~mag\,arcsec$^{-2}$. 
The uncertainty for each background measurement was estimated by
calculating the standard deviation of four independent background
measurements: the value derived from the 2MASS $J$-band profile, 
the NSA Sersic value, and two independent
$r^{1/4}$ profile values (the first using the $0.10$ arcsec~pix$^{-1}$
\emph{Astrodrizzle} images and ELLIPSE, and the other using the images 
in the original $0.128$ arcsec~pix$^{-1}$ resolution with no geometric 
distortion correction and ELLIPROF). 
In eight cases, the standard deviation was high due to an outlier in
the four independent measurements.
These cases were recalculated without the outlier (though the outlier
was still considered when determining the background measurement
itself).
In five cases, the standard deviation was lower than expected
based on the empirical estimates of the uncertainty from the $r^{1/4}$ fits
(which were performed first), often in cases where one of the independent 
measurements had been excluded as an outlier. 
In these cases the empirical uncertainty was used instead of the
standard deviation. 
Overall, the average uncertainty for our background measurements is
$0.13$~e$^{-}$s$^{-1}$pix$^{-1}$ in the native $0.128$~arcsec~pix$^{-1}$ 
resolution. This compares favorably with 
the standard deviation of the differences between the different background
estimates, or
$0.11$~e$^{-}$s$^{-1}$pix$^{-1}$.
The variation in background levels between different observations is
$0.5$~e$^{-}$s$^{-1}$pix$^{-1}$. 
These values are consistent with the published range of values for
WFC3/IR from zodiacal light \citep{Pirzkal2014} and the He emission.
Final estimated background levels are listed in Table \ref{tbl:obs}.

\subsection{Neighbor Subtraction}
\label{sec:nearsub}

Many of the target galaxies have nearby neighbors in the field of view.
Due to their size and position, these galaxies could not be masked
without significantly distorting the final surface brightness profile
fits. 
Instead, we iteratively modeled these galaxies using ELLIPROF. 
An initial model for the main galaxy was created and subtracted before fitting
the neighbor galaxy. The model of the neighbor galaxy was then subtracted 
from the original image, and the target galaxy fitted again. 
This procedure was repeated until the models for both galaxies converged and
the residual structure in the background was minimized.
Although we used PyRAF's ELLIPSE and BMODEL for fitting the final surface
brightness profiles and generating the models, we found that ELLIPROF did a
much better job with the iterative fitting of multiple galaxies because,
unlike ELLIPSE, it extrapolates the galaxy models to the full field of view.
The iterative fitting of neighbor galaxies was performed both as part of
the measurement of the instrumental background and the final surface 
brightness measurements of the target galaxies.
 
\subsection{Mask Creation}
\label{sec:masking}

After the background values were measured and any neighbor galaxy models 
generated, we subtracted both from the images and created masks for other
objects in the field of view. 
Due to the size of the galaxies, we required an iterative
procedure to identify and mask sources within the bright interior of
each galaxy. 
The first iteration of the mask was generated by using the CTX image
extension produced by \emph{Astrodrizzle}, which identifies bad pixels in 
the image. 
Next, we fed the mask and residual image $R_{0}$ (the image with background
and nearby neighbors subtracted) into SExtractor \citep{Bertin1996}
to detect
objects in the image.  
The first SExtractor iteration was used to identity large, 
bright objects in the outer regions of the galaxy image. 
SExtractor identified an object if at least six contiguous pixels 
deviated from the median grid value by more than $4\sigma$.
Since the center of the galaxy was always identified as an object, 
no objects within 5~arcsec of the center were masked.
All remaining pixels were combined with the previous
iteration of the mask to form the new residual image.
ELLIPSE was then run on the image with the SExtractor objects masked, 
generating a set of isophotes that were used by the PyRAF task BMODEL to
create a model galaxy image.
BMODEL does not generate values outside the largest fitted
isophote, so we extrapolated the model by fitting a power law to the
outer five isophotes.
The new galaxy model was then subtracted from the initial residual image 
to form an updated residual image $R_{1}$ for subsequent object identification
and masking.

Next, we fed the $R_{1}$ residual image and mask into SExtractor a
second time, this time with the goal of identifying bright
extended sources closer to the target galaxy center that were missed 
during the first SExtractor run.
For the second iteration, the detection threshold was set at $11\sigma$, and
the minimum area was held at six pixels. 
Objects found within 5~arcsec of the center were ignored to prevent
the center of the galaxy from being masked.  
Objects masked by this iteration were added to the previous mask
to create the penultimate version.

The new mask was used to create an updated galaxy model and residual
image $R_{2}$, which was analyzed with SExtractor a final time.
With the brightest sources in the interior masked, the galaxy model was
finally good enough so that fainter point sources close to the galaxy
center could be identified.
We set the SExtractor detection threshold to $4\sigma$ and the
minimum object area to five pixels, and objects within 5~arcsec 
of the galaxy center were ignored.
The last iteration of masking could introduce erroneous masking around
the edge of the largest isophote fitted by ELLIPSE.
To avoid this, we combined the first and last iterations of the mask;
the first iteration was used for the region outside the largest
isophote, and the last iteration was used for the region within.
Finally, to make sure we masked all the light from the objects, we 
expanded the footprint of each masked object by smoothing
the point source mask using a $\sigma = 0.6$-pixel Gaussian and masking
any fractional pixel values.

After the background was subtracted, neighboring galaxies removed,
and the object mask created, we ran each galaxy through ELLIPSE one
last time to generate a final galaxy model using BMODEL and a final 
residual image $R_{f}$.
The galaxy images, isophotal models, and residual images are shown
in Figure~\ref{fig:combined}.
The fourth panel in Figure~\ref{fig:combined} (and subsequent figures
in the Appendix) shows the magnified central 
region of the original image to show the nuclear structure including 
dust and other central features.
The isophotes were generated using CMODEL, a correction to
the BMODEL task by \citet{Ciambur2015} that correctly utilizes the
higher-order moments when generating the model.

\section{Photometric Measurements}
\label{sec:results}

\subsection{Surface Brightness Profiles}
\label{sec:profiles}

\begin{figure*}[!tbp]
  \centering\offinterlineskip
  \begin{minipage}[b][13.55cm][t]{0.56\textwidth}
    \includegraphics[width=\textwidth]{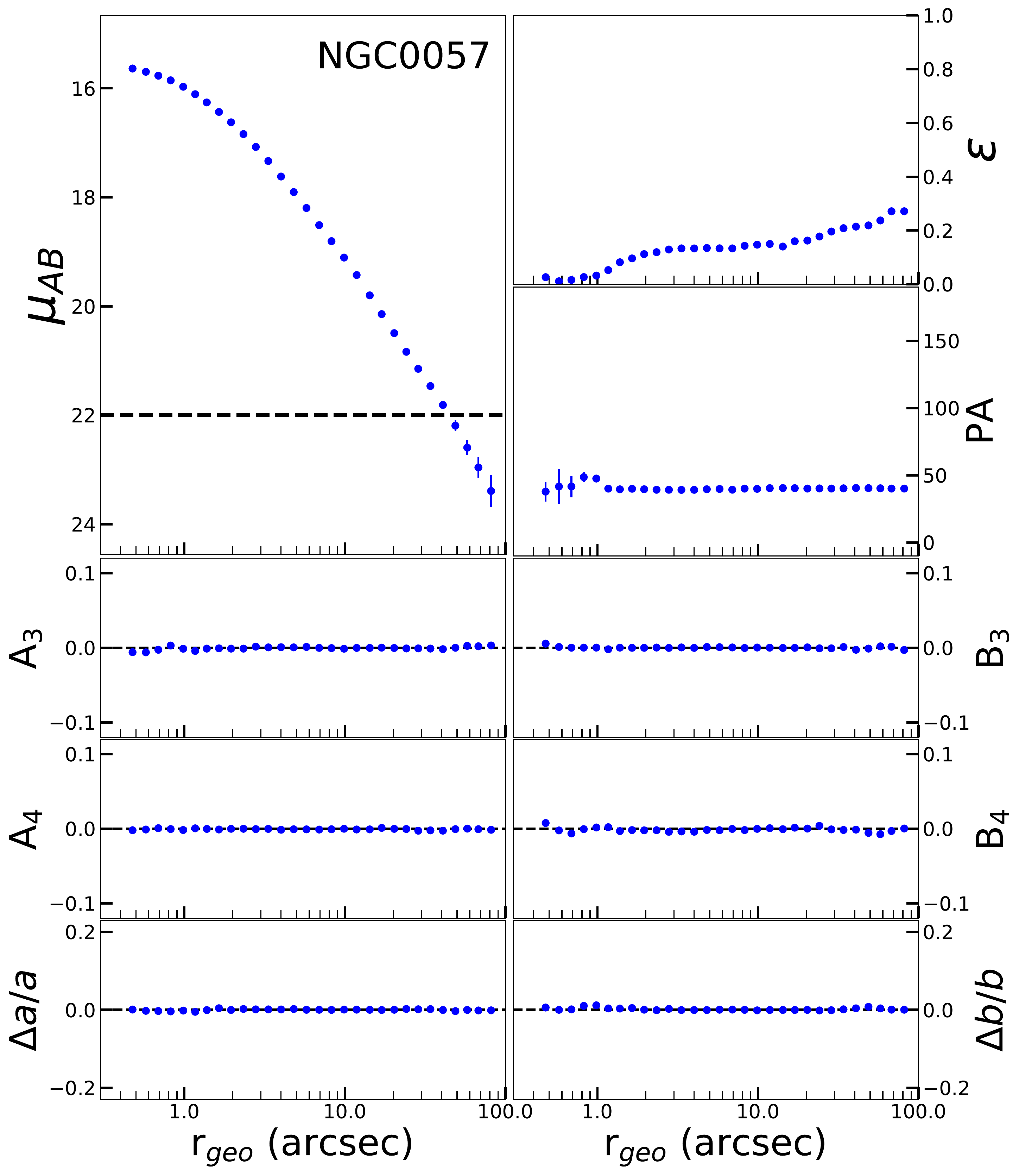}
  \end{minipage}
  \begin{minipage}[b][13.45cm][t]{0.41\textwidth}
    \includegraphics[width=\textwidth,trim=0 0 0 0]{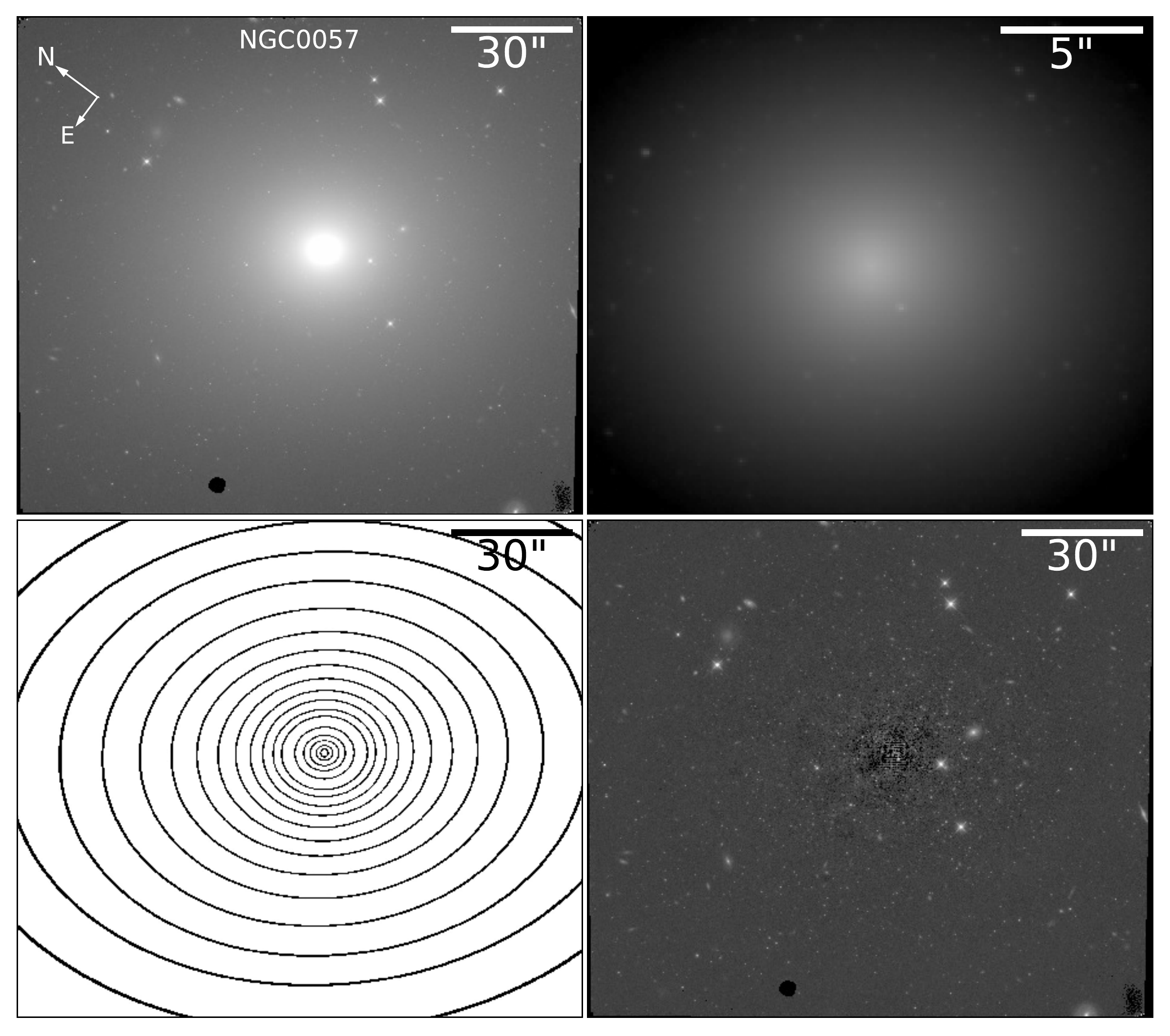}
    \vspace{0.1cm}
    \caption{\small Images and radial plots for NGC~57. The scale is
      $1$ arcsec = $370$ pc. The images are (\emph{Upper Left}) the
      F110W band image, (\emph{Lower Left}) the isophotes derived
      from the surface brightness profile model
      (semi-major axis increases by 15\% between isophotes),
      (\emph{Upper Right}) a closer view of the center of the F110W
      image (scaled to make central features, such as dust, more
      visible), and (\emph{Lower Right}) the residual image with galaxy
      models subtracted. The radial plots in the left panel show isophotal 
      parameters as a function of the
      geometric mean radius, $r_{geo} = \sqrt{ab}$, where $a$ is the
      semi-major axis and $b$ is the semi-minor axis. The horizontal black
      dotted line in the surface brightness profile shows the
      background measurement. 
      Radial plots and images for the remainder of the sample
      are displayed in the Appendix.}
    \label{fig:combined}
  \end{minipage}\\
\end{figure*}

The surface brightness profiles fitted using ELLIPSE are shown in
Figure~\ref{fig:combined} for NGC~57.
The profiles and images for the other 34 galaxies can be found in the
Appendix.
In addition to the surface brightness profile, eight parameters are
plotted for each galaxy as a function of radius that describe the
shape of the isophotes: position angle, ellipticity $\varepsilon$, the
higher order deviations from ellipticity $A_{3}$, $B_{3}$, $A_{4}$,
and $B_{4}$, and finally the isophotal drifts from the center of the
galaxy along the semi-major and semi-minor axes, $\Delta a$ and
$\Delta b$, respectively.
The higher order deviations from ellipticity used here are defined as
\begin{equation}
  \label{eq:an_bn}
  A_{n} = \frac{a_{n}}{a \nabla I}\ , \:\:\:\:\:\: B_{n} = \frac{b_{n}}{a \nabla I}
\end{equation}
where $a_{n}$ and $b_{n}$ are defined in Equation\,\ref{eq:isophote},
$\nabla I$ is the intensity gradient at a particular isophote, and $a$
is the semi-major axis.
Intensity- and luminosity-weighted values for ellipticity, PA, 
and $B_{4}$ are tabulated in Table \ref{tbl:par_shift_mod}.
In addition to overall luminosity-weighted values, the behavior of
these parameters has been examined within three different radial regimes:
an inner regime extending from $0.4$ arcsec to $0.1r_{s}$, 
an intermediate regime extending from $0.1r_{s}$ to $r_{s}$, 
and an outer regime comprised of all points beyond $r_{s}$.
The parameter $r_{s} = 7.5$ kpc was chosen as a representative physical
scale based on typical $R_e$ values from 2MASS and NSA fits to the galaxies 
in our sample.

\subsection{Radial Variations of Isophotal Parameters}
\label{sec:par_shift}

\begin{deluxetable*}{ccccrrrrccc} 
\centering
\tabletypesize{\scriptsize}
\tablecaption{Parameter Shifts and Luminosity-weighted Values}
\tablewidth{0pt}
\tablehead{
\colhead{Galaxy} & \colhead{$r_{s}$} & 
\colhead{$\Delta PA_{tot}$} &
\colhead{$ \langle \Delta r \rangle_{L}$} & 
\colhead{$ \langle \epsilon \rangle_{L}$} &
\colhead{$ \langle PA \rangle_{{L}}$} &
\colhead{$\langle B_{4} \rangle_{I}$} &
\colhead{$\langle B_{4} \rangle_{L}$} &
\colhead{Structural} & \colhead{Type} &
\colhead{Group} \\
\colhead{} & \colhead{(arcsec)} & 
\colhead{(Deg)} & 
\colhead{} & \colhead{} & 
\colhead{(Deg)} & \colhead{$\times 100$} &
\colhead{$\times 100$} &
\colhead{Features} & \colhead{} &
\colhead{} \\
\colhead{(1)} & \colhead{(2)} & 
\colhead{(3)} & \colhead{(4)} & 
\colhead{(5)} & \colhead{(6)} & 
\colhead{(7)} & \colhead{(8)} & 
\colhead{(9)} & \colhead{(10)} &
\colhead{(11)}
}
\startdata
NGC~0057 & $21.3$ & \phn$5$ & $0.0023$ & $0.17 \pm 0.002$ & $40.2 \pm 0.5$ & $0.055 \pm 0.062$ & $-0.095 \pm 0.014$ & & E & 1 \\
NGC~0315 & $22.0$ & \phn$7$ & $0.0049$ & $0.27 \pm 0.001$ & $44.3 \pm 0.2$ & $-1.1 \pm 0.041$ & $-0.95 \pm 0.018$ & D & E & 6B \\
NGC~0383 & $21.7$ & $13$ & $0.0045$ & $0.19 \pm 0.001$ & $138.9 \pm 0.3$ & $0.17 \pm 0.013$ & $0.3 \pm 0.013$ & DN & E-S0 & 29 \\
NGC~0410 & $21.7$ & \phn$0$ & $0.013$\phn & $0.24 \pm 0.001$ & $35.8 \pm 0.9$ & $-0.0093 \pm 0.048$ & $0.99 \pm 0.02$\phn & & E & 29B \\
NGC~0507 & $22.2$ & $71$ & $0.03$\phn\phn & $0.16 \pm 0.002$ & $52.1 \pm 0.9$ & $0.12 \pm 0.049$ & $0.4 \pm 0.02$\phn & & E-S0 & 35B \\
NGC~0533 & $19.9$ & $10$ & $0.0068$ & $0.27 \pm 0.001$ & $49.3 \pm 0.2$ & $-0.097 \pm 0.048$ & $-0.013 \pm 0.012$ & & E & 3B \\
NGC~0545 & $20.9$ & $22$ & $0.017$\phn & $0.29 \pm 0.002$ & $57.2 \pm 0.7$ & $-0.4 \pm 0.06$\phn & $-0.072 \pm 0.02$\phn & N & E-S0 & 35B \\
NGC~0547 & $20.9$ & $36$ & $0.014$\phn & $0.14 \pm 0.001$ & $98.8 \pm 1.4$ & $-0.71 \pm 0.099$ & $-0.25 \pm 0.019$ & N & E & 35 \\
NGC~0665 & $20.7$ & $74$ & $0.019$\phn & $0.28 \pm 0.002$ & $102.4 \pm 1.4$ & $0.16 \pm 0.053$ & $1.2 \pm 0.061$ & D & S0 & 4B \\
NGC~0708 & $22.4$ & $82$ & $0.018$\phn & $0.32 \pm 0.002$ & $39.8 \pm 0.2$ & $0.14 \pm 0.026$ & $-0.024 \pm 0.031$ & DN & E & 39B \\
NGC~0741 & $20.9$ & $36$ & $0.071$\phn & $0.16 \pm 0.001$ & $88.0 \pm 1.1$ & $0.14 \pm 0.051$ & $0.044 \pm 0.021$ & N & E & 5B \\
NGC~0777 & $21.4$ & \phn$1$ & $0.0028$ & $0.16 \pm 0.001$ & $148.6 \pm 0.8$ & $-0.25 \pm 0.05$\phn & $-0.26 \pm 0.011$ & & E & 7B \\
NGC~0890 & $27.8$ & \phn$3$ & $0.0079$ & $0.39 \pm 0.002$ & $53.7 \pm 0.3$ & $-0.75 \pm 0.05$\phn & $-1.9 \pm 0.02$\phn & & E-S0 & 1 \\
NGC~1016 & $16.3$ & $10$ & $0.0068$ & $0.06 \pm 0.001$ & $42.8 \pm 1.0$ & $-0.14 \pm 0.043$ & $-0.17 \pm 0.012$ & & E & 8B \\
NGC~1060 & $23.0$ & $23$ & $0.012$\phn & $0.22 \pm 0.001$ & $74.8 \pm 0.4$ & $-0.094 \pm 0.046$ & $-0.46 \pm 0.012$ & & E-S0 & 12B \\
NGC~1129 & $20.9$ & $81$ & $0.043$\phn & $0.17 \pm 0.002$ & $61.7 \pm 0.9$ & $-0.14 \pm 0.028$ & $-0.047 \pm 0.026$ & NC & E & 33B \\
NGC~1167 & $22.0$ & $17$ & $0.014$\phn & $0.17 \pm 0.002$ & $71.2 \pm 0.8$ & $-1.1 \pm 0.083$ & $-0.17 \pm 0.033$ & & S0 & 3B \\
NGC~1272 & $20.0$ & $87$ & $0.0075$ & $0.05 \pm 0.002$ & $1.2 \pm 7.3$ & $-0.076 \pm 0.056$ & $-0.14 \pm 0.014$ & & E & 117 \\
NGC~1453 & $27.4$ & $11$ & $0.0087$ & $0.17 \pm 0.001$ & $30.1 \pm 0.2$ & $-0.32 \pm 0.061$ & $0.0084 \pm 0.011$ & & E & 12B \\
NGC~1573 & $23.8$ & \phn$4$ & $0.0046$ & $0.29 \pm 0.001$ & $31.7 \pm 0.1$ & $0.41 \pm 0.054$ & $0.22 \pm 0.011$ & & E & 15B \\
NGC~1600 & $24.3$ & \phn$5$ & $0.0082$ & $0.32 \pm 0.001$ & $8.8 \pm 0.1$ & $1.8 \pm 0.095$ & $-0.79 \pm 0.015$ & & E & 16B \\
NGC~1684 & $24.4$ & $10$ & $0.041$\phn & $0.29 \pm 0.002$ & $90.5 \pm 0.7$ & $1.5 \pm 0.047$ & $1.4 \pm 0.059$ & DN & E & 11B \\
NGC~1700 & $28.4$ & $11$ & $0.01$\phn\phn & $0.29 \pm 0.001$ & $90.6 \pm 0.3$ & $0.054 \pm 0.051$ & $0.95 \pm 0.014$ & & E & 4B \\
NGC~2258 & $26.2$ & $32$ & $0.012$\phn & $0.24 \pm 0.001$ & $150.8 \pm 1.2$ & $0.066 \pm 0.056$ & $-0.18 \pm 0.045$ & N & S0 & 3B \\
NGC~2274 & $21.0$ & $14$ & $0.039$\phn & $0.11 \pm 0.001$ & $165.0 \pm 0.2$ & $-0.42 \pm 0.055$ & $-0.15 \pm 0.013$ & & E & 6B \\
NGC~2513 & $21.9$ & $21$ & $0.019$\phn & $0.21 \pm 0.001$ & $168.8 \pm 0.7$ & $0.53 \pm 0.038$ & $0.03 \pm 0.012$ & D & E & 4B \\
NGC~2672 & $25.2$ & $33$ & \nodata & $0.15 \pm 0.002$ & $136.8 \pm 0.6$ & $0.16 \pm 0.042$ & $-0.34 \pm 0.023$ & N & E & 3B \\
NGC~2693 & $20.8$ & $64$ & $0.018$\phn & $0.27 \pm 0.002$ & $161.3 \pm 1.3$ & $-1 \pm 0.028$ & $-0.039 \pm 0.015$ & DN & E & 1 \\
NGC~4914 & $20.8$ & \phn$1$ & $0.014$\phn & $0.39 \pm 0.001$ & $154.8 \pm 0.1$ & $1.6 \pm 0.065$ & $0.53 \pm 0.016$ & & E & 1 \\
NGC~5322 & $45.2$ & \phn$0$ & $0.014$\phn & $0.34 \pm 0.001$ & $92.1 \pm 0.3$ & $1.1 \pm 0.046$ & $-0.71 \pm 0.016$ & D & E & 8B \\
NGC~5353 & $37.6$ & $26$ & $0.039$\phn & $0.50 \pm 0.003$ & $143.5 \pm 0.8$ & $-2 \pm 0.047$ & $-0.0085 \pm 0.051$ & DN & S0 & 12B \\
NGC~5557 & $30.3$ & $19$ & $0.011$\phn & $0.17 \pm 0.001$ & $90.5 \pm 0.6$ & $-0.13 \pm 0.043$ & $-0.41 \pm 0.015$ & & E & 4B \\
NGC~6482 & $25.2$ & \phn$1$ & $0.0078$ & $0.22 \pm 0.002$ & $64.2 \pm 0.5$ & $-0.64 \pm 0.048$ & $1.1 \pm 0.025$ & & E & 3B \\
NGC~7052 & $22.3$ & \phn$3$ & $0.025$\phn & $0.48 \pm 0.001$ & $63.4 \pm 0.1$ & $0.1 \pm 0.018$ & $-0.23 \pm 0.013$ & D & E & 1 \\
NGC~7619 & $28.7$ & \phn$7$ & $0.0045$ & $0.23 \pm 0.001$ & $36.8 \pm 0.2$ & $1.2 \pm 0.073$ & $0.57 \pm 0.013$ & & E & 12B \\
\enddata

\tablecomments{Column (1): galaxy name. Column (2): Scale of 7.5 kpc
  in arcsec on image. Column (3): Maximum change in PA over full
  radial range. Column (4): Luminosity-weighted central drift,
    defined as
    $\langle \Delta r \rangle_{L} = \langle \Delta r_{cen} / r_{geo}
    \rangle_{L}$ over the full radial range, where $\Delta r_{cen}$ is
    defined in Section \ref{sec:drift}. Column (5):
  Luminosity-weighted ellipticity. Column (6): Luminosity-weighted
  PA. Column (7): Intensity-weighted $B_{4}{\times}100$. Column (8):
  Luminosity-weighted $B_{4}{\times}100$.  Column (9): Features: D
  indicates dust in the center, N indicates presence of nearby
  neighbors, C indicates a double nucleus. Column (10): Morphological
  type from Hyperleda. Column (11): Number of group members in the
  2MASS HDC group catalog \citep{Crook2007}; ``B'' indicates brightest
  group galaxy. Central drift values could not be measured for
  NGC~2672.}

\label{tbl:par_shift_mod}
\end{deluxetable*}

\subsubsection{Ellipticity}
\label{sec:ell_shift}

\begin{figure}
  \centering
  \includegraphics[width=0.45\textwidth]{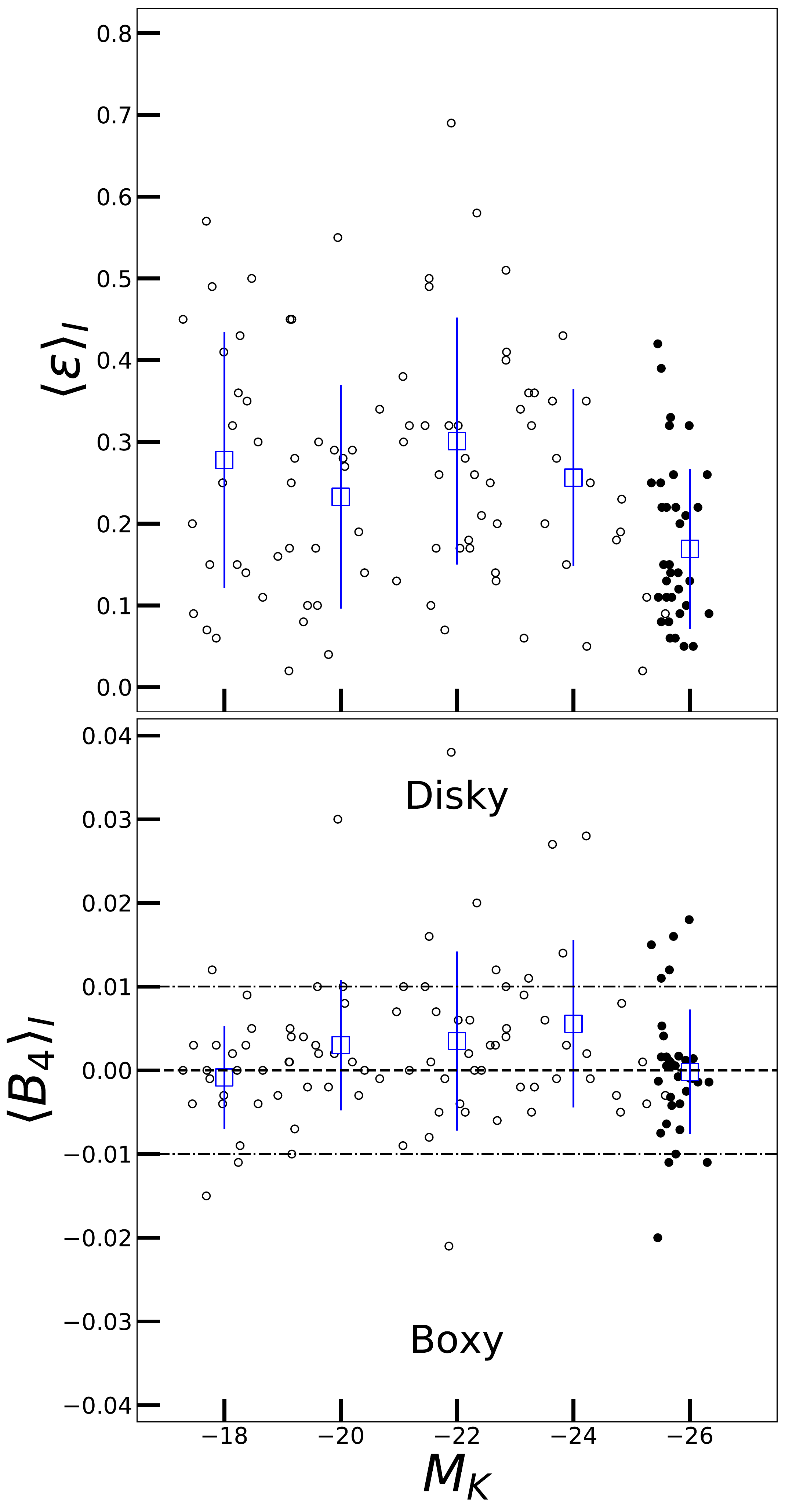}
  \caption{Intensity-weighted mean ellipticity and $B_{4}$ for each
    galaxy, plotted against $K$-band magnitude, which scales with mass. 
    Only isophotes with
    semi-major axis $1\arcsec < a < 7.5$ kpc were used.  Open circles
    are values from the ACS Virgo Cluster Survey
    \citep{Ferrarese2006}. Filled circles are values for the MASSIVE
    galaxies presented in this paper. Blue open squares are the
    average values after sorting the combined samples into 
    magnitude bins 2 mag wide. Error bars were generated from the
    scatter in each bin, and span a range of $2\sigma$.}
  \label{fig:par_mag}
\end{figure}

In their study of early-type galaxies from the ACS Virgo Cluster
Survey (ACSVCS), \citet{Ferrarese2006} concluded that there was no
trend in ellipticity with increasing stellar mass.
When combined with our sample, we find that early-type galaxies are
rounder at the high-mass end (Fig.~\ref{fig:par_mag}).
Both the average intensity-weighted ellipticity and the scatter in
ellipticity decrease with increasing mass.
\citet{Emsellem2011} found that 
slow rotators with $10^{11} M_{\odot} < M < 10^{11.5} M_{\odot}$ 
have lower ellipticities than slow rotators with $M < 10^{11} M_{\odot}$.
\citet{Emsellem2011} connected this to an observed trend that the
specific angular momentum of a slow rotator is inversely correlated
with stellar mass.
The observed behavior of ellipticity with stellar mass for our sample
is in agreement with that of \citet{Emsellem2011}, but for a sample
that begins, rather than ends, at ${\sim}10^{11.5} M_{\odot}$.

Over the radial profiles of the galaxies in this sample, isophotes tend
to be rounder towards the centers of the galaxies (see
Fig.~\ref{fig:shifts}).
Only six galaxies without central dust have higher ellipticities near
their centers than they do at larger radii.
Of these, only one (NGC~1129) has an increase in ellipticity larger
than 20\% from the value at $7.5$ kpc, and this increase is likely the
result of ELLIPSE attempting to fit isophotes to NGC~1129's double
nucleus.
In contrast, 18 galaxies without central dust have rounder interior
isophotes, and four galaxies (NGC~410, NGC~665, NGC~890, and NGC~2258)
have a decrease in ellipticity larger than $0.20$ from the value at
$7.5$ kpc.

\subsubsection{Isophotal Rotation}
\label{sec:isophotal_twists}

The degree of isophotal twisting (or change in PA) over the full
radial regime is listed in Table~\ref{tbl:par_shift_mod}.
Separate isophotal twists were also calculated for the three radial
regimes. 
Within a given radial region, the $\Delta PA$ was computed by taking
the difference of the minimum and maximum value of the PA.
The minimum and maximum values were computed by averaging the absolute
minimum and maximum PA with the two adjacent isophotes.
The isophotal twist is the difference between these characteristic
minimum and maximum values in each radial regime.
The sum of the PA twists in each radial regime sometimes overestimates
the overall twist, so we also calculated the PA twist over the
full radial range using the same method.
Using the full radial range, $37\%$ (13) of our galaxies feature a change
in PA of more than ${\sim}20^{\circ}$.

Two galaxies, NGC~1272 and NGC~2672, have significant isophotal twists
beyond $7.5$ kpc. 
NGC~2672 has a large nearby neighbor that distorts the measurement of
its outer isophotes.
The outer parts of NGC~1272 are very round (see Fig.~22), so the change in PA is probably insignificant.

Five galaxies in our sample (NGC~507, NGC~665, NGC~708, NGC~1129 and NGC~1272)
have isophotal twists at radii between $0.75$ and $7.5$ kpc.
NGC~665 is a barred S0 galaxy; its profile has discontinuities in the
ellipticity and position angle at the boundaries between the inner spheroid
and nuclear dust disk, the stellar bar, and the outer disk.
The interior of NGC~708, the cD galaxy in Abell 262, is obscured by an
extended, irregular dust structure stretching ${\sim}5$ to 10 arcsec
from the center, which creates a large and discontinuous twist of its
isophotes.
NGC~507 has an isophotal twist of $70^{\circ}$ outside of ${\sim}10$ 
arcsec without any sign of dust; \citet{Lauer2005} also observed a change 
in PA in the nucleus, inside 0.4~arcsec.
It also has a bright nearby companion to the north, just outside the WFC3/IR
field of view, which may explain the isophotal distortion.
NGC~1129, which appears to be a merger remnant with a double-peaked nucleus
(see Section \ref{sec:n1129main}), has isophotes that
twist by almost $83^{\circ}$ outside of $2$ arcsec.
This is consistent with \citet{Peletier1990}, who found a rotation of
$90^{\circ}$ over the full radial range for NGC~1129.
The very large twists in these galaxies suggest that they
are either triaxial with a favorable viewing orientation or
they have disrupted isophotes, potentially due to a
recent major merger or interaction.

Isophotal rotation measurements inside $0.75$ kpc seen in two of our galaxies
(NGC~665 and NGC~2258) are not significant due to
the presence of nuclear dust and round isophotes, respectively.

There are a few cases that show a modest, systematic twist 
of less than $20^{\circ}$ over their radial profiles.
NGC~1060, NGC~1167, NGC~1453, NGC~1573, NGC~1600, and NGC~2513 all
display consistent shifts in PA that do not have inflection points or 
discontinuities, possibly because they are triaxial, rather than because
of recent interactions with neighboring galaxies. 

\subsubsection{Isophotal Central Drift}
\label{sec:drift}

In addition to looking for twisting isophotes, we have examined our
sample for signs of drifting isophotal centers 
(Table \ref{tbl:par_shift_mod}). 
The drift $\Delta r_{cen} = \sqrt{\Delta a \Delta b}$ was calculated by
taking the luminosity-weighted average of the shift in central
position over each radial regime.
The measured drift is relative to the center of the innermost
isophote.
A significant drift of the isophotal centers may indicate that the
galaxy's spheroidal component and central black hole do not lie at the 
bottom of the potential well created by the overall stellar mass 
distribution of the galaxy.

Most of the galaxies in our sample with significant central drift 
$\Delta r_{cen} / r_{geo} \ge 0.1$ outside of $0.4$ arcsec are
galaxies with dust in their centers, which obscures or distorts 
the true centers of the interior isophotes.
Many galaxies feature potentially significant drifts deep in their
interiors (within $0.4$ arcsec), but these isophotes have been
excluded from our analysis due to being within two times the FWHM
of the PSF of the center.
Three galaxies, NGC~741, NGC~665, and NGC~1129, have significant drifts in their
outer isophotes that are not explained by the presence of dust.
The central drift in NGC~741 occurs around $10$ arcsec, and is likely
due to multiple interactions the galaxy is undergoing with nearby
neighbors.
The central drift observed in NGC~665 accompanies the sudden change in
ellipticity and $B_{4}$ that marks the transition between the stellar
bar and outer disk.
NGC~1129 shows multiple signs of a recent major merger.
If the interaction is recent enough, the outermost stellar orbits
may not necessarily have settled around the new center, which would
explain the displacement of NGC~1129's outermost isophotes.

\begin{figure*}[h]
  \centering
  \includegraphics[width=0.95\textwidth]{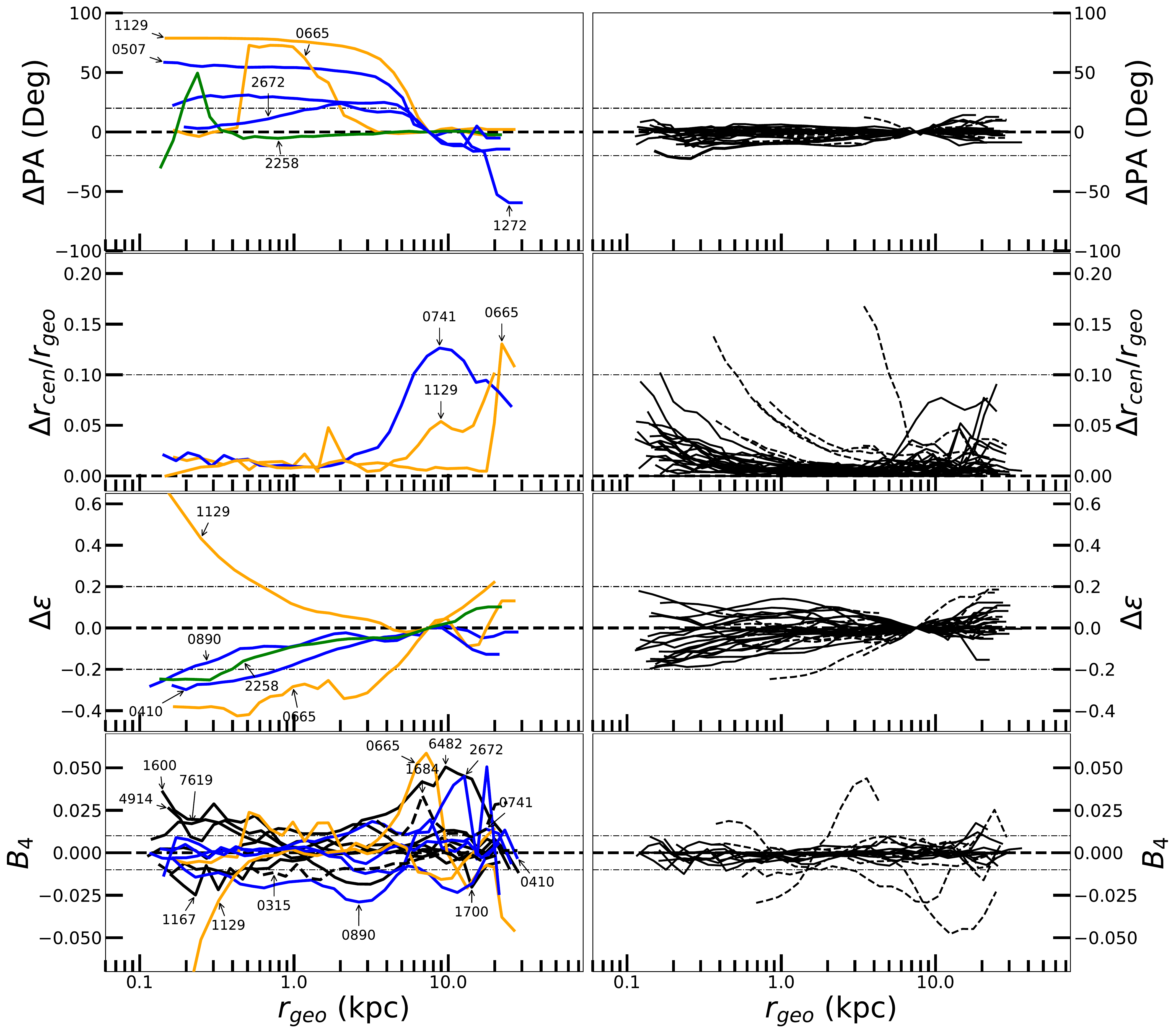}
  \caption{Parameter shifts as a function of geometric mean radius
    $r_{geo} = \sqrt{ab}$. The top panel shows the change in PA from
    the PA at a reference radius of $7.5$ kpc. The second panel shows
    the central drift $\Delta r_{cen} / r_{geo}$. The third
    panel shows the change in ellipticity from the ellipticity at the
    reference radius. The lower panel shows $B_{4}$. The line
      colors in the left plots indicate the number of parameters that
      have significant shift over their radial range. Blue, green, and
      gold lines indicate that multiple isophotes have a significant
      shift in one, two, or three parameters, respectively (not
      including the shift in $B_{4}$). Black lines indicate galaxies
    that only have a significant shift in $B_{4}$. The plots on the
    left display the profiles of the galaxies with a significant shift
    in the plotted parameter, while plots on the right show the
    profiles of the rest of the galaxies that do not have a
    significant shift. Dashed lines refer to dusty galaxies, which,
    for the most part, were not counted as having significant shifts
    due to the presence of central dust. The dot-dashed lines indicate
    the (arbitrary) limits used for deciding which galaxies to
    highlight in the left panels.}
  \label{fig:shifts}
\end{figure*}

\subsubsection{Boxy and Disky Deviations from Elliptical Isophotes}
\label{sec:b4_rot}

We also measured the boxiness or diskiness as a function of radius,
with luminosity and intensity-weighted mean values listed in Table~2.
Figure~\ref{fig:par_mag} displays the intensity-weighted ellipticity
and $B_{4}$ parameters as a function of the $K$-band magnitude, which
is a proxy for mass, using data from our sample and from the ACSVCS
sample \citep{Ferrarese2006}.
The accepted paradigm for early type galaxies is that high mass
galaxies commonly have boxy isophotes, while their lower mass
counterparts typically have disky isophotes \citep{Ferrarese1994,
  vandenBosch1994, Lauer1995, Rest2001, Lauer2005, Ferrarese2006},
where the boxiness or diskiness of a galaxy is determined by whether
the parameter $B_{4}$ is positive (disky) or negative (boxy).
Whether the deviation is significant enough to call a galaxy boxy or
disky is typically given by the condition that 
$\lvert B_{4}\,{\times}\,100 \rvert \gtrsim 1$ \citep{Jedrzejewski1987}. 

Using an intensity-weighted mean, which gives a higher weighting to
the inner isophotes, our sample contains five disky and four boxy
galaxies, where the remaining 26 galaxies have $B_{4}$ values 
consistent with purely elliptical profiles.
The average intensity-weighted $B_{4}$ for our sample is
$B_{4}= -0.00002 \pm 0.0077$, consistent with zero.
If we use a luminosity-weighted mean, which gives heavier weighting to
the outer isophotes (near a radius of $7.5$ kpc), our sample 
contains five disky and two boxy galaxies and has a mean
$B_{4} = -0.00009 \pm 0.0063$, also consistent with zero.
For comparison, the high mass ($M_{K} < -24.5$)
galaxies from the ACSVCS have a mean intensity weighted 
$B_{4} = -0.001 \pm 0.0028$.
Both our measures of mean $B_{4}$ are consistent with 
\citet{Ferrarese2006}.
We do see, however, that the mean $B_{4}$ not only shifts to slightly
more negative values when outer isophotes are given heavier weighting,
but the scatter also decreases compared to weighting schemes that
prioritize the interior isophotes. 
Overall, our sample suggests that high mass early-type galaxies, while
slightly boxy on average, 
often show a range of $B_4$ values, even within a single galaxy.
Galaxies that appear to be significantly disky or boxy with intensity
weighting may not appear so when weighted by luminosity because of
radial variations in $B_4$.
Galaxies with extreme mean $B_{4}$ values often exhibit the greatest
variation in $B_{4}$ over their full radial profile.
This can be seen in the bottom panels of Figure~\ref{fig:shifts}.
The galaxies with boxy or disky isophotes rarely maintain a consistent
$B_{4}$ value over their full profile, and instead tend to have large
variations in the value of $B_{4}$.
Only two galaxies in our sample, NGC~315 and NGC~890, have
consistently boxy isophotes over their full radial profile, and only
one galaxy, NGC~1684, has consistently disky isophotes.

\begin{figure}
  \centering
  \includegraphics[width=0.45\textwidth]{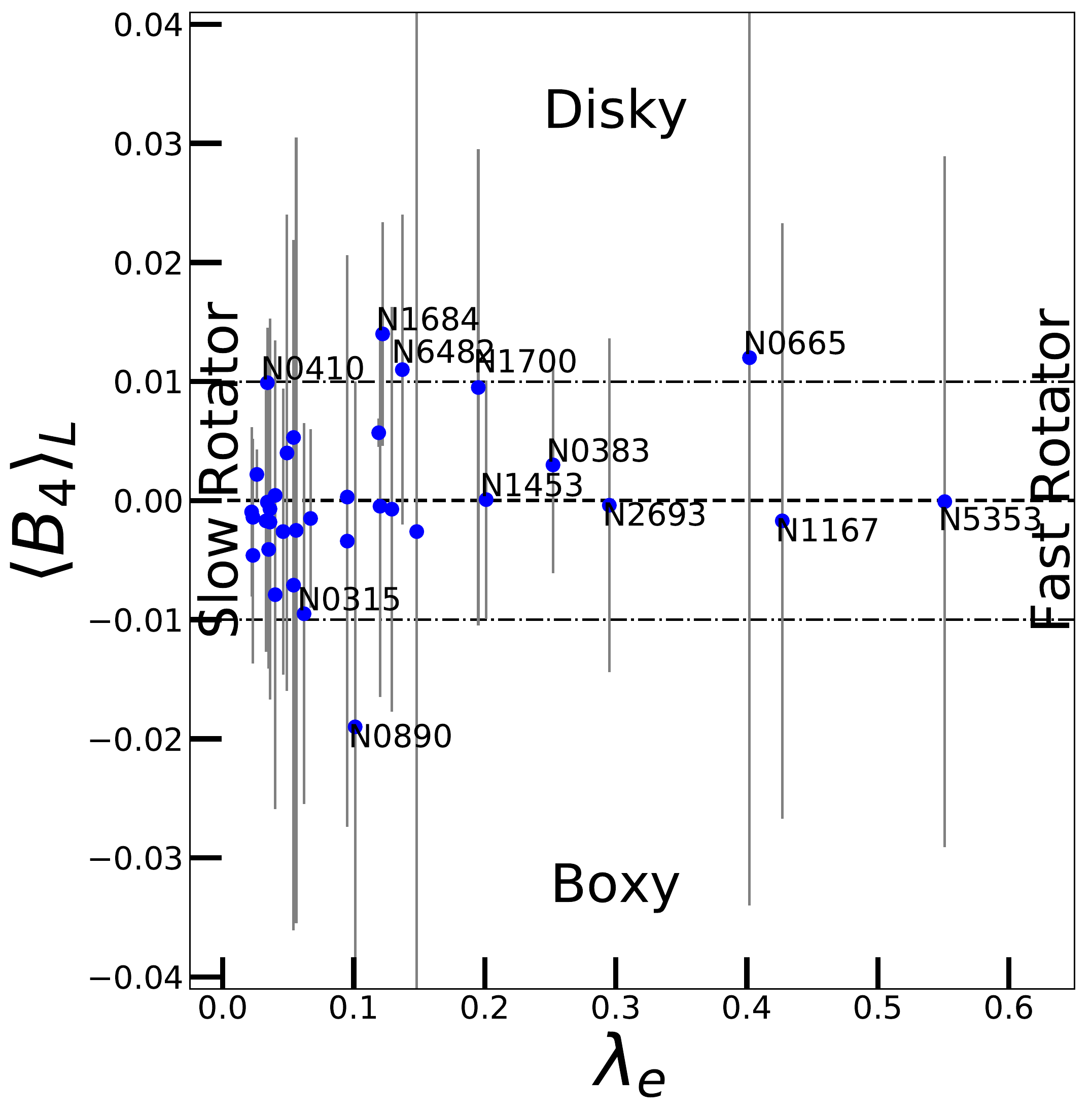}
  \caption{Plot of luminosity-weighted $B_{4}$ against the
    dimensionless spin parameter rotation criterion $\lambda_{e}$. 
    Gray bars are not error bars; they
    indicate the full range of $B_{4}$ values found in a galaxy's radial 
    profile.}
  \label{fig:b4_rot}
\end{figure}

A second aspect of the typical early type galaxy paradigm is that fast
rotators tend to be disky, while slow rotators are preferentially
boxy.
In Figure~\ref{fig:b4_rot}, we examine the relationship between
$B_{4}$ and rotation in our sample 
using the fast rotation criterion $\lambda_{e}\,{>}\,0.2$ from Paper
VII \citep{Veale2017-vii} for comparison. The dimensionless parameter
$\lambda$, the ``spin parameter,'' quantifies the rotation of a galaxy within the circularized radius $r$, and is given by
\begin{equation}
  \label{eq:lambda}
  \lambda(<r) \equiv \frac{\langle r \lvert V \rvert \rangle}{\langle r
    \sqrt{V^{2} + \sigma^{2}} \rangle} ,\,\, \lambda_e \equiv \lambda(< r_e),
\end{equation}
where $V$ is the stellar velocity and
$\sigma$ is the velocity dispersion.
All averages are weighted by luminosity. 
The value of $\lambda$ when the averages are calculated within the
kinematically determined effective radius $r_{e}$ is $\lambda_{e}$.
The spin parameter criterion for identifying fast rotators is $\lambda_{e}\,{>}\,0.2$
\citep{Veale2017-vii,Lauer2012}. 
Figure~\ref{fig:b4_rot} shows the luminosity-weighted average $B_{4}$
within $7.5$ kpc plotted against $\lambda_{e}$.
Of the six fast rotators in our sample, only one is significantly disky;
the others are consistent with purely elliptical profiles.
One other galaxy, NGC~1700, which lies on the boundary 
between fast and slow rotators,
is significantly disky.
The remaining 28 galaxies are slow rotators, and range from
significantly boxy to significantly disky.
Most slow rotators have slightly negative $B_{4}$ values, 
where 19 have $B_{4}\,{<}\,0$
and the remaining nine have $B_{4}\,{>}\,0$.
Overall, fast rotators have an average $B_{4} = 0.0022 \pm 0.0046$, while slow
rotators have an average $B_{4} = -0.0007 \pm 0.0063$, consistent with zero.
The slow rotators are purely elliptical on average, but individually
they can range from significantly disky to significantly boxy, with no
clear preference for either extreme.

\subsubsection{Connections Between Parameter Shifts}
\label{sec:conn}

The radial variation in the isophotal parameters are shown
for each galaxy individually in the Appendix.
To look for trends in variations for the sample as a whole, we plotted
the changes in PA, isophotal centers, ellipticity, and $B_{4}$ values
together in Figure~\ref{fig:shifts} as a function of radius.
Because most galaxies only vary slightly, we separated those with
significant variation from the rest for clarity.
Galaxies with any significant isophotal shifts caused by dust were
excluded. 
Of the remaining 26 galaxies, 9 have a significant shift in PA,
center, or ellipticity.
Two of these galaxies have shifts in at least two parameters.
NGC~665 and NGC~1129 feature significant shifts in all three
parameters. NGC~665 is a barred S0 galaxy with discontinuities at the
boundaries between the stellar bar and the central regions and outer disk;
NGC~1129 has a double-peaked 
nucleus indicating that it has probably experienced a recent merger.
No obvious patterns emerge, and Figure~\ref{fig:shifts} shows that the
galaxies in this sample exhibit a rich variety of profiles, with
individual galaxies having photometric structures that are highly
individualistic and specific to their formation histories.

\subsection{Dust Morphologies}
\label{sec:dust}

Dust features are common in the interiors of early-type
galaxies \citep{Ferrarese2006, Lauer2005, Tran2001, Tomita2000}. 
Of the 35 galaxies in our study, ten have
clearly visible dust features in their centers (Table~2).
Several previous studies found dust in about half of the early-type
galaxies.
Of the galaxies brighter than $B_{T} = 12.5$ mag in the sample of
\citet{Ferrarese2006}, 42\% were found to have dust.
\citet{Lauer2005} examined a heterogeneous sample of 77 archival 
\hst\ images
and found dust in $49\%$ of their sample.
\citet{Tran2001} examined a set of 68 early-type galaxies 
with radial velocities below $3400$ km s$^{-1}$, absolute
$V$-band magnitude less than $-18.5$, and absolute galactic latitude
above 20 degrees. 
Within this sample, $43\%$ were found to have dust.
\citet{Tomita2000} surveyed a set of 25 early-type galaxies in the \hst\ archive
imaged with WFPC2 in both $V$ and $I$-band, and found dust in $56\%$
of the galaxies.
\citet{Ferrari1999} examined a sample of 22 early-type galaxies with $B_{T} > 13$
mag and found dust in $75\%$ of their sample.
\citet{vanDokkum1995} looked at an archival \hst\ sample of 64 early-type galaxies
with $V$-band data and found a dust incidence rate of $48\%$.
Our sample has a lower incidence of dust (${\sim}29\%$) than these other studies.
Dust extinction in the near-IR is ${\sim}15\%$ as large as at optical 
wavelengths, which could explain why this study finds dust less frequently
than previous optical studies.
Our lower dust incidence rate may also be due to the higher mass of
this sample, as no previous studies have consistently probed galaxies
with $M_{*}\,{\gtrsim}\,10^{11.5}M_{\odot}$.

The dust features in our sample are most easily classified into three
groups:
\begin{itemize}
  \item G1: Irregular structures, such as wisps and lanes (NGC~708).
  \item G2: Medium to large, clumpy, irregular disks with typical
    radii of ${\sim}1.5$ kpc (NGC~383, NGC~665, NGC~1684, NGC~5353).
  \item G3: Small, regular, smooth disks with typical radii of
    ${\sim}0.5$ kpc (NGC~315, NGC~2513, NGC~2693, NGC~5322, NGC~7052).
\end{itemize}
It has been suggested that these groups form a dust settling sequence,
where the dust is first pulled into the center, forming the structures
in G1, then after a few orbits it settles into a more regular pattern
as in G2, and after further orbital times the rings become smoother
and more regular, finally settling into the structures seen in G3
\citep{Tran2001}.
These small disks then feed the central black hole, thereby emptying
from the inside out until the center is dust free.
In this scenario, the irregularity of the first group would decay
quickly into the more regular structures of the other two, thereby
requiring a source for the dust seen in G1.
Several authors have argued that an external source is required for
this dust \citep{Goudfrooij1994, vanDokkum1995, deKoff2000, Tran2001}.  
\citet{Ferrarese2006} suggest that the potential evolution of the
dust from G1 to G2 to G3 means that even the regular, smooth dust
rings require an external origin.
\citet{Lauer2005} argued for a local origin from evolved stars. 

We compared the occurrence of dust in our sample with the warm gas
mass in Table 1 of \citet{Pandya2017}, which has data for $12$ of 
our $35$ galaxies, of which four contain central dust.
Our comparison shows that dusty and non-dusty galaxies have similar
distributions of warm ionized gas mass.
We also compared our sample to the molecular hydrogen mass in Table 2
of \cite{Davis2016}, which contained four dusty and three non-dusty
galaxies.
Galaxies with dust in our sample have a higher molecular hydrogen
mass than non-dusty galaxies by a factor of ${\sim}5$.

\subsection{NGC~1129}
\label{sec:n1129main}

\begin{figure}
  \centering
  \includegraphics[width=0.45\textwidth]{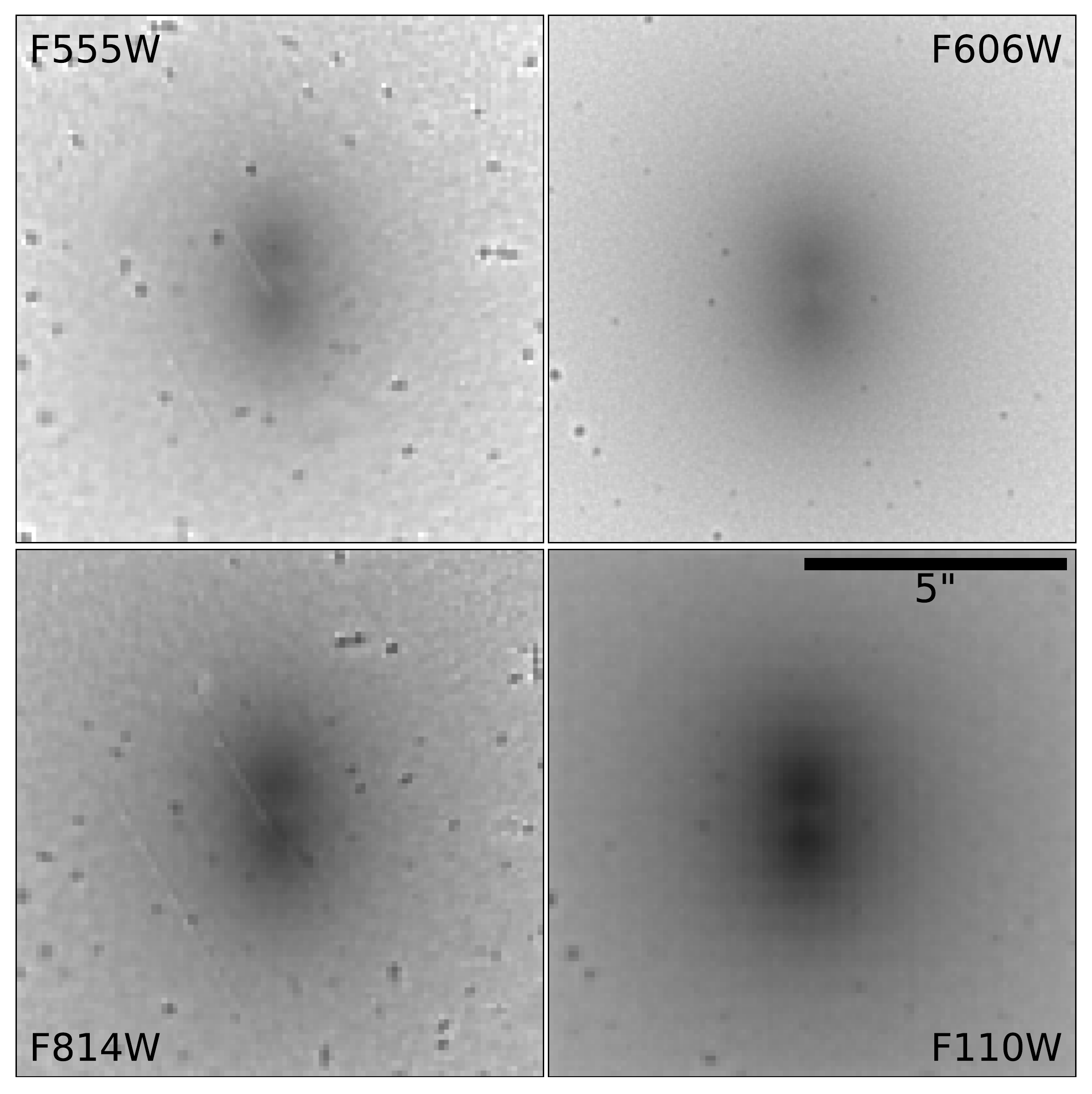}
  \caption{Images of the core of NGC~1129 in multiple wavelengths at
    the same spatial scale. The two peaks have close to the same
    brightness in all filters, and the dip in brightness between them
    does not become more or less pronounced in different wavelength
    filters. Images are shown with the same orientation and scaling, 
    and have been sharpened to enhance the visibility of the two nuclei.}
  \label{fig:n1129_nuc}
\end{figure}

The core of NGC~1129 has two nearly-identical peaks separated by 
${\sim}$1 arcsec.
To determine if this was a double nucleus or a single nucleus bisected
by a dust lane, we compared our F110W images with archival \hst\
images taken by ACS in the F435W and F606W bands,\footnote{Program GO-13698,
  PI J. Lyman} and by WFPC2 in the F555W and F814W
bands.\footnote{Program GO-6810, PI D. Geisler}
The images in Figure~\ref{fig:n1129_nuc} show no evidence of differential
extinction as a function of wavelength, leading us to conclude that
NGC~1129 has two nearly-identical nuclei with no sign of dust. 
The presence of a double nucleus supports the case that NGC~1129 has
recently undergone a major merger and the nuclei of the two galaxies
have yet to coalesce.
This would explain the extreme twisting in the outer
isophotes and the inflection point in the ellipticity, which would
naturally occur where the unmixed outer components of the progenitor
galaxies no longer dominate the light profile. 
However, we cannot rule out the possibility that the core of 
NGC~1129 has a stellar torus seen edge-on \citep{Lauer2002}.

\section{Summary}
\label{sec:summary}

Our photometric analysis of 35 of the most massive galaxies within $80$
Mpc with $M_{K}\,{<}\,{-}25.5$ leads us to several interesting conclusions
related to the evolutionary histories of these galaxies, and sets the
stage for measuring the distances and black hole masses for these
galaxies. In this study, we find that:

\begin{enumerate}
\item The ellipticities of the galaxies in our sample encompass a wide
  range of values, from a minimum intensity-weighted value of $0.08$
  to a maximum of $0.42$.
  When combined with the ACSVCS sample, our results suggest that
  elliptical galaxies become rounder as their stellar mass increases
  (Fig.~\ref{fig:par_mag}).
  Most galaxies in our sample are rounder within $7.5$~kpc of the
  center than they are farther out (Fig.~\ref{fig:shifts}).

\item In our sample, we find that $37\%$ ($13$) of the galaxies have
  isophotal rotations greater than $20^{\circ}$ over their full radial
  range (Table \ref{tbl:par_shift_mod}). 
  Galaxies that are triaxial or that have recently been disrupted by
  mergers are likely to show significant rotations of their position
  angles.
  There is probably no single explanation for the twists; in some cases, the
  PA measurements are not meaningful because the isophotes are round.
  Excluding twists due to bars, dust, or neighbors, the three galaxies
  that display significant outer isophotal twisting (NGC~507,
  NGC~1129, and NGC~1272) appear to be the result of recent mergers.

\item 
  Most of the galaxies in our sample do not have isophotes with center 
  drifts greater than $10\%$ of the semi-major axis length for that
  isophote (Fig.~\ref{fig:shifts}).
  Those that do are barred S0 galaxies, are interacting with neighbors, or
  have recently merged.

\item The average intensity-weighted boxiness of our sample is
  $B_{4}\,{=}\, {-}0.00002\,{\pm}\,0.0077$ 
  and is consistent with previous findings
  by \citet{Ferrarese2006}, who found $B_{4}\,{\sim}\,0$ at the high mass
  end of their sample.
  The addition of our higher mass sample supports the conclusion of
  \citet{Ferrarese2006} that $B_{4}$ decreases towards zero
  from positive (diskier) values as stellar mass increases. 
  When segregated by luminosity-weighted spin parameter (Fig.~\ref{fig:b4_rot}),
  we find that all the fast rotators are either significantly disky
  ($B_{4}\,{>}\,0.01$), or have isophotes that are close to elliptical
  ($0.01\,{>}\,B_{4}\,{\gtrsim}\,0$).
  Overall, fast rotators are slightly disky on average, but with a
  luminosity-weighted average $B_{4} = 0.0022 \pm 0.0046$
  that is consistent with zero.
  Slow rotators are purely elliptical on average, with a
  luminosity-weighted average $B_{4} = {-}0.0007 \pm 0.0063$ also consistent
  with zero.
  Slow rotators occupy a wide range of $B_{4}$ values, from
  approximately $-0.02$ to $0.015$, with most being clustered near
  $B_{4}\,{\sim}\,0$.
  There are roughly the same number of significantly disky slow
  rotators as significantly boxy ones; in aggregate
  the slow rotators display a preference for elliptical isophotes. 
  The lack of boxy galaxies among the fast rotators suggest that
  fast rotators are more likely to be disky, but galaxies with a low
  spin parameter may or may not be boxy.
  Individual galaxies with significantly boxy or disky $B_4$
  values also show complex radial variations and a wide range in $B_4$, 
  making it difficult to interpret the measurement of any particular
  galaxy as being overall boxy or disky (Fig.~\ref{fig:shifts}).

\item Central dust is slightly less prevalent (${\sim}29\%$) in our sample
  than in previous studies of early-type galaxies, which found roughly
  half of the galaxies contained central dust, but
  dust that would be visible in optical images may not be visible in our
  F110W IR images.
  Most of the central dust in our sample takes the form of dust disks,
  though smooth compact disks (G2) and irregular extended disks (G3)
  occur in roughly equal numbers.

\end{enumerate}
  
Our high-resolution WFC3/IR images of 35 MASSIVE galaxies reveal a
wide variety of behaviors in isophotal parameters.
Some broad trends, such as diskiness and ellipticity decreasing at higher
stellar masses,
can be seen in the data.
These descriptions, however, belie the degree to which individual
galaxies vary in these parameters over their full radial range.
$B_{4}$ values vary substantially in massive elliptical galaxies and
the galaxies with the most extreme $B_{4}$ values often exhibit the
greatest variation in $B_{4}$ over their full radial profile.
The breadth of isophotal parameter variations that we report in this
paper suggests that more nuanced descriptions of the shapes of massive
early-type galaxies should be used in galaxy formation studies, beyond
simple classifications by a single parameter such as a global $B_{4}$,
PA, or ellipticity.

\section*{Acknowledgements}
\label{acknowledgements}

The authors thank John Lucey for his 2MASS profiles and Gabriel
Brammer for his specialized Python program.
C.F. Goullaud would like to thank Melanie Veale and Tom Zick for
insightful discussions.
The MASSIVE survey is supported in part by NSF AST-1411945, NSF
AST-1411642, HST-GO-14219, and HST-AR-14573.
This study was based on observations made with the NASA/ESA 
\emph{Hubble Space Telescope}, 
obtained at the Space Telescope Science Institute, which is operated by the 
Association of Universities for Research in Astronomy, Inc., under NASA 
contract NAS 5-26555. These observations are associated with program GO-14219.

\bibliography{references}

\appendix

\section{WFC3/IR Images and Radial Photometric Parameter Profiles}
\label{sec:images}

Images and radial plots for the full sample are shown below (except
for NGC~57, which is shown as Figure~\ref{fig:combined}).
The images in the right panels are: (\emph{Upper Left}) the F110W band
image, (\emph{Lower Left}) the isophotes derived from the surface
  brightness profile model (semi-major axis increases by 15\% between
isophotes), (\emph{Upper Right}) a closer view of the center of the
F110W image (scaled to make central features, such as dust, more
visible), and (\emph{Lower Right}) the residual image with galaxy
models subtracted.
The radial plots in the left panels show the isophotal parameters as a 
function of
the geometric mean radius, $r_{geo} = \sqrt{ab}$, where $a$ is the
semi-major axis and $b$ is the semi-minor axis.
The horizontal black dotted line in the surface brightness profile
shows the background measurement.
Open symbols denote isophotes that are affected by dust.

\begin{figure*}[!tbp]
  \centering\offinterlineskip
  \begin{minipage}[b][13.55cm][t]{0.56\textwidth}
    \includegraphics[width=\textwidth]{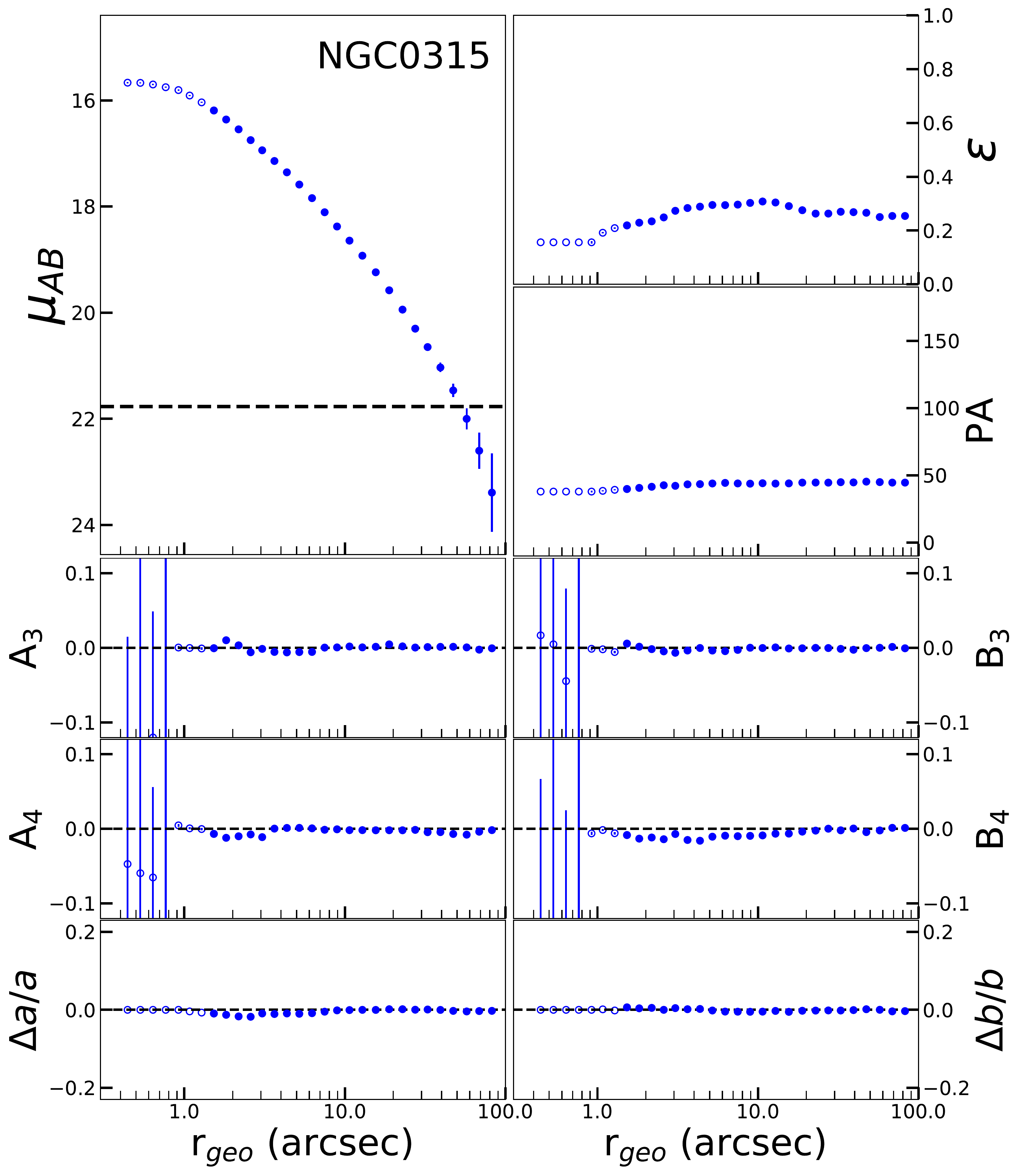}
  \end{minipage}
  \begin{minipage}[b][13.45cm][t]{0.41\textwidth}
    \includegraphics[width=\textwidth,trim=0 0 0 0]{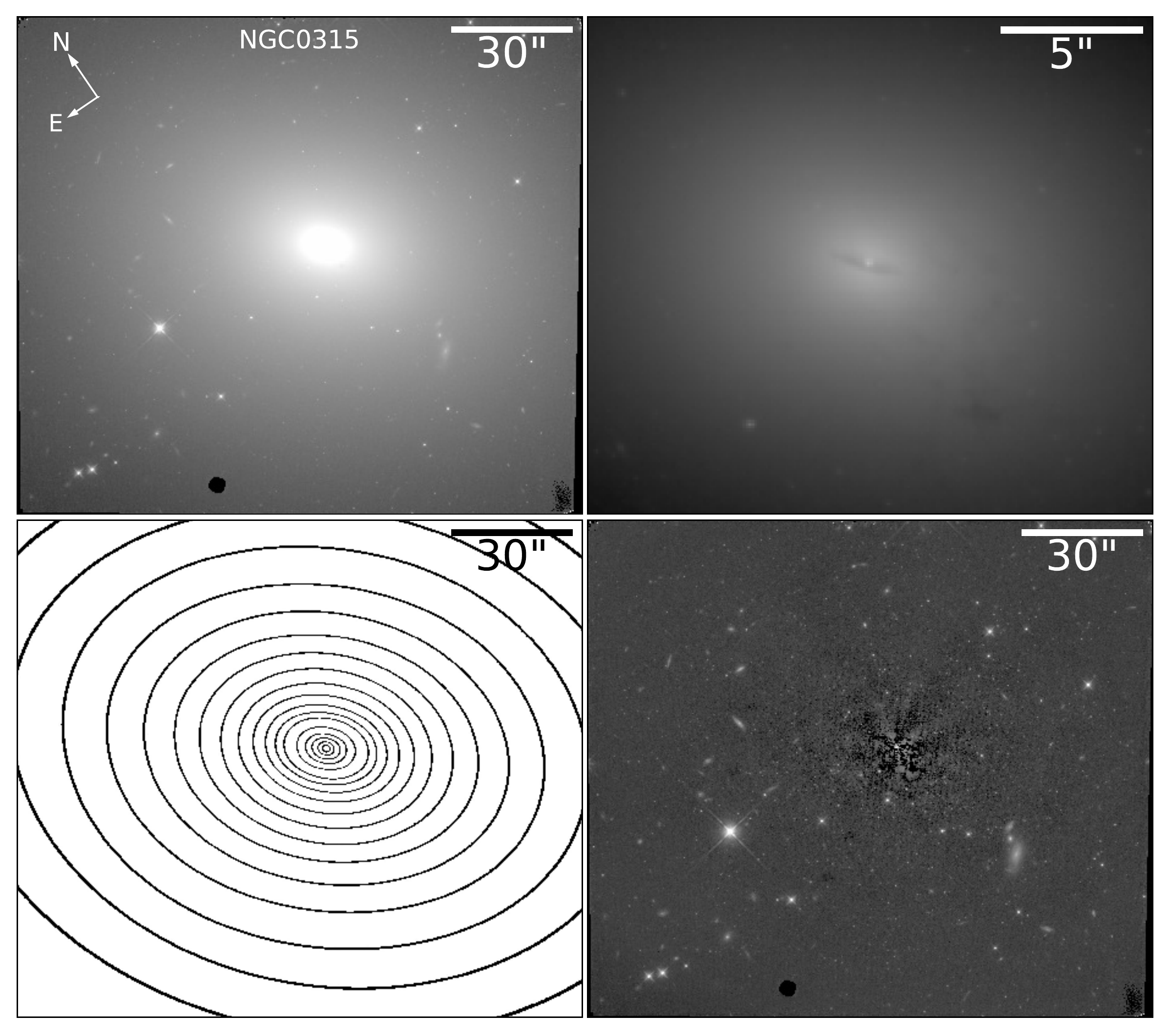}
    \caption{\small NGC~315 contains a small (${\sim}1.5$ arcsec radius)
      compact nuclear dust disk that obscures the interior
      isophotes. The isophotes between ${\sim}1$--$10$ arcsec show
      significant boxiness. This boxy region corresponds with an
      ellipticity that increases with radius. Outside ${\sim}10$
      arcsec, $B_{4}$ returns to zero and the ellipticity decreases
      with increasing radius. \\
      Scale: $1$ arcsec = $341$ pc. }
  \end{minipage}\\
  \vspace{-1.3cm}
  \begin{minipage}[b][13.55cm][t]{0.56\textwidth}
    \includegraphics[width=\textwidth]{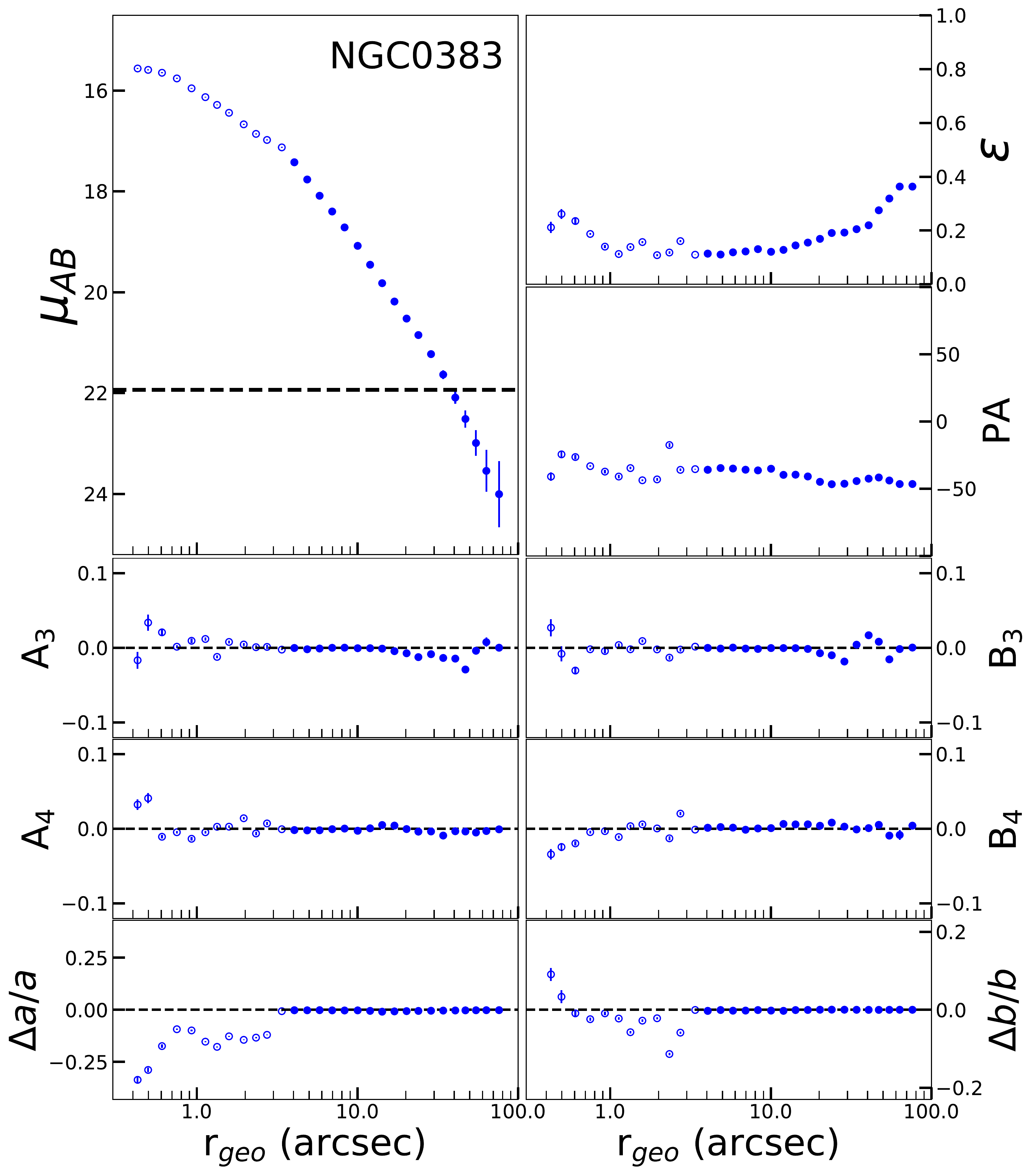}
  \end{minipage}
  \begin{minipage}[b][13.45cm][t]{0.41\textwidth}
    \includegraphics[width=\textwidth,trim=0 0 0 0]{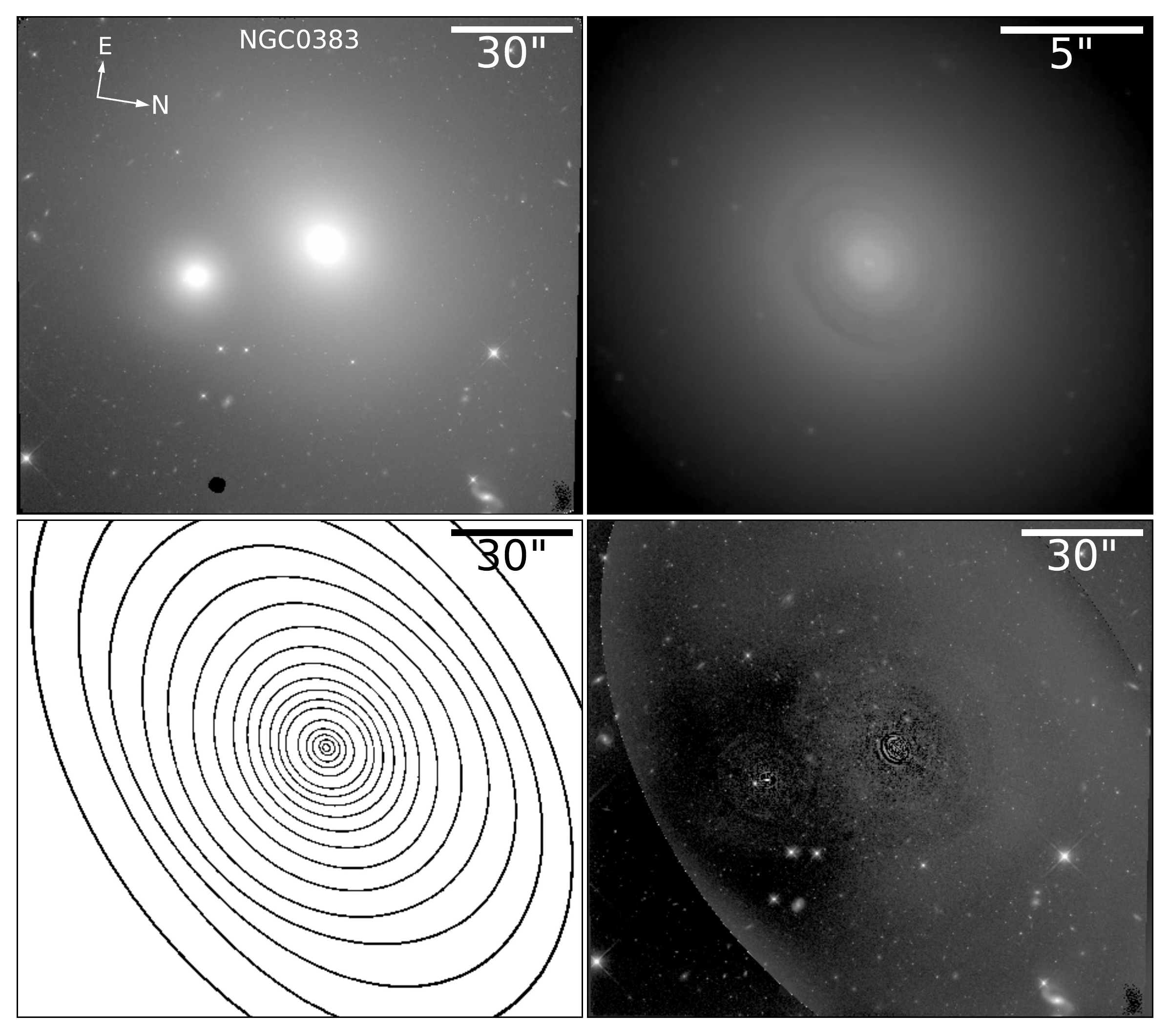}
    \caption{\small NGC~383 contains a ${\sim}4$
      arcsec radius nuclear dust disk, which obscures isophotes within that
      radius. The outermost isophotes, beyond ${\sim}40$--$50$ arcsec,
      have an ellipticity that increases sharply with increasing
      radius. Inside that radius the ellipticity increases with radius.
      NGC~383 has a large nearby neighbor
      located $33$ arcsec southwest of the center of NGC~383. \\
      Scale: $1$ arcsec = $346$ pc. }
  \end{minipage}\\
\end{figure*}

\begin{figure*}[!tbp]
  \centering\offinterlineskip
  \begin{minipage}[b][13.55cm][t]{0.56\textwidth}
    \includegraphics[width=\textwidth]{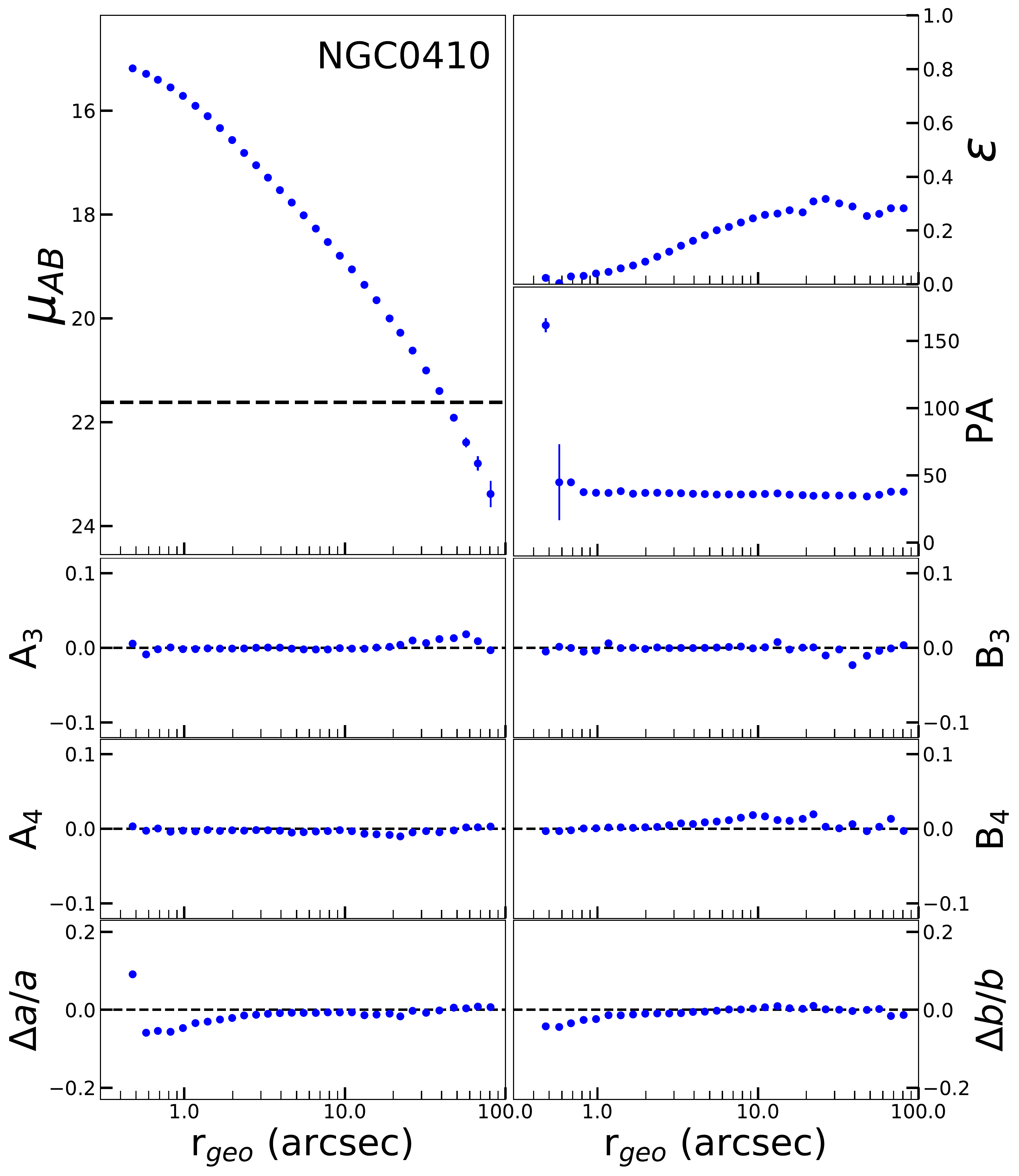}
  \end{minipage}
  \begin{minipage}[b][13.45cm][t]{0.41\textwidth}
    \includegraphics[width=\textwidth,trim=0 0 0 0]{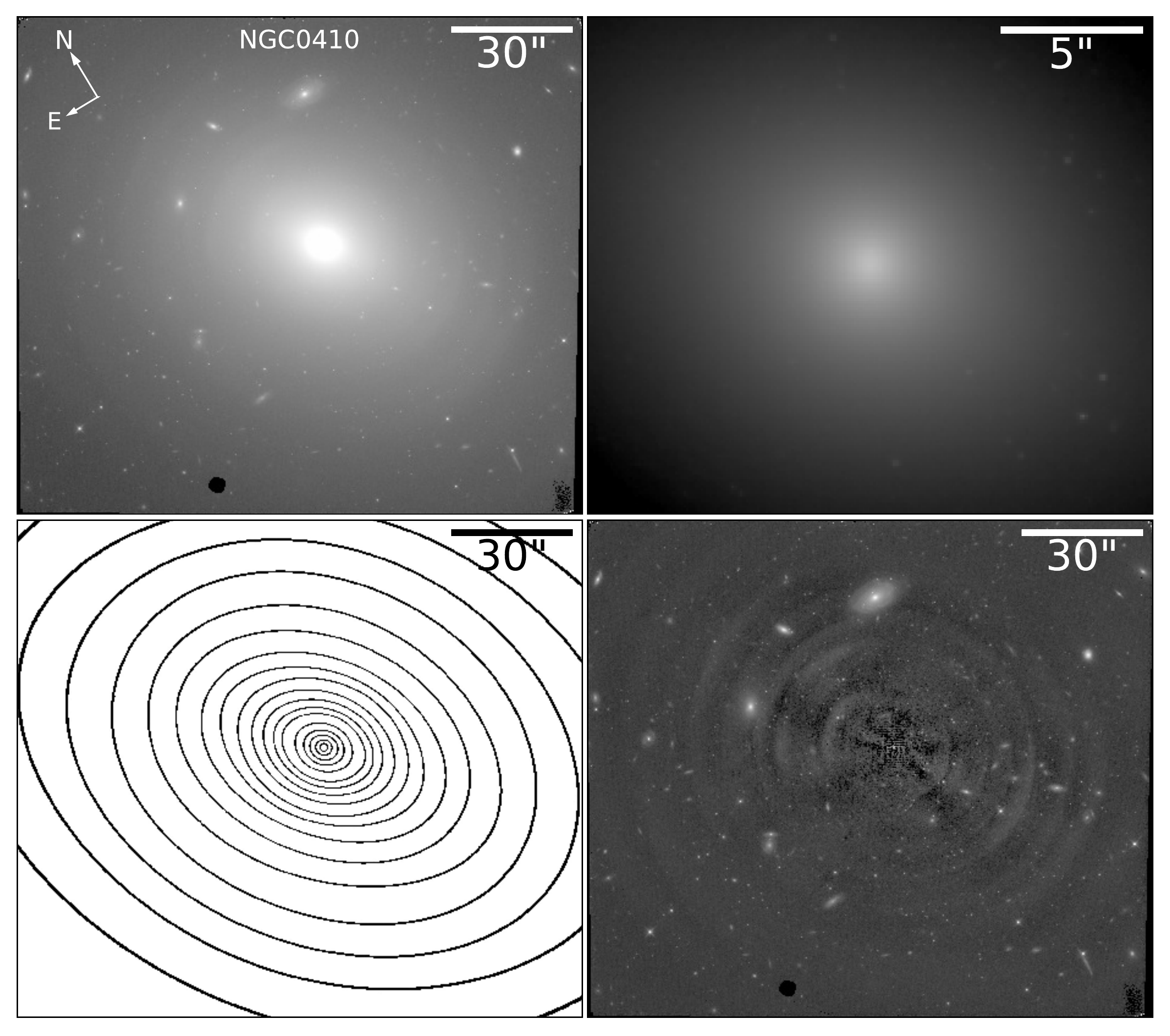}
    \caption{\small NGC~410 has disky isophotes between
      ${\sim}10$--$30$ arcsec. Outside of this range the isophotes
      go back to $B_{4}{\approx}0$. The ellipticity also peaks at
      ${\sim}30$ arcsec, beyond which it is approximately constant. \\
      Scale: $1$ arcsec = $346$ pc.  }
  \end{minipage}\\
  \vspace{-1.3cm}
  \begin{minipage}[b][13.55cm][t]{0.56\textwidth}
    \includegraphics[width=\textwidth]{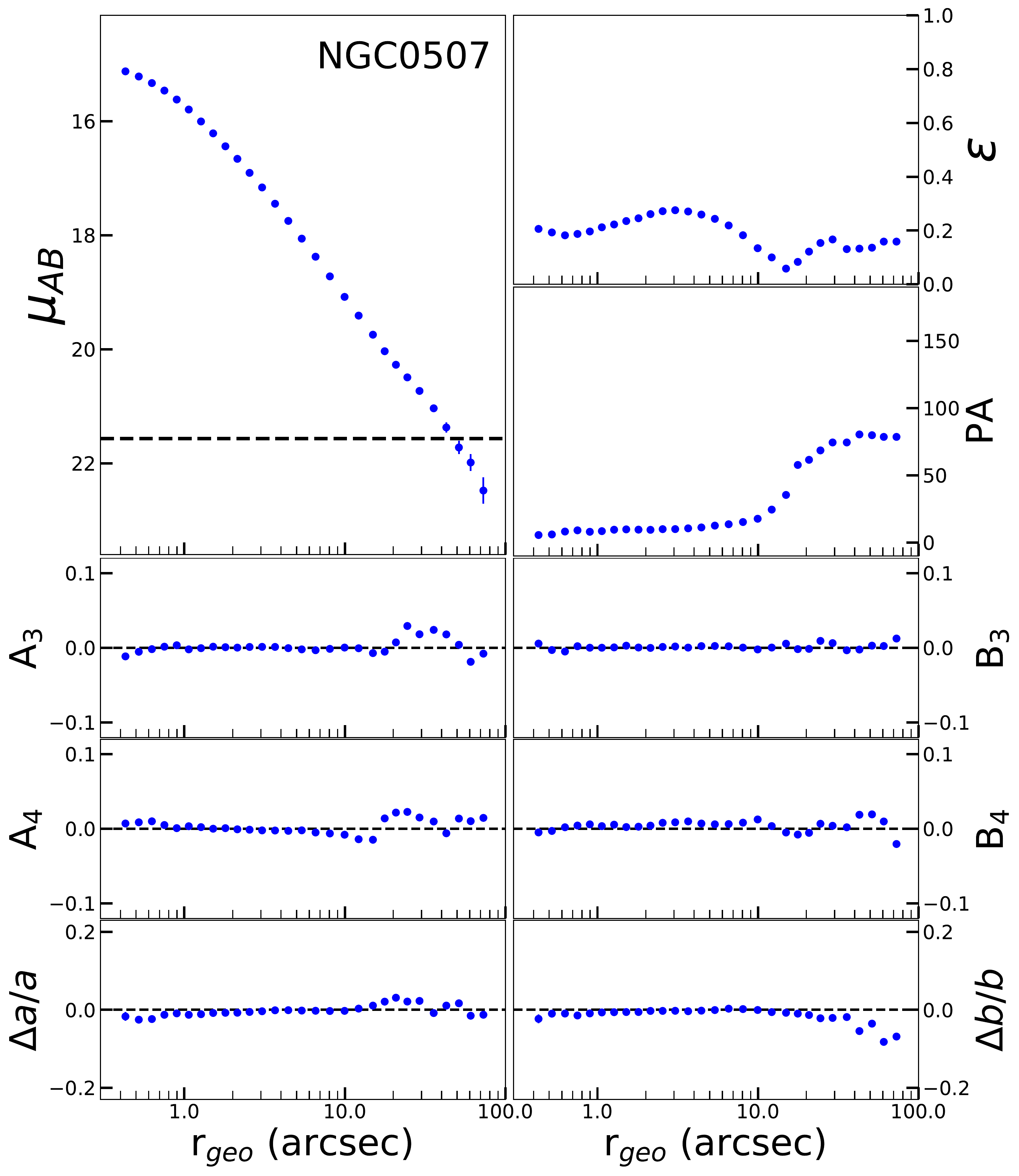}
  \end{minipage}
  \begin{minipage}[b][13.45cm][t]{0.41\textwidth}
    \includegraphics[width=\textwidth,trim=0 0 0 0]{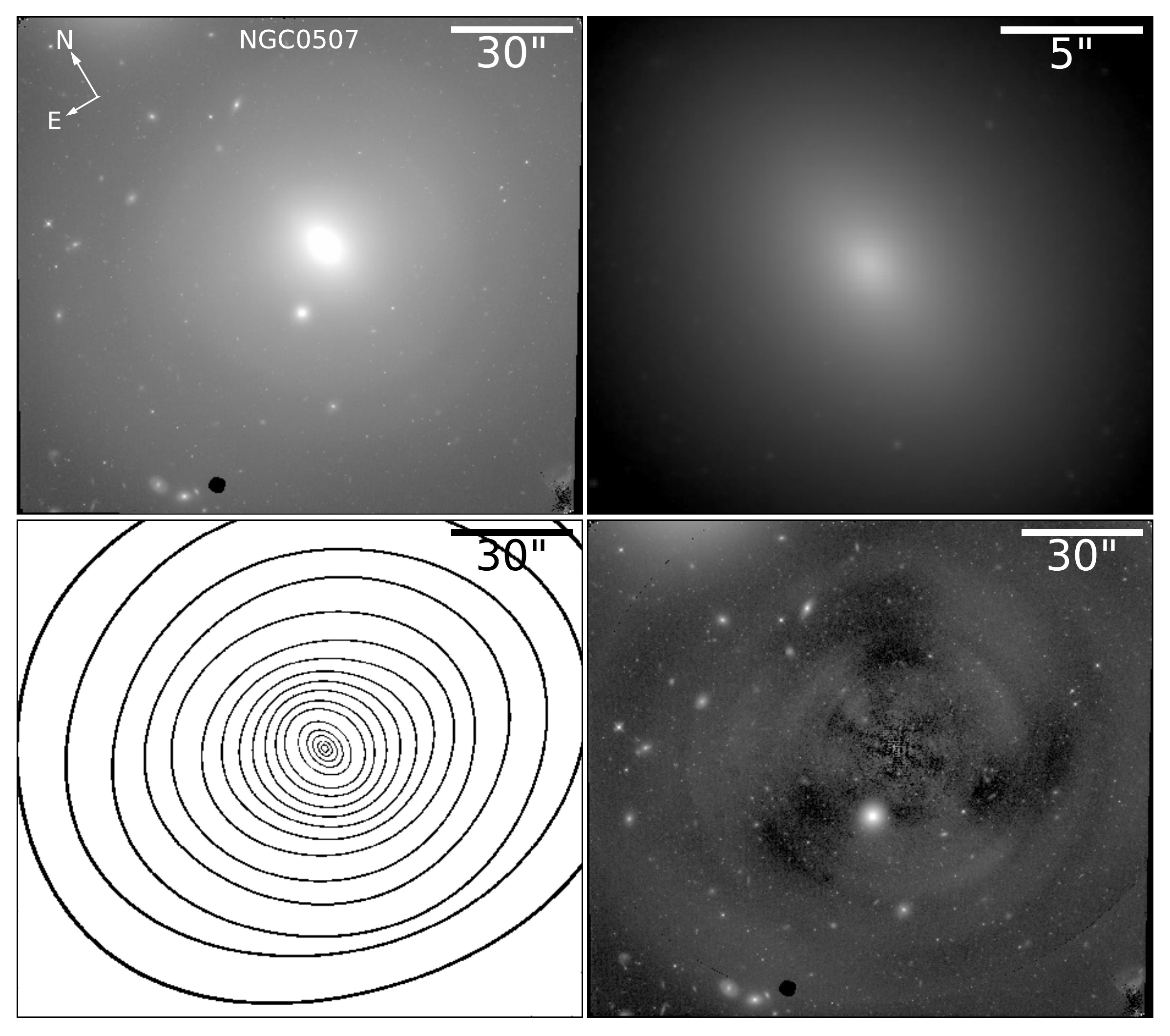}
    \caption{\small NGC~507 has one of the most dramatic PA twists in
      our sample. The PA changes by $71^{\circ}$ between
      ${\sim}5$--$40$ arcsec. The ellipticity also varies significantly,
      increasing up to its peak at ${\sim}2$ arcsec before becoming
      nearly circular at ${\sim}15$ arcsec (also the radius at which
      the PA twist is steepest), and finally rising to a moderate
      value for the outer isophotes. The isophotes between
      ${\sim}3$--$15$ arcsec are disky, and isophotes beyond that
      radius have variable $B_{4}$ values. NGC~507 has an
      off-frame companion due North. \\
      Scale: $1$ arcsec = $338$ pc.  }
  \end{minipage}\\
\end{figure*}

\begin{figure*}[!tbp]
  \centering\offinterlineskip
  \begin{minipage}[b][13.55cm][t]{0.56\textwidth}
    \includegraphics[width=\textwidth]{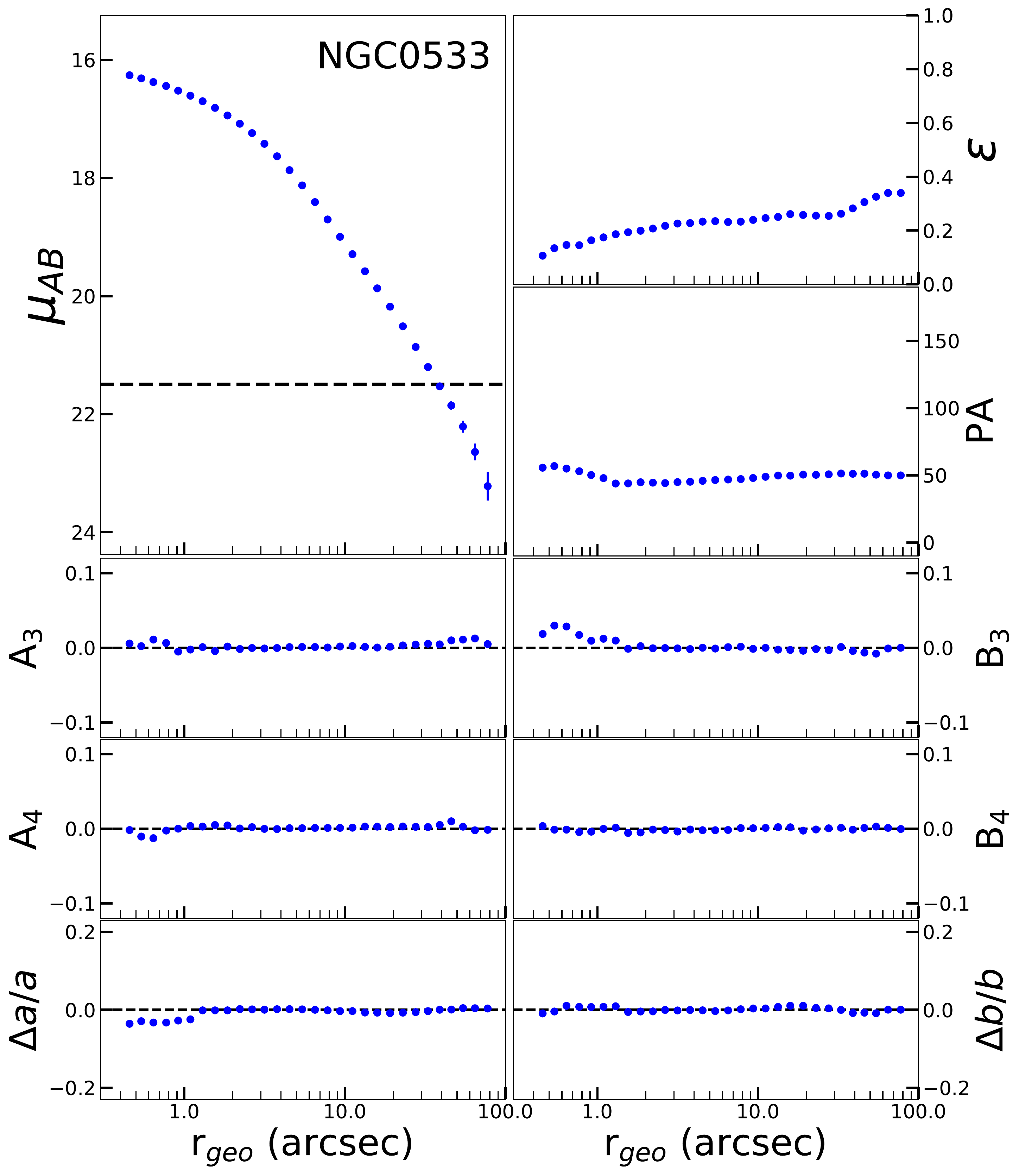}
  \end{minipage}
  \begin{minipage}[b][13.45cm][t]{0.41\textwidth}
    \includegraphics[width=\textwidth,trim=0 0 0 0]{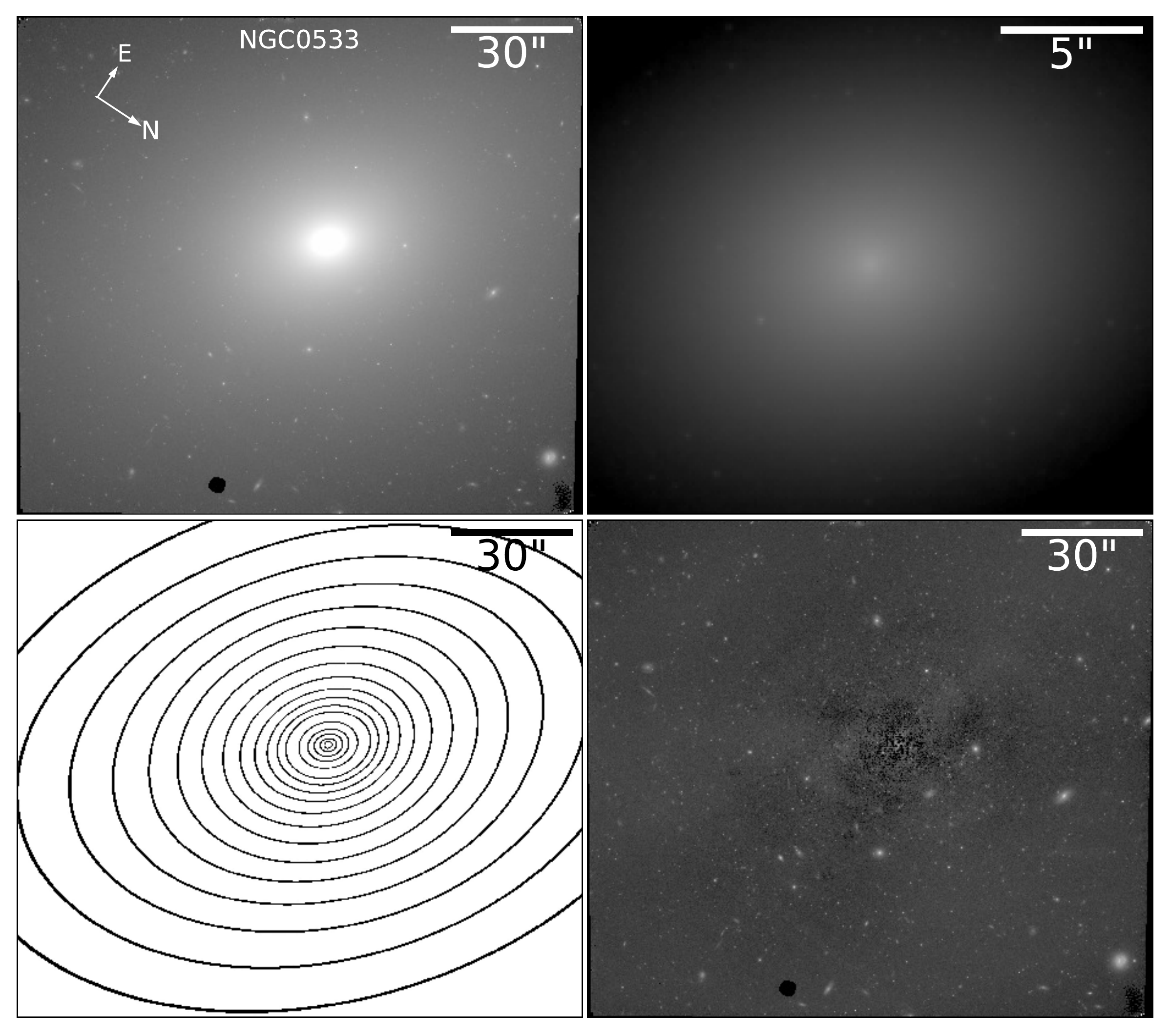}
    \caption{\small NGC~533 has ellipticity that increases with radius;
     the other parameters are nearly constant.\\
      Scale: $1$ arcsec = $378$ pc.  }
  \end{minipage}\\
  \vspace{-1.3cm}
  \begin{minipage}[b][13.55cm][t]{0.56\textwidth}
    \includegraphics[width=\textwidth]{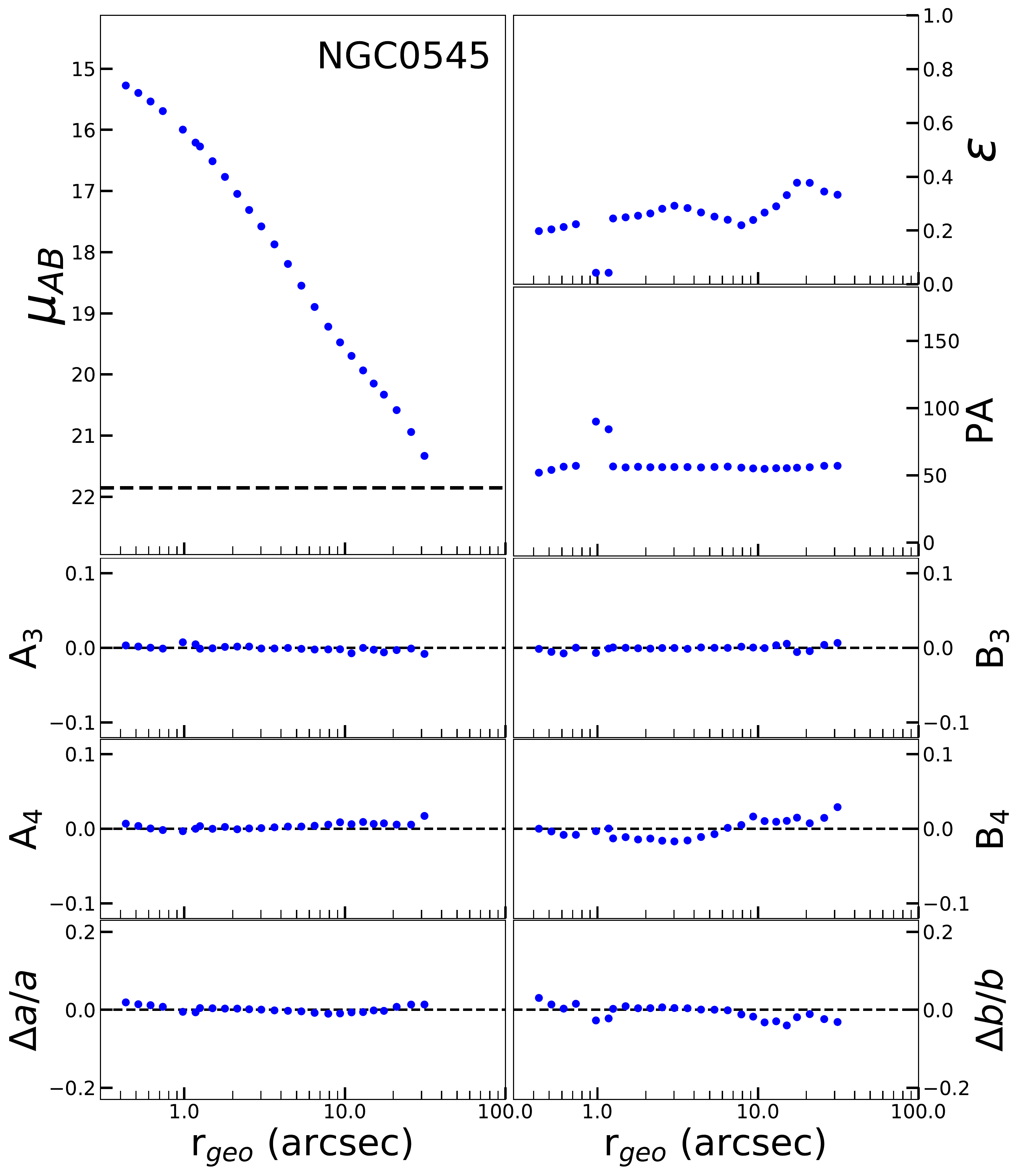}
  \end{minipage}
  \begin{minipage}[b][13.45cm][t]{0.41\textwidth}
    \includegraphics[width=\textwidth,trim=0 0 0 0]{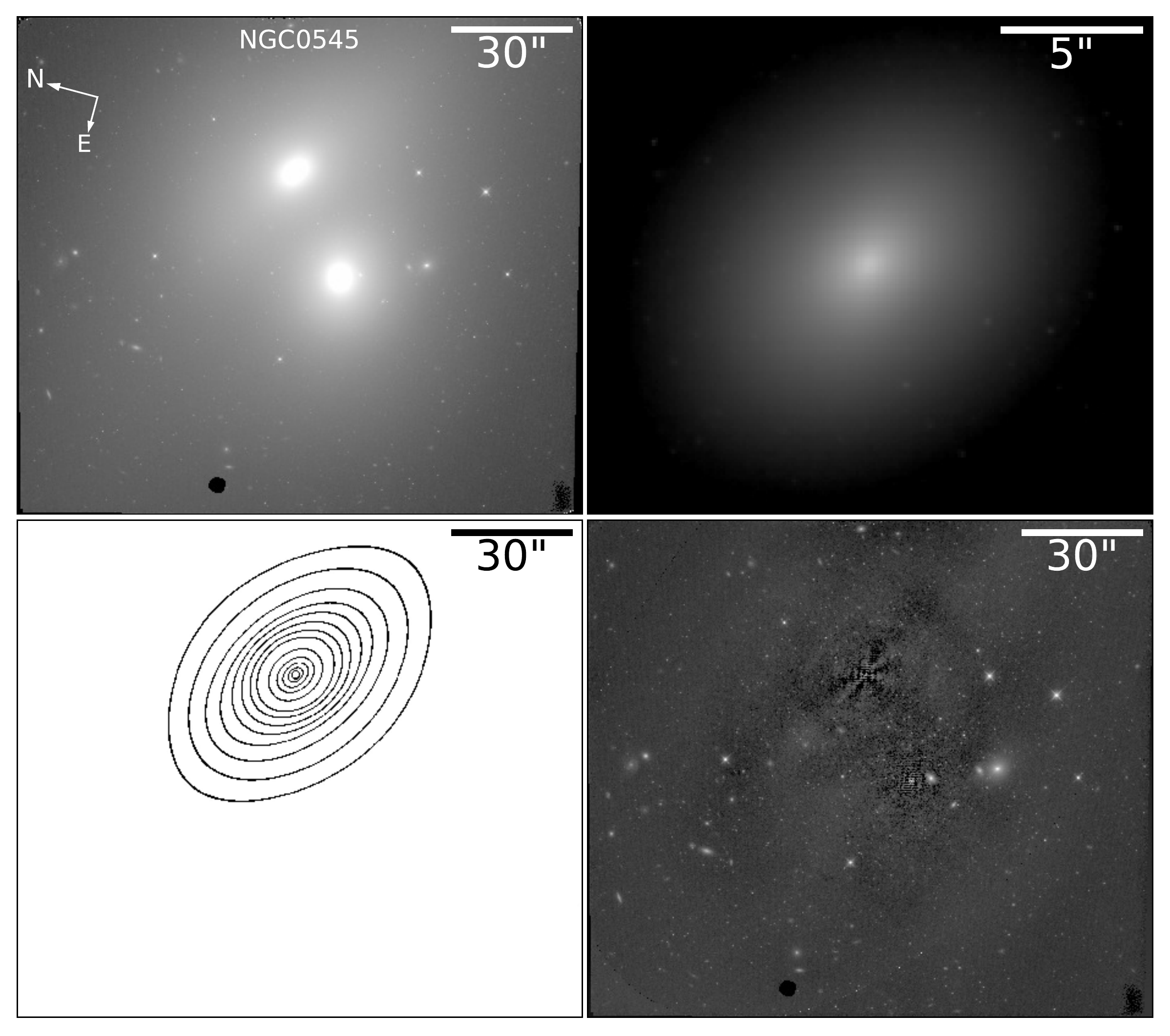}
    \caption{\small NGC~545 (the northern galaxy of the pair) 
      has significantly disky isophotes beyond
      ${\sim}10$ arcsec. Within this radius, the isophotes are 
      significantly boxy. The beginning of this transition
      occurs at a local minima in ellipticity. 
      It is difficult to
      separate the isophotes of the two galaxies beyond ${\sim}40$
      arcsec, so those isophotes have been removed from our
      analysis. NGC~545 has two isophotes at ${\sim}1$ arcsec where
      ELLIPSE failed to generate good fits. \\
      Scale: $1$ arcsec = $358$ pc.  }
  \end{minipage}\\
\end{figure*}

\begin{figure*}[!tbp]
  \centering\offinterlineskip
  \begin{minipage}[b][13.55cm][t]{0.56\textwidth}
    \includegraphics[width=\textwidth]{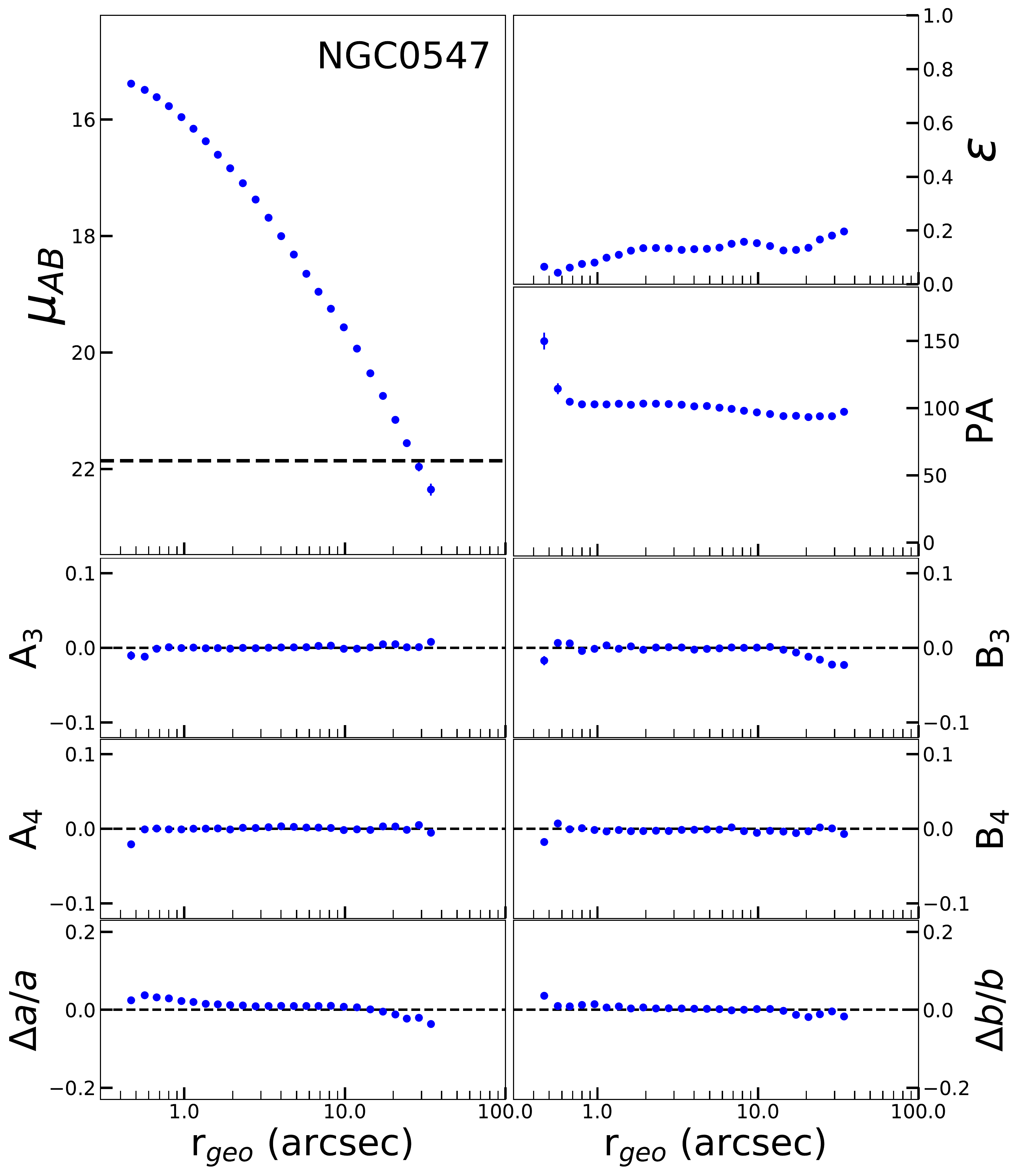}
  \end{minipage}
  \begin{minipage}[b][13.45cm][t]{0.41\textwidth}
    \includegraphics[width=\textwidth,trim=0 0 0 0]{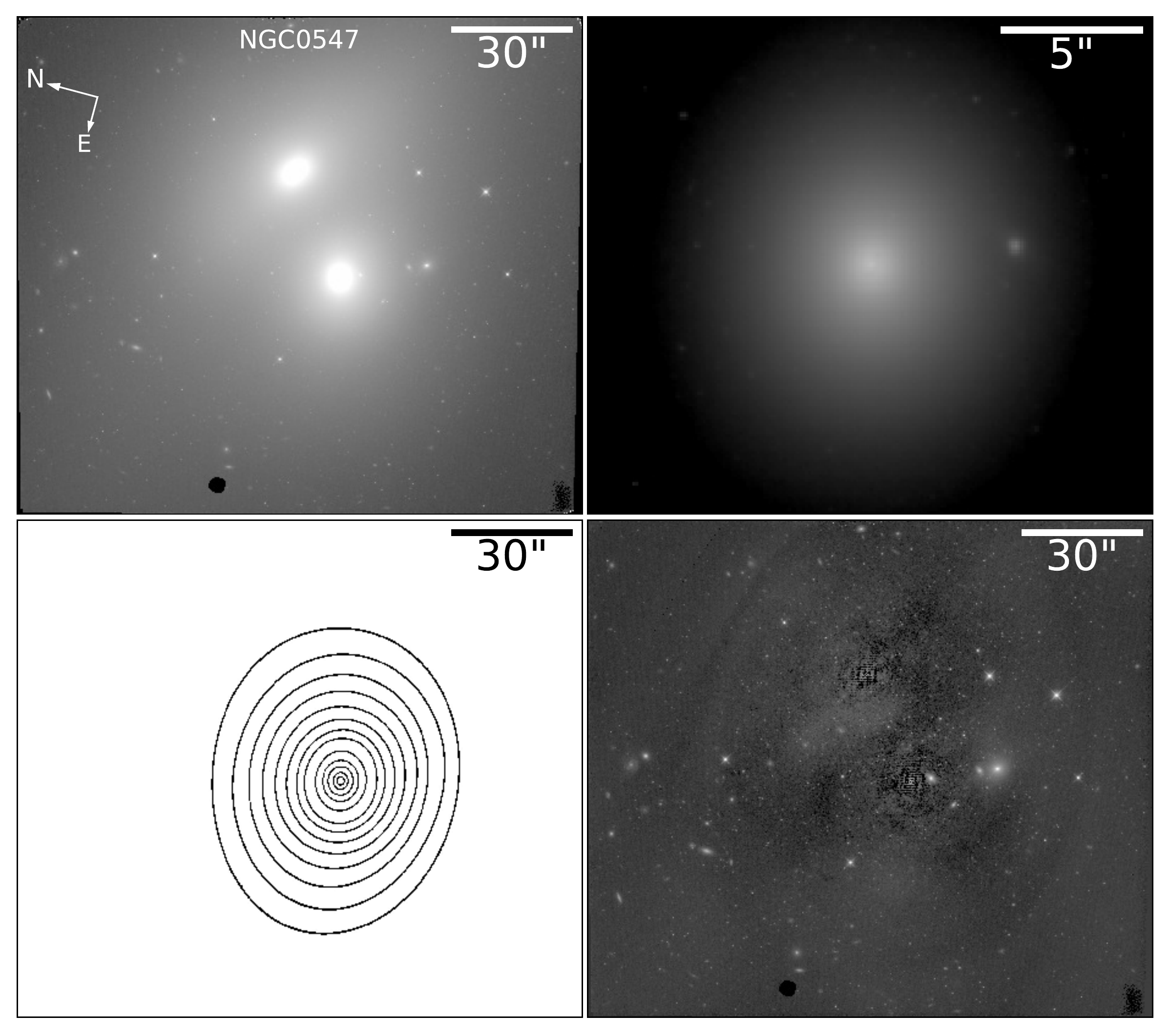}
    \caption{\small NGC~547 (the southern galaxy) 
      has the beginnings of a PA twist within
      ${\sim}0.8$ arcsec, though the degree of the twist is obscured
      by the PSF. 
      It is difficult to separate the isophotes of
      the two galaxies beyond ${\sim}40$
      arcsec, so those isophotes have been removed from our analysis. \\
      Scale: $1$ arcsec = $358$ pc. }
  \end{minipage}\\
  \vspace{-1.3cm}
  \begin{minipage}[b][13.55cm][t]{0.56\textwidth}
    \includegraphics[width=\textwidth]{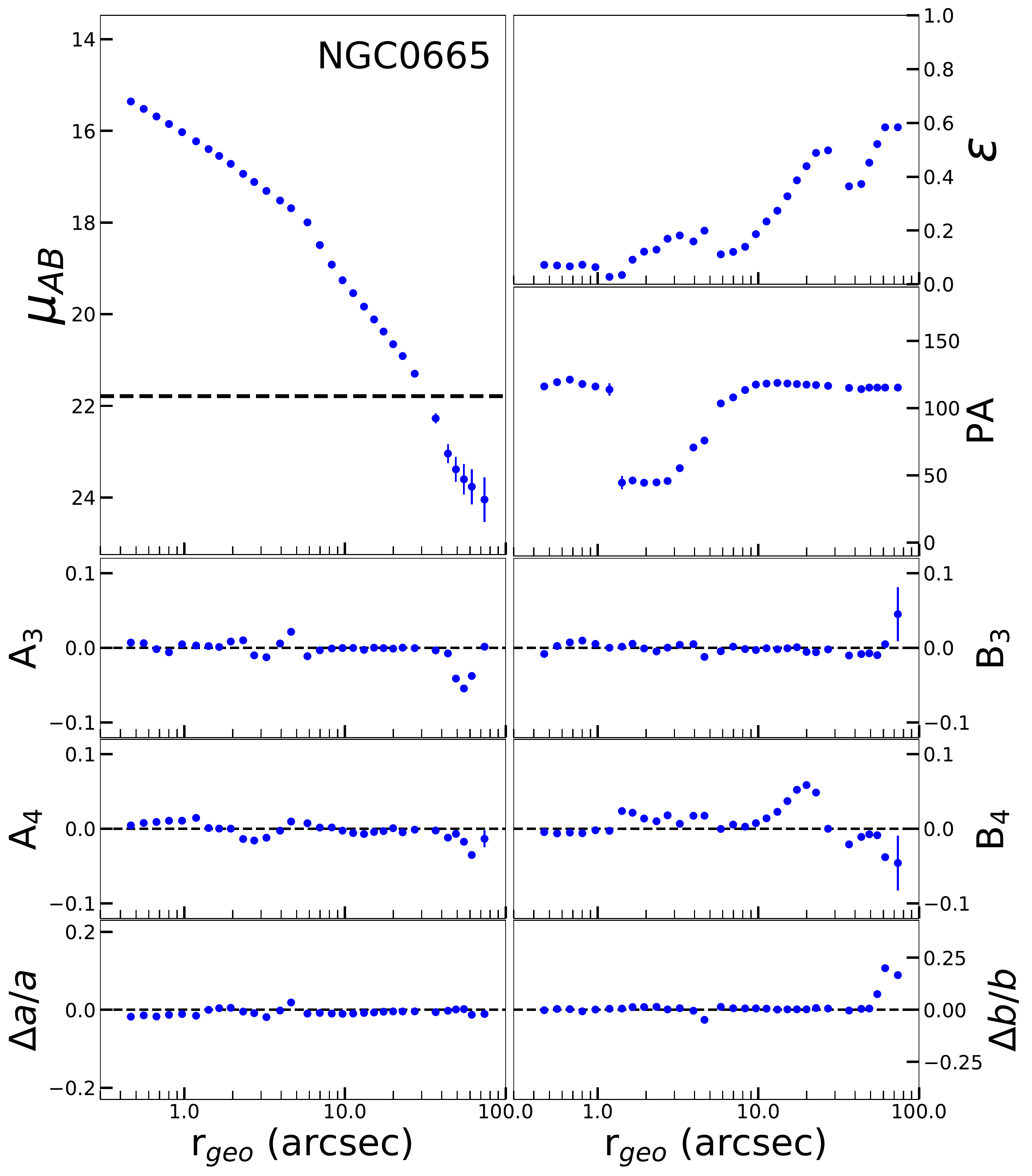}
  \end{minipage}
  \begin{minipage}[b][13.45cm][t]{0.41\textwidth}
    \includegraphics[width=\textwidth,trim=0 0 0 0]{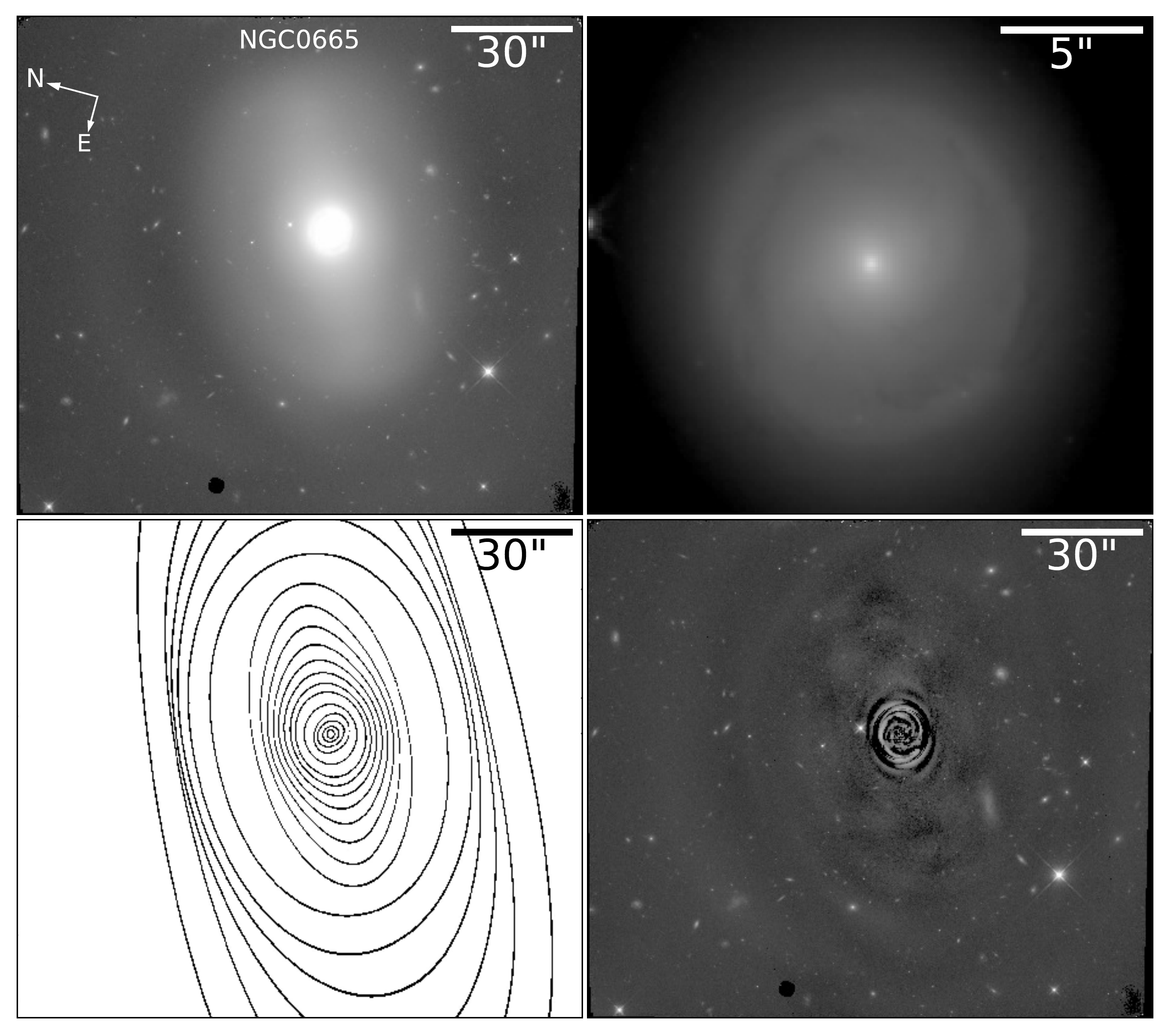}
    \caption{\small NGC~665 is a barred S0 galaxy with a
      bright nuclear peak and a dust disk which extends to a radius 
      of ${\sim}4.5$ arcsec. 
      The stellar bar extends out to ${\sim}40$ arcsec. 
      The bright nucleus inside ${\sim}1$ arcsec has a stable ellipticity 
      and PA. The nuclear dust disk
      has a rising ellipticity and fairly stable PA. The PA rotates
      by $74^{\circ}$ as the dust disk transitions to the stellar bar,
      and the ellipticity continues to rise. The outermost isophotes
      are very boxy and have displaced centers. \\
      Scale: $1$ arcsec = $362$ pc. }
  \end{minipage}\\
\end{figure*}

\begin{figure*}[!tbp]
  \centering\offinterlineskip
  \begin{minipage}[b][13.55cm][t]{0.56\textwidth}
    \includegraphics[width=\textwidth]{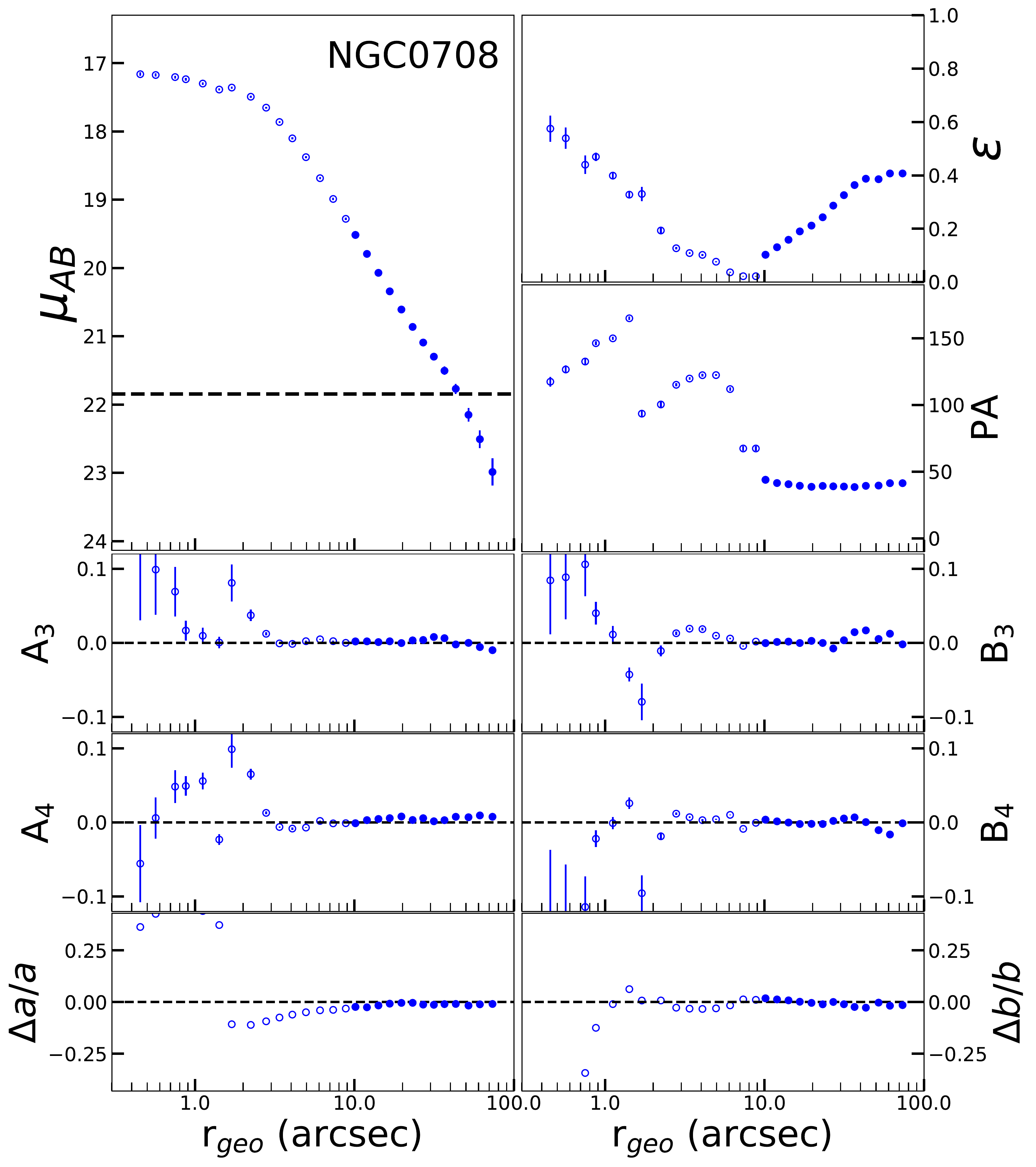}
  \end{minipage}
  \begin{minipage}[b][13.45cm][t]{0.41\textwidth}
    \includegraphics[width=\textwidth,trim=0 0 0 0]{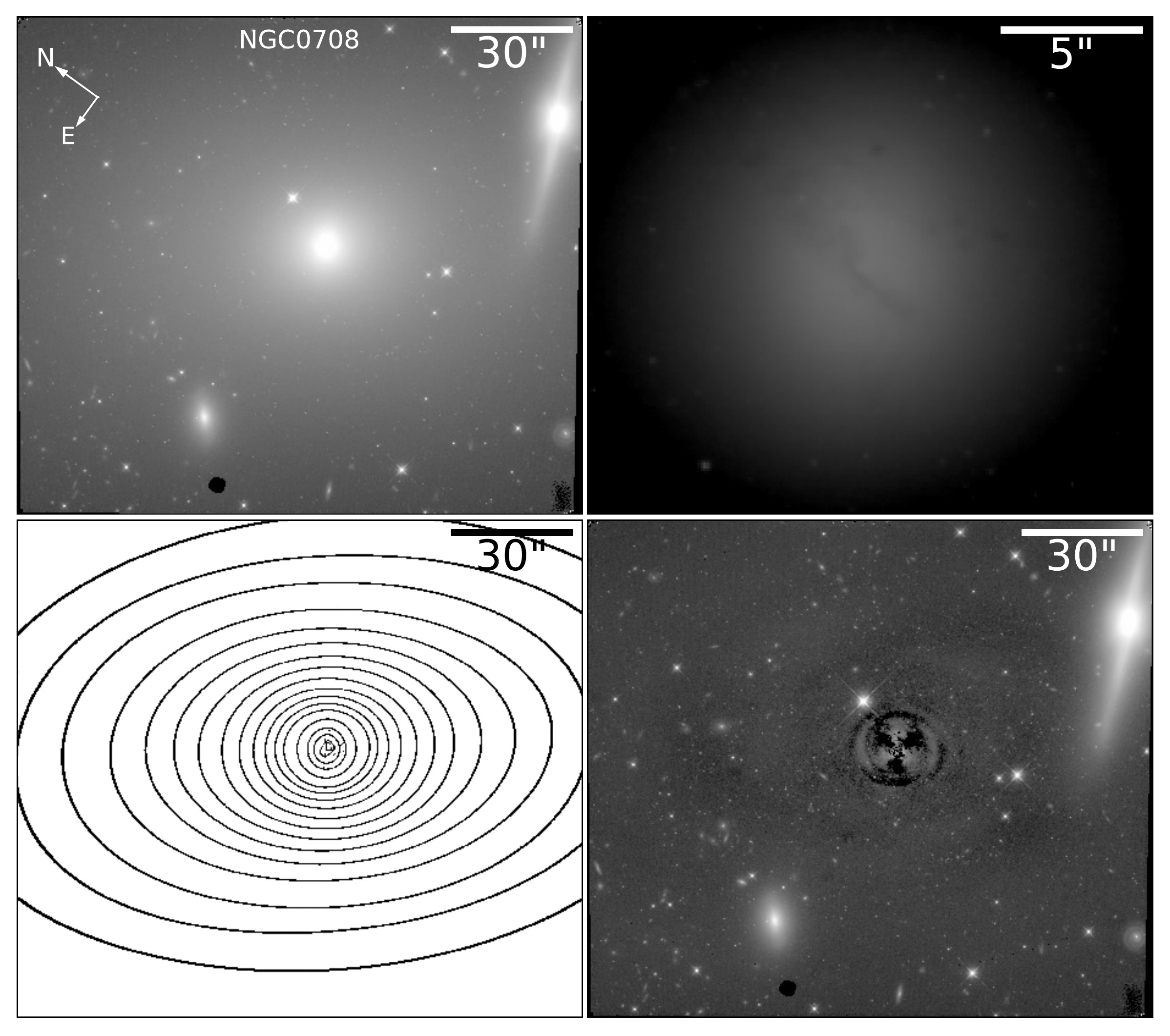}
    \caption{\small NGC~708 (the cD galaxy in Abell 262) has a large 
      irregular dust structure,
      including a dust lane across the center, that obscures the
      central ${\sim}9$ arcsec. NGC~708 has two smaller companions in the 
      frame. \\
      Scale: $1$ arcsec = $335$ pc.  }
  \end{minipage}\\
  \vspace{-1.3cm}
  \begin{minipage}[b][13.55cm][t]{0.56\textwidth}
    \includegraphics[width=\textwidth]{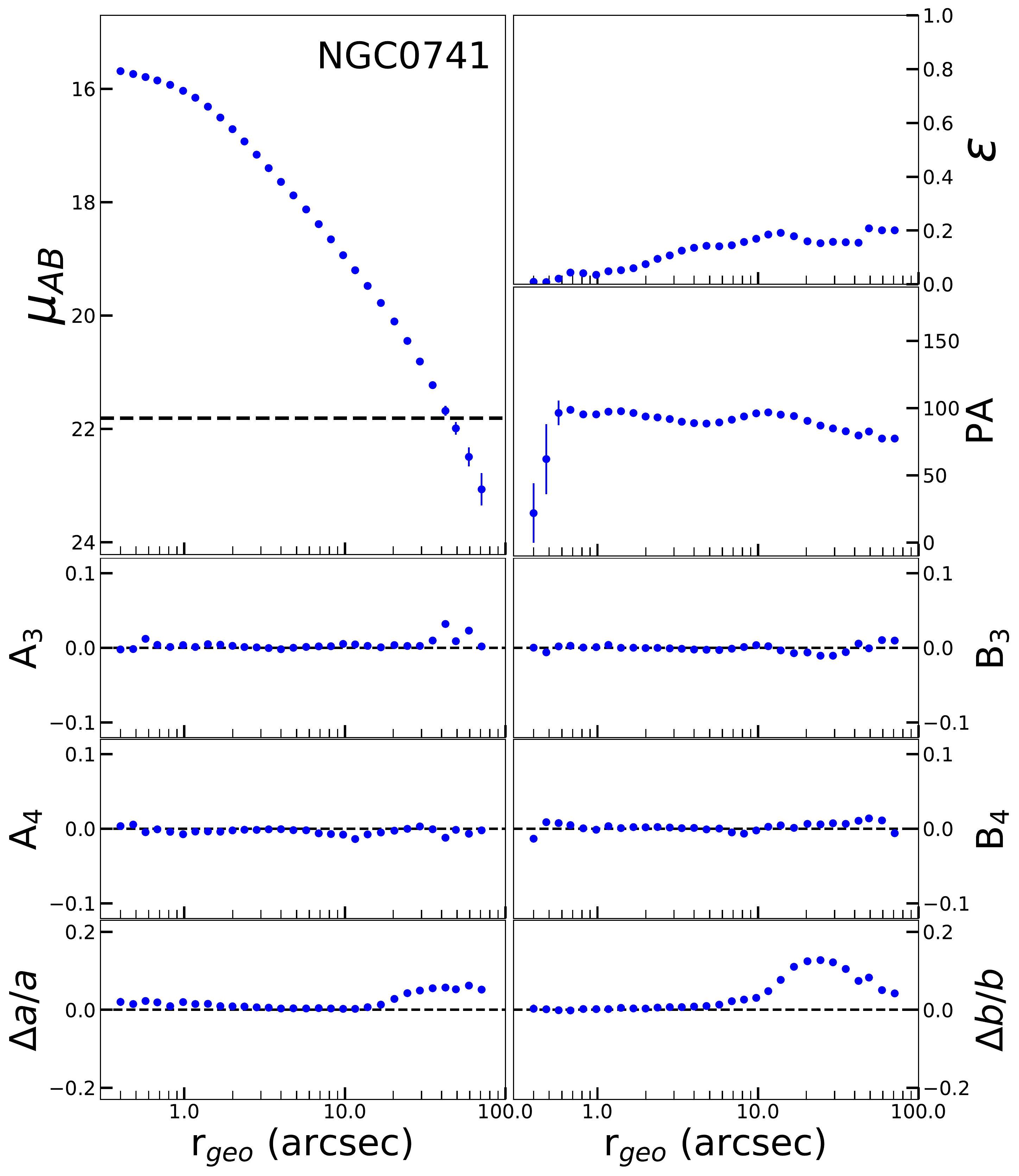}
  \end{minipage}
  \begin{minipage}[b][13.45cm][t]{0.41\textwidth}
    \includegraphics[width=\textwidth,trim=0 0 0 0]{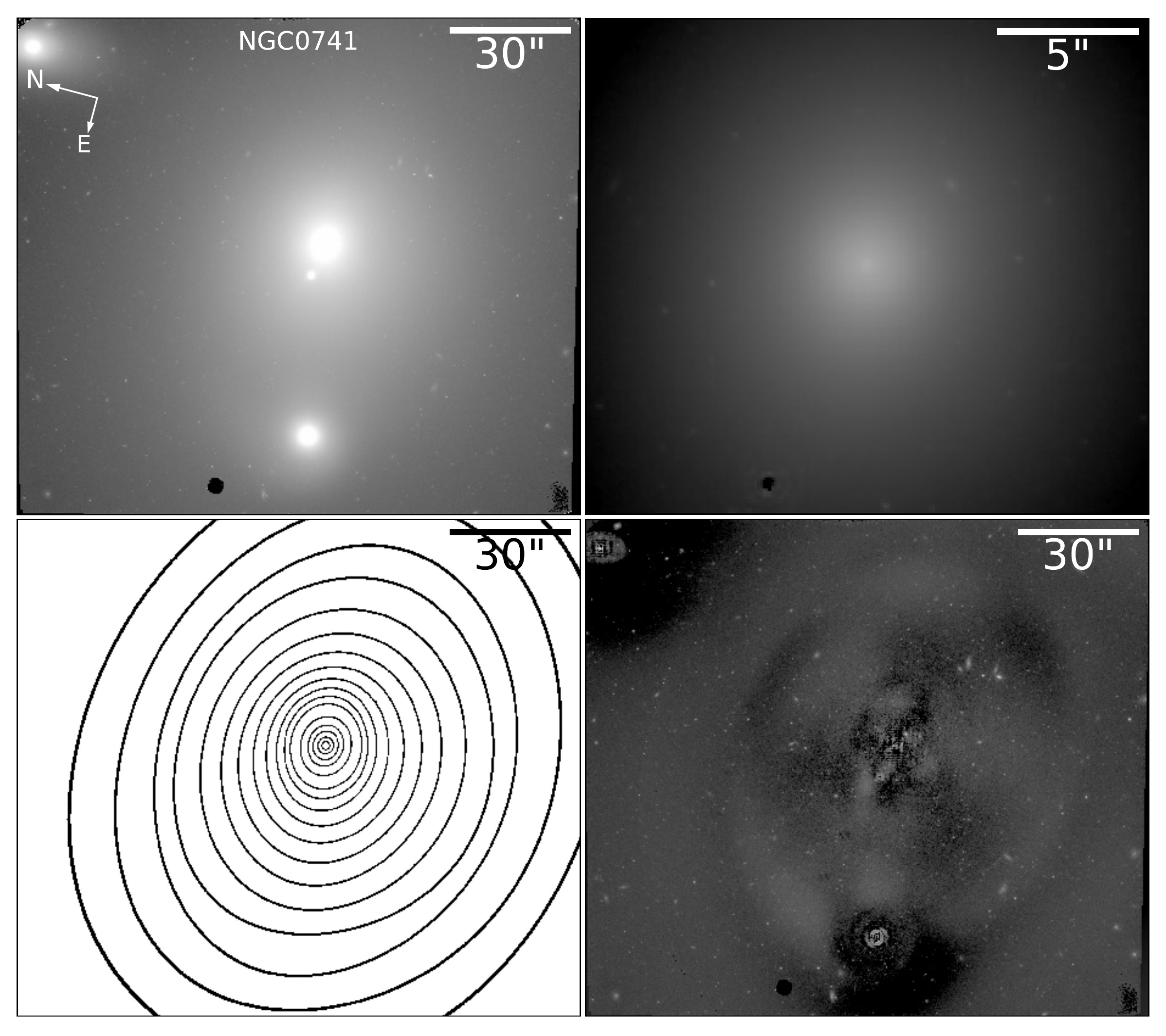}
    \caption{\small NGC~741 has a significant central drift to the
      south for isophotes starting at ${\sim}30$ arcsec. 
      It has three smaller companions in
      the frame. Two are east of the center, and one is north. \\
      Scale: $1$ arcsec = $358$ pc.  }
  \end{minipage}\\
\end{figure*}

\begin{figure*}[!tbp]
  \centering\offinterlineskip
  \begin{minipage}[b][13.55cm][t]{0.56\textwidth}
    \includegraphics[width=\textwidth]{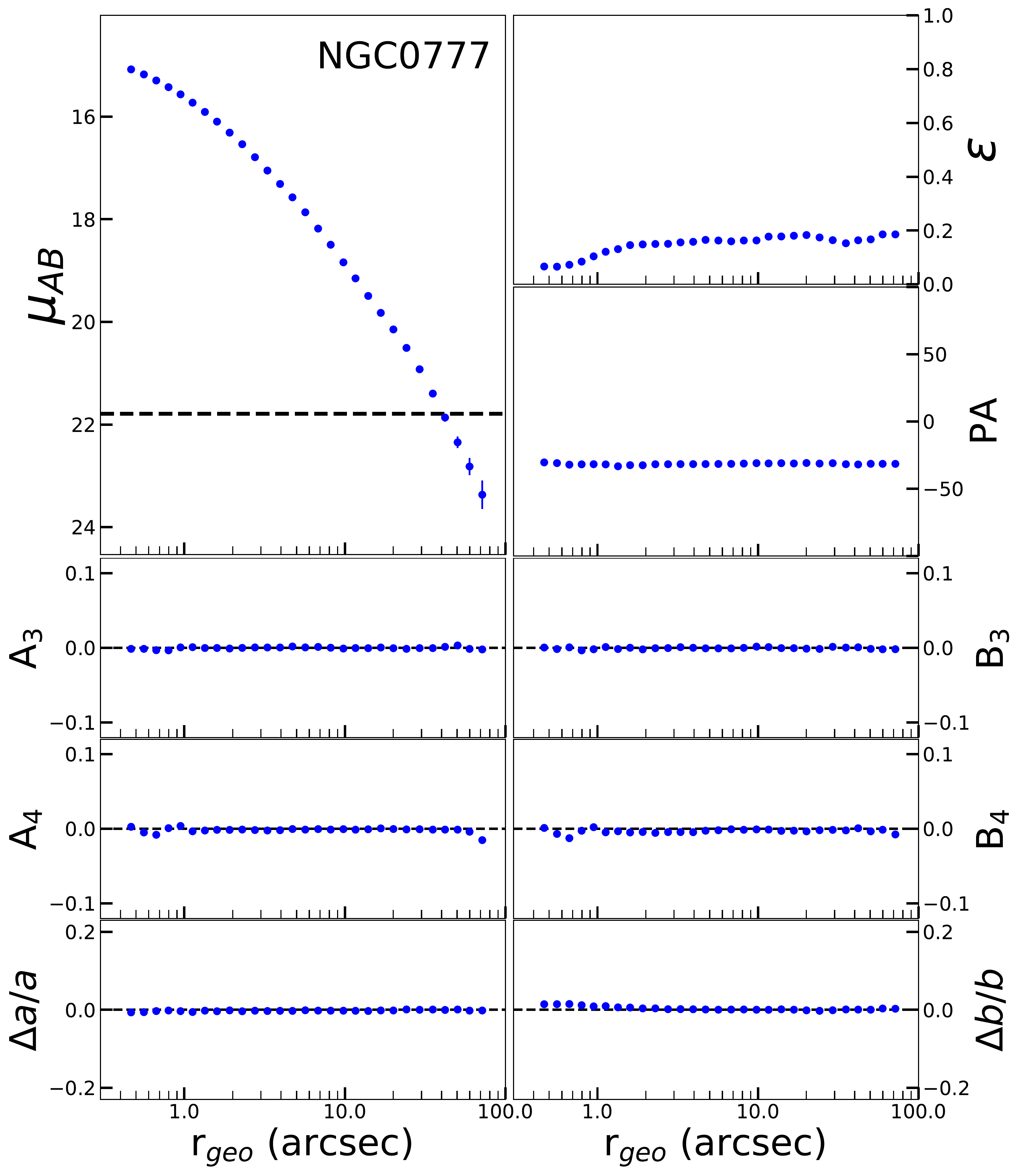}
  \end{minipage}
  \begin{minipage}[b][13.45cm][t]{0.41\textwidth}
    \includegraphics[width=\textwidth,trim=0 0 0 0]{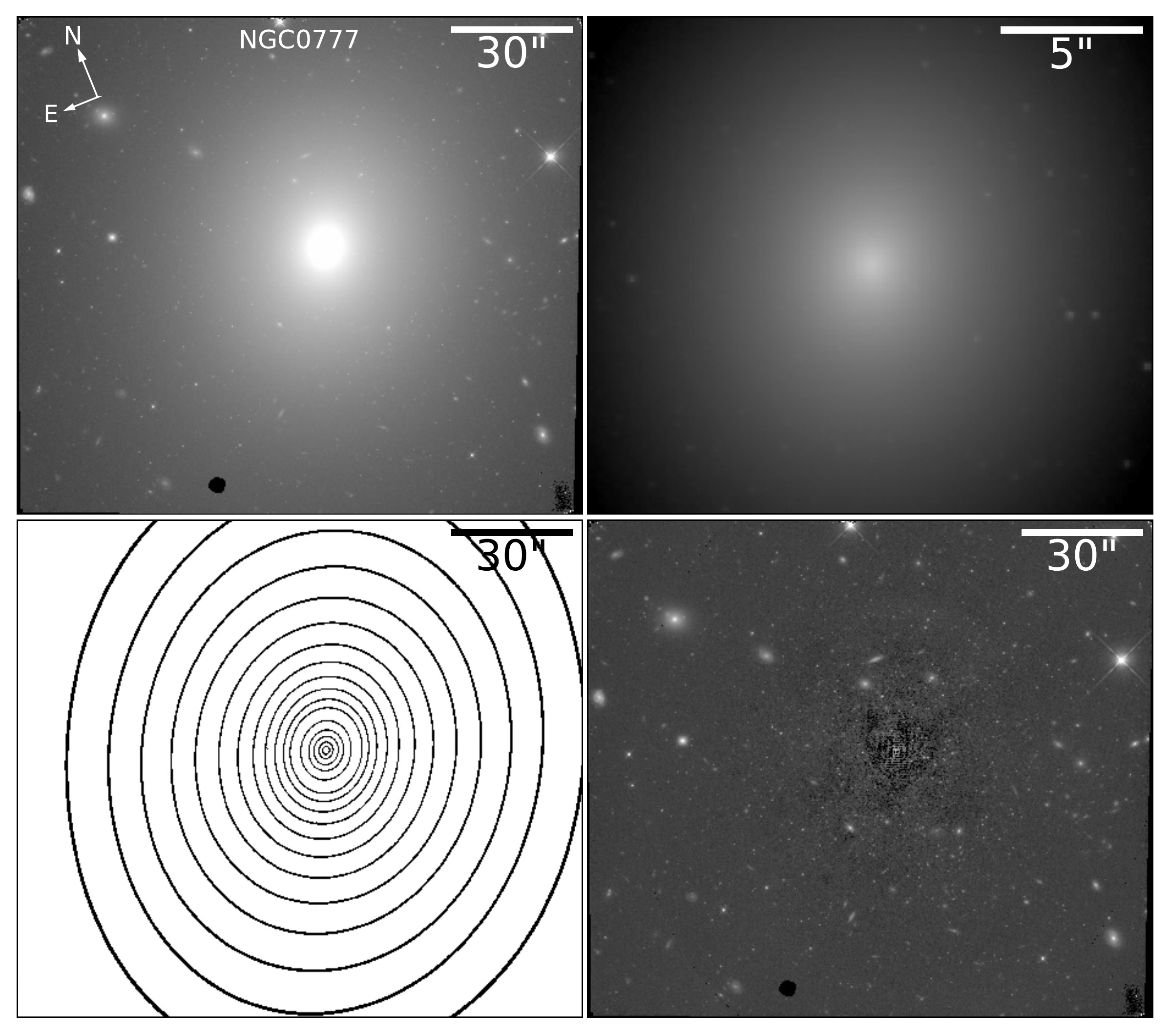}
    \caption{\small NGC~777 has slightly increasing ellipticity with radius;
      the other parameters are all nearly constant.\\
      Scale: $1$ arcsec = $350$ pc.  }
  \end{minipage}\\
  \vspace{-1.3cm}
  \begin{minipage}[b][13.55cm][t]{0.56\textwidth}
    \includegraphics[width=\textwidth]{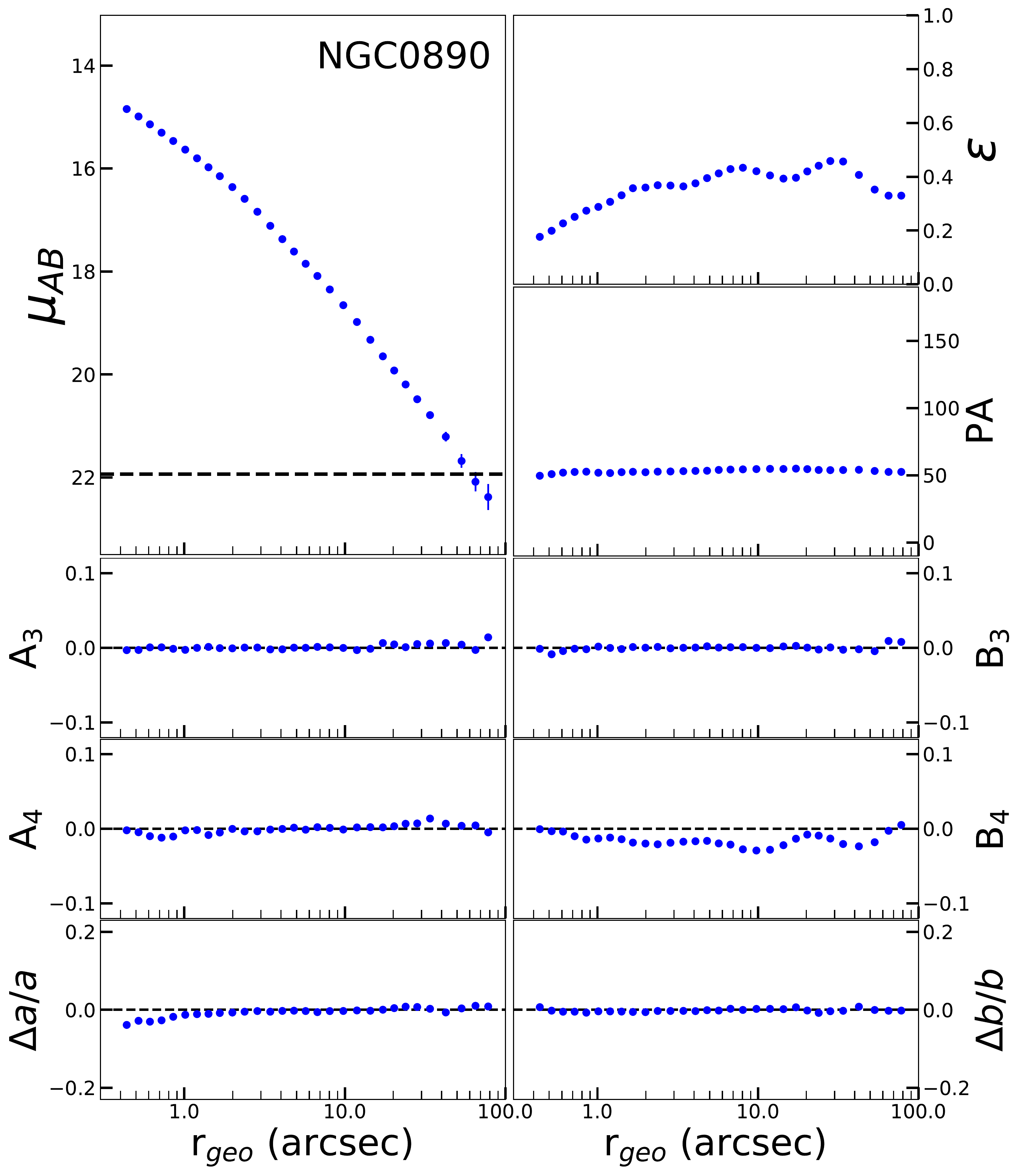}
  \end{minipage}
  \begin{minipage}[b][13.45cm][t]{0.41\textwidth}
    \includegraphics[width=\textwidth,trim=0 0 0 0]{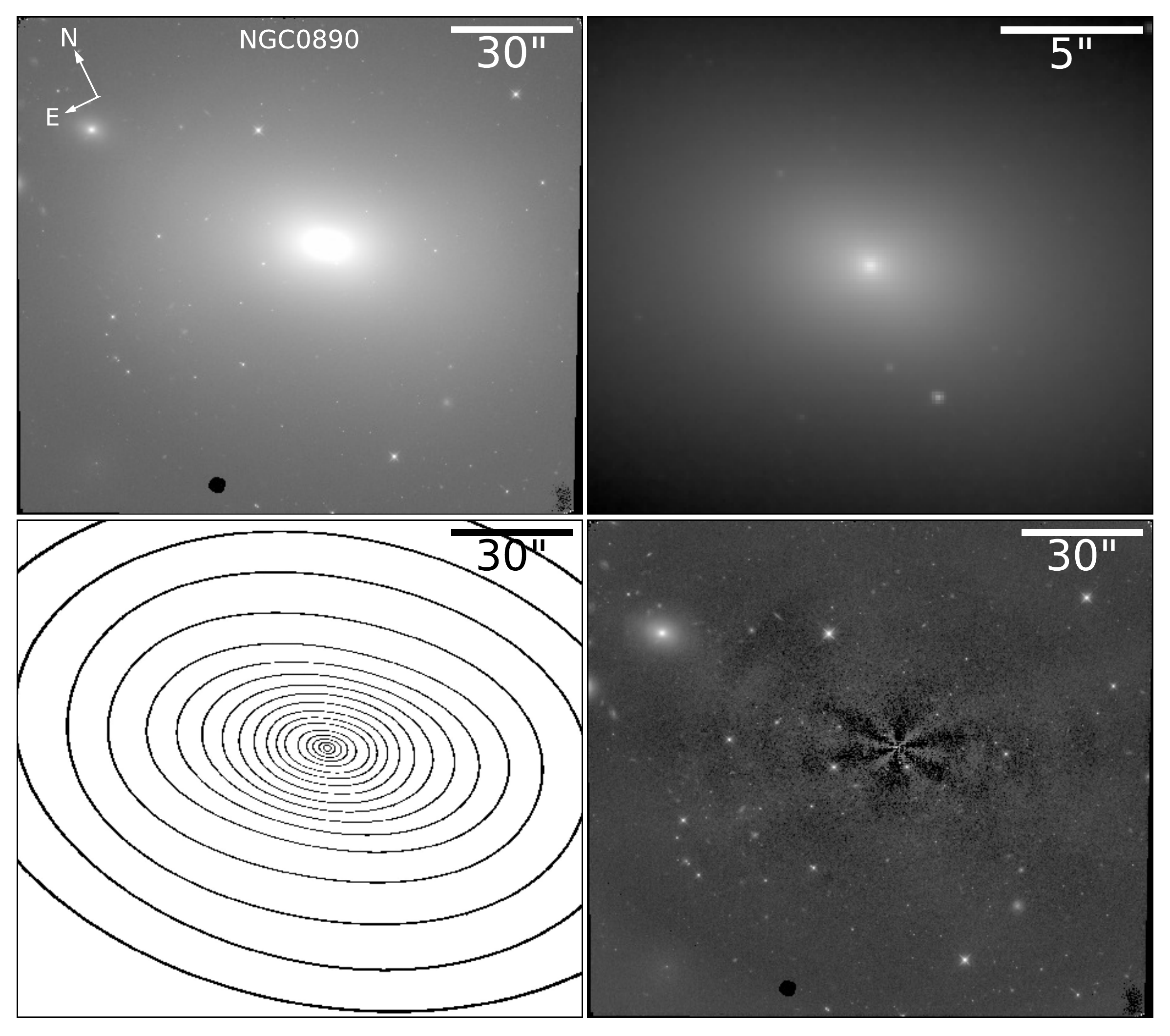}
    \caption{\small NGC~890 has a varying ellipticity that rises out to
      ${\sim}35$ arcsec.
      The isophotes are boxy, but they become more elliptical 
      outside ${\sim}35$ arcsec.\\
      Scale: $1$ arcsec = $270$ pc. }
  \end{minipage}\\
\end{figure*}

\begin{figure*}[!tbp]
  \centering\offinterlineskip
  \begin{minipage}[b][13.55cm][t]{0.56\textwidth}
    \includegraphics[width=\textwidth]{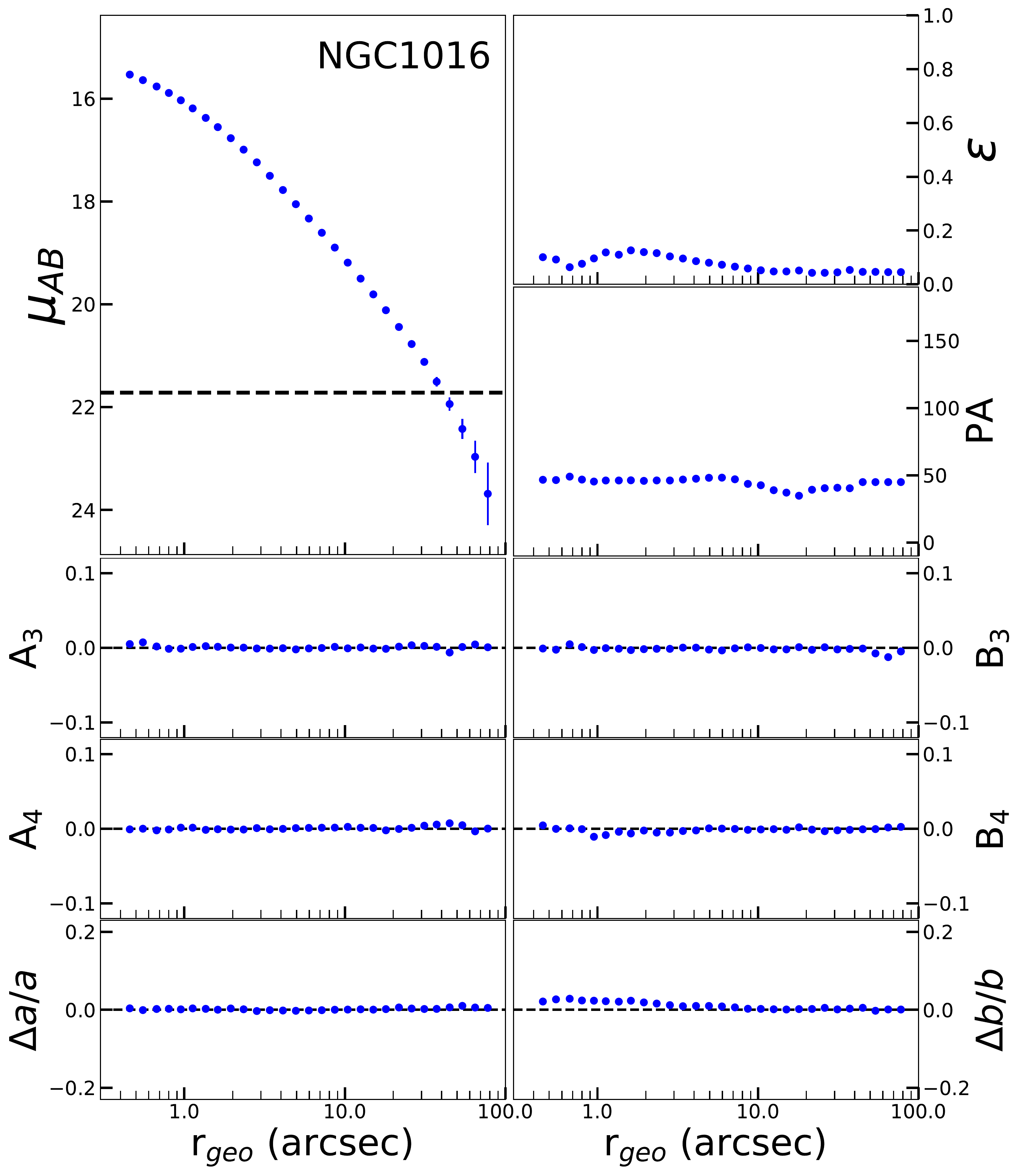}
  \end{minipage}
  \begin{minipage}[b][13.45cm][t]{0.41\textwidth}
    \includegraphics[width=\textwidth,trim=0 0 0 0]{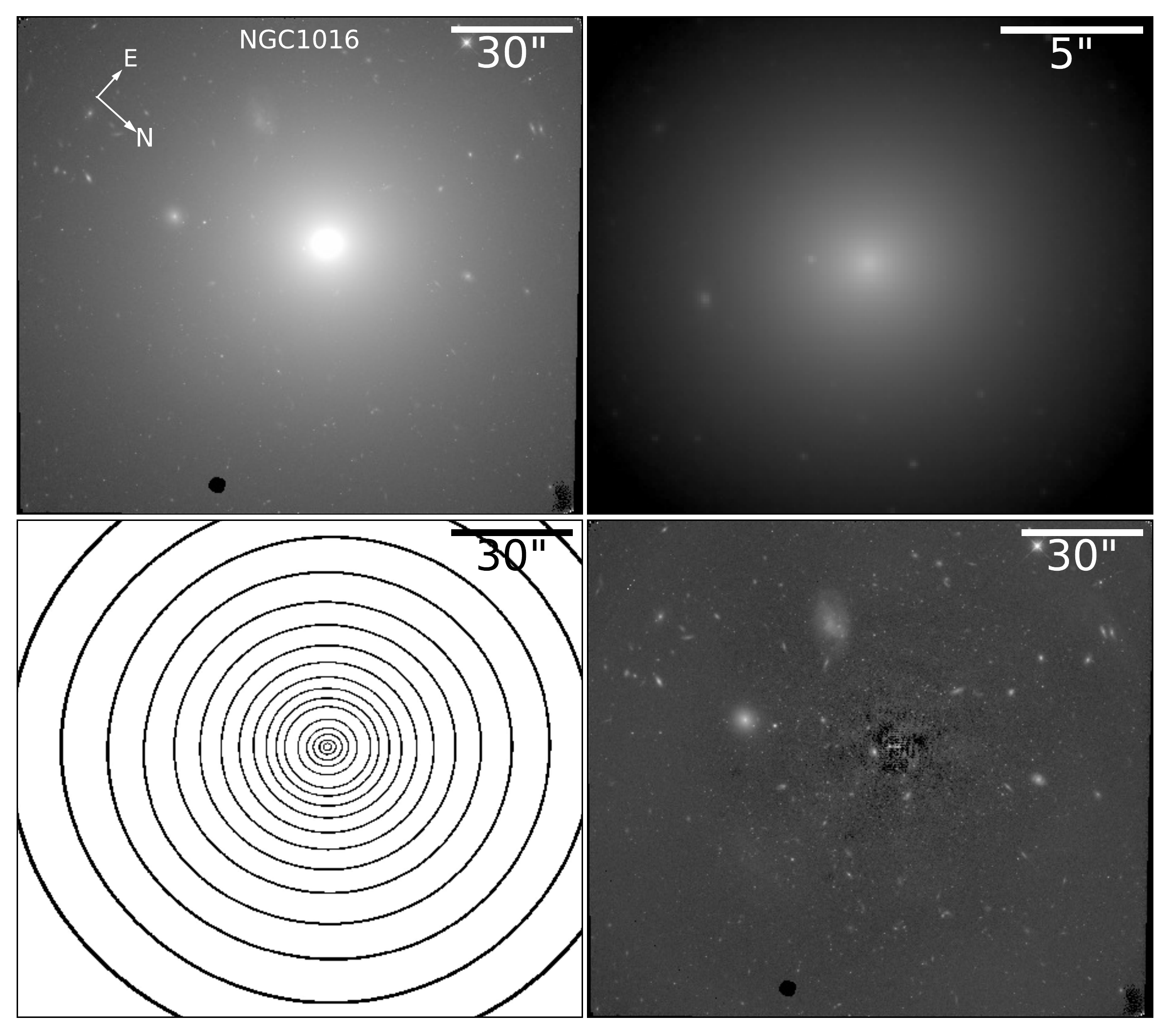}
    \caption{\small NGC~1016 is very round with nearly constant parameters.\\
      Scale: $1$ arcsec = $462$ pc.  }
  \end{minipage}\\
  \vspace{-1.3cm}
  \begin{minipage}[b][13.55cm][t]{0.56\textwidth}
    \includegraphics[width=\textwidth]{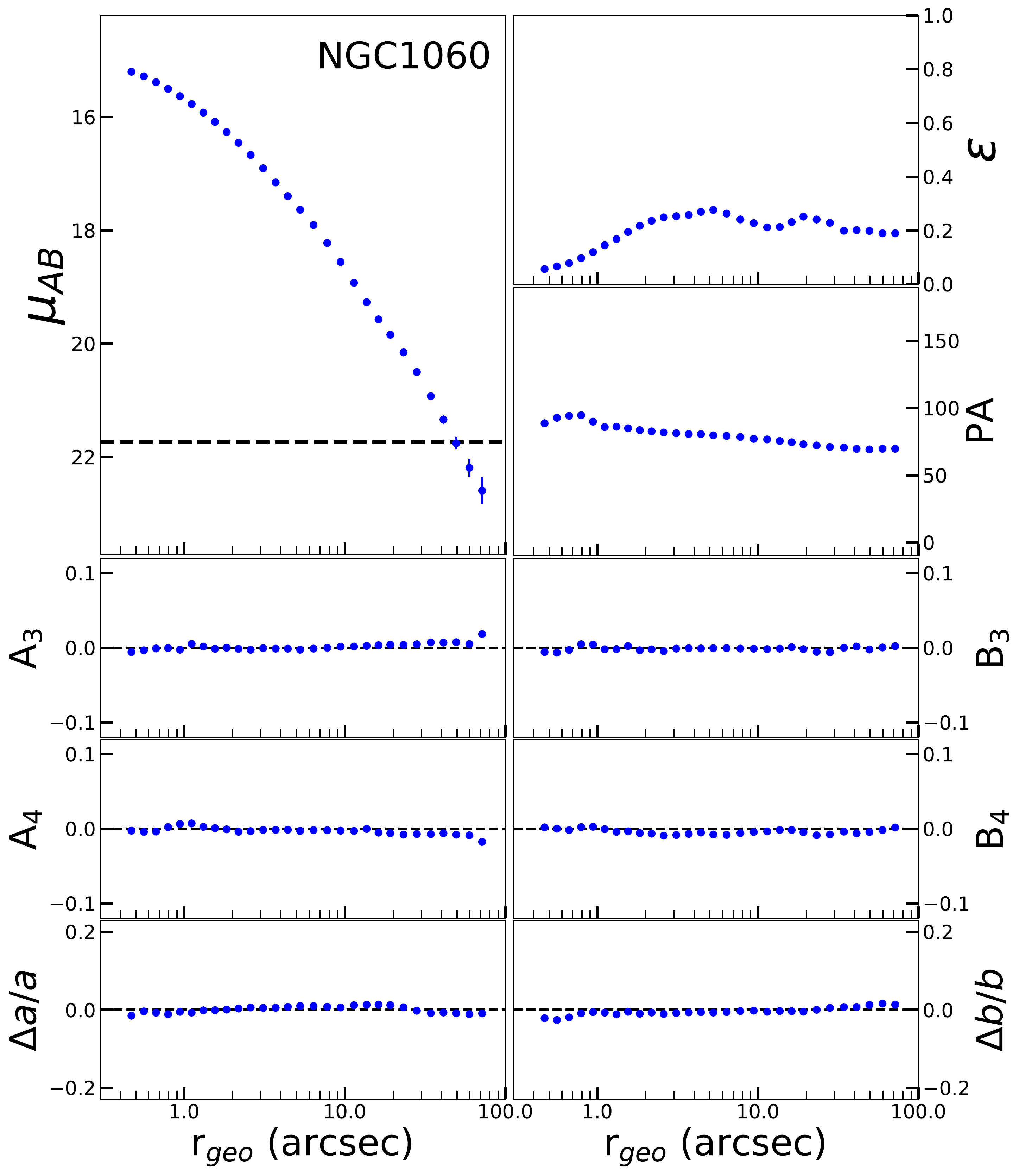}
  \end{minipage}
  \begin{minipage}[b][13.45cm][t]{0.41\textwidth}
    \includegraphics[width=\textwidth,trim=0 0 0 0]{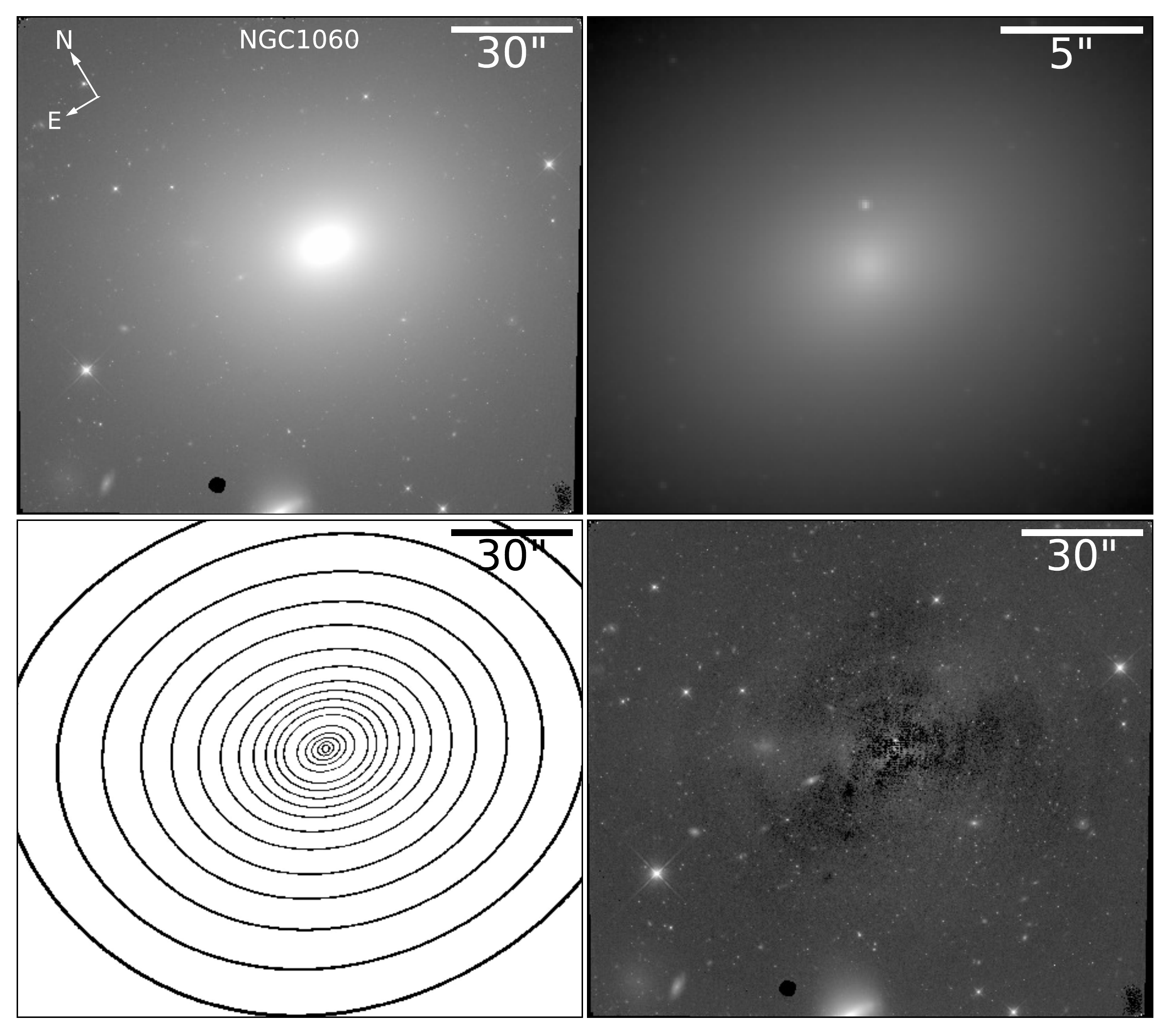}
    \caption{\small NGC~1060 has an ellipticity
      profile that rises as the radius increases,
      peaks at ${\sim}5$ arcsec, and then undergoes a gradual decrease
      for isophotes beyond that
      radius. NGC~1060 has a small companion to the southeast.\\
      Scale: $1$ arcsec = $327$ pc.  }
  \end{minipage}\\
\end{figure*}

\begin{figure*}[!tbp]
  \centering\offinterlineskip
  \begin{minipage}[b][13.55cm][t]{0.56\textwidth}
    \includegraphics[width=\textwidth]{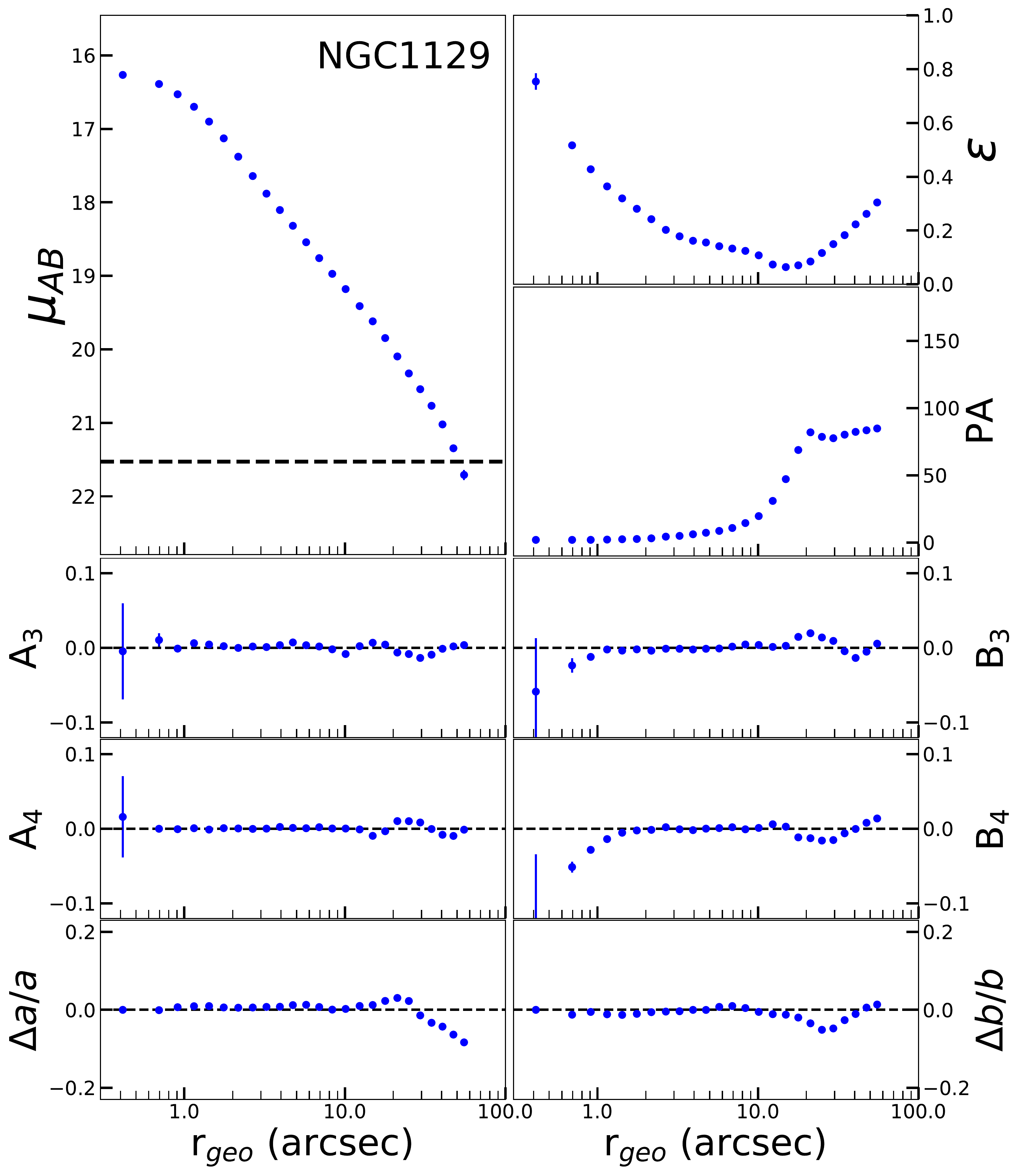}
  \end{minipage}
  \begin{minipage}[b][13.45cm][t]{0.41\textwidth}
    \includegraphics[width=\textwidth,trim=0 0 0 0]{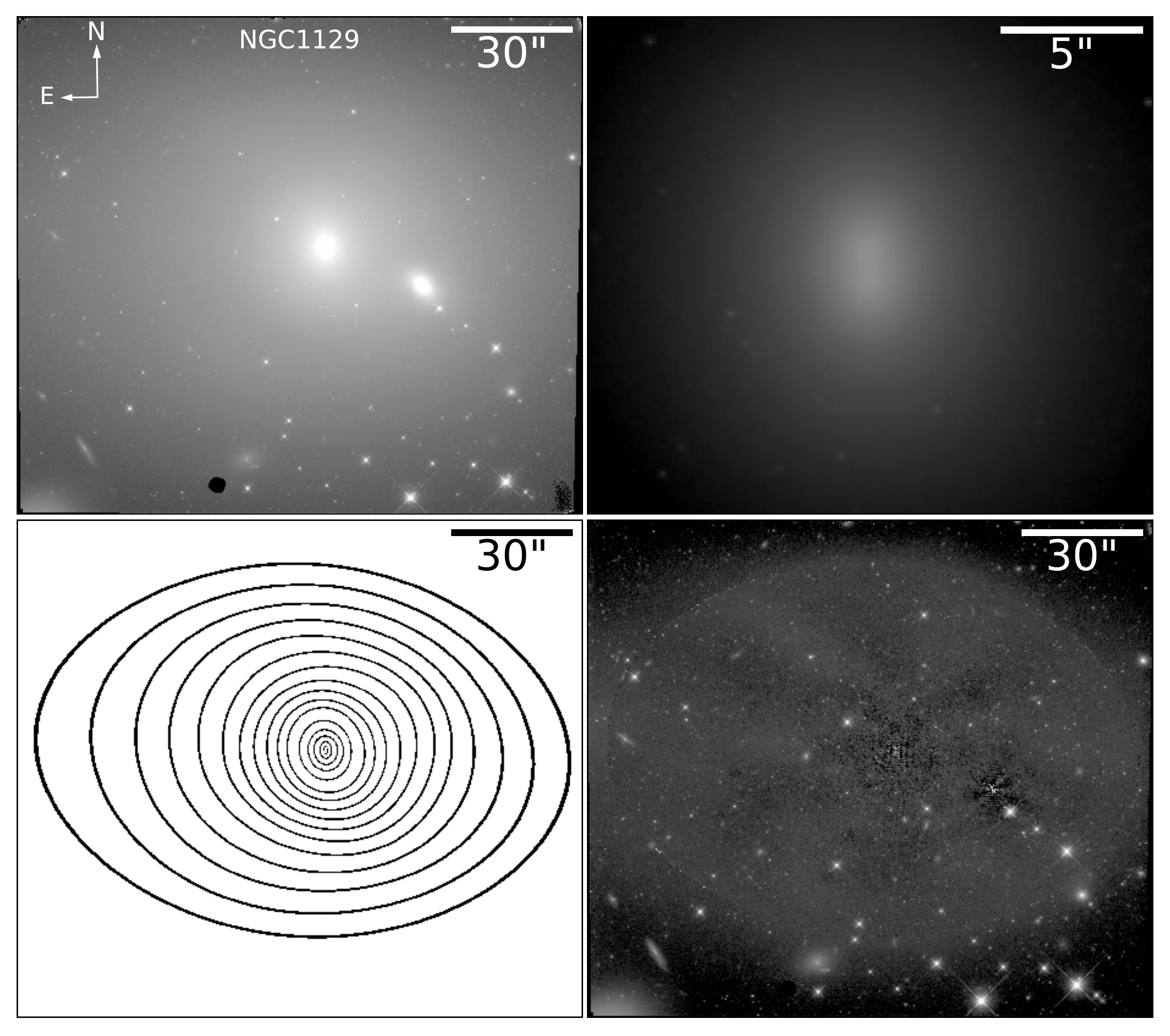}
    \caption{\small NGC~1129 has arguably the most interesting
      isophotes in our sample. The behavior of each parameter is split
      into two regimes. The inner regime extends to ${\sim}15$ arcsec,
      and the outer regime consists of all isophotes beyond that. The
      inner region has a steeply decreasing ellipticity and increasing PA
      out to ${\sim}15$ arcsec. These isophotes are neither boxy
      nor disky. The behavior of the isophotes abruptly changes in
      the outer region. The ellipticity begins 
      increasing, the PA stabilizes, and the isophotes immediately
      become boxy, before becoming more disky for the outermost four
      isophotes. NGC~1129 has a double-peaked nucleus, suggesting it has
      undergone a recent merger. NGC~1129 has a small nearby companion
      $25$ arcsec southwest from the center. \\ 
      Scale: $1$ arcsec = $358$ pc.}
  \end{minipage}\\
  \vspace{-1.3cm}
  \begin{minipage}[b][13.55cm][t]{0.56\textwidth}
    \includegraphics[width=\textwidth]{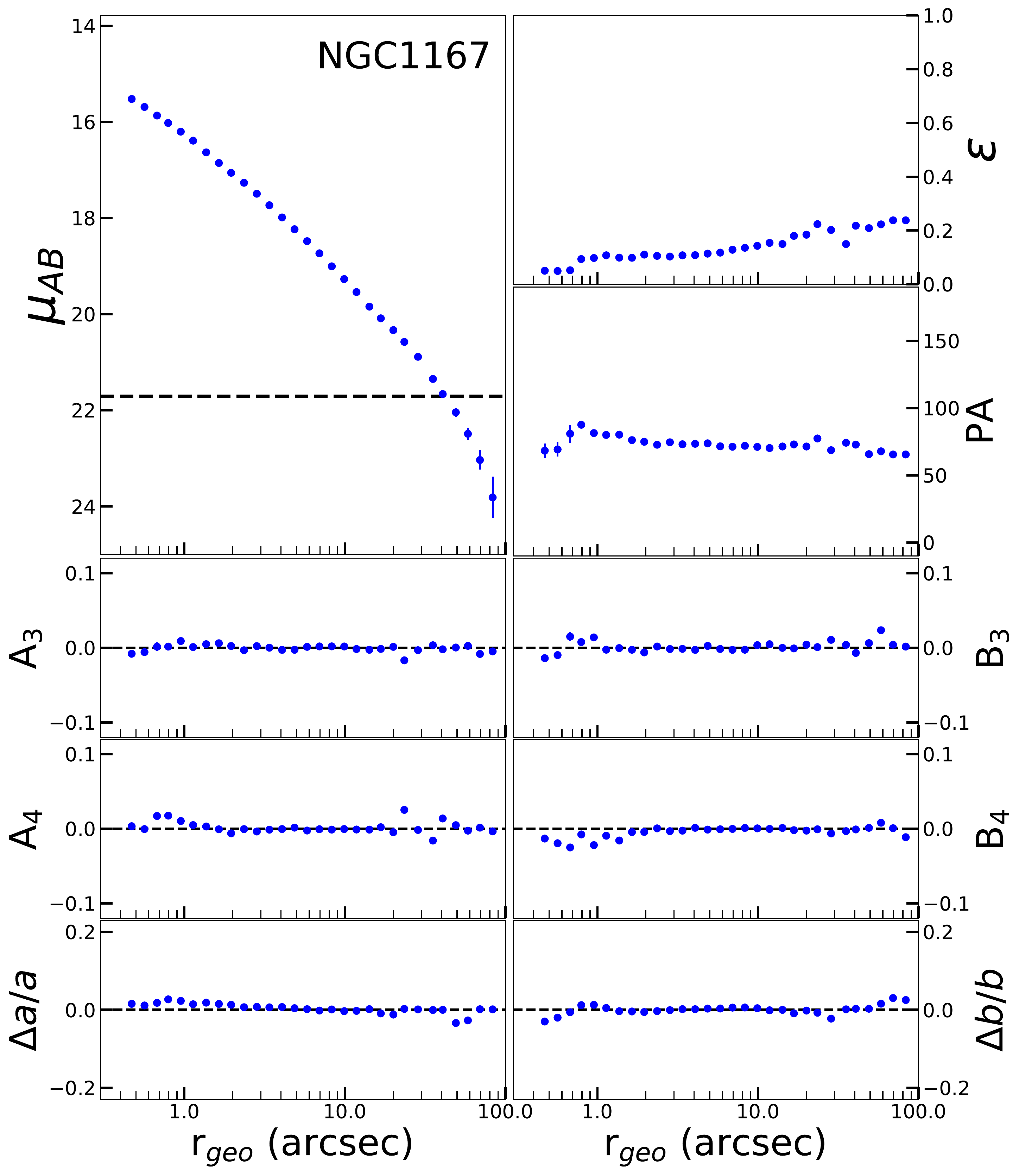}
  \end{minipage}
  \begin{minipage}[b][13.45cm][t]{0.41\textwidth}
    \includegraphics[width=\textwidth,trim=0 0 0 0]{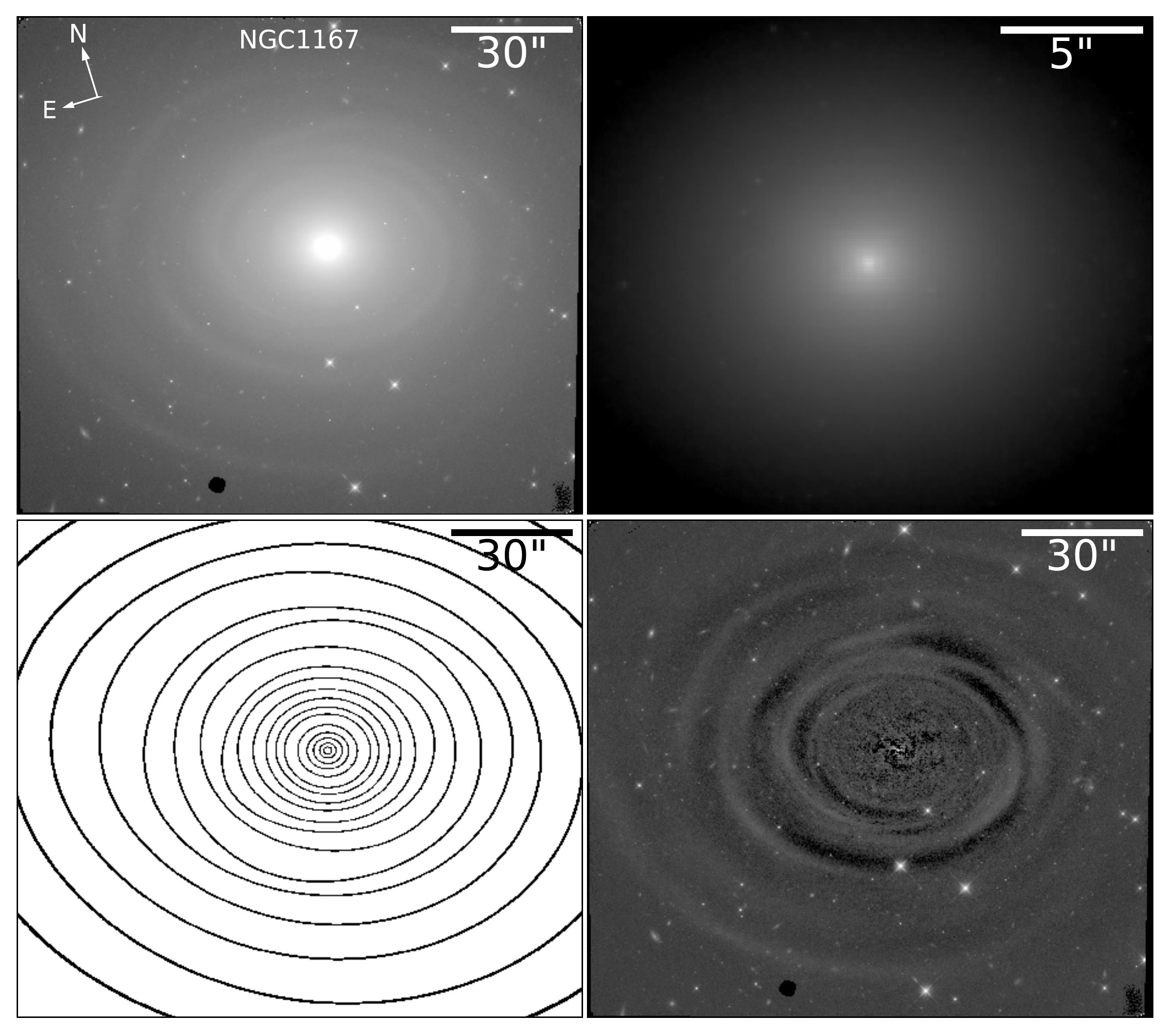}
    \caption{\small NGC~1167 is a S0 disk galaxy. Unlike NGC~0665, which has
      a strong bar, 
      the NGC~1167 isophotes do not exhibit any special behavior beyond some
      boxiness within ${\sim}1.5$ arcsec. \\
      Scale: $1$ arcsec = $340$ pc.  }
  \end{minipage}\\
\end{figure*}

\begin{figure*}[!tbp]
  \centering\offinterlineskip
  \begin{minipage}[b][13.55cm][t]{0.56\textwidth}
    \includegraphics[width=\textwidth]{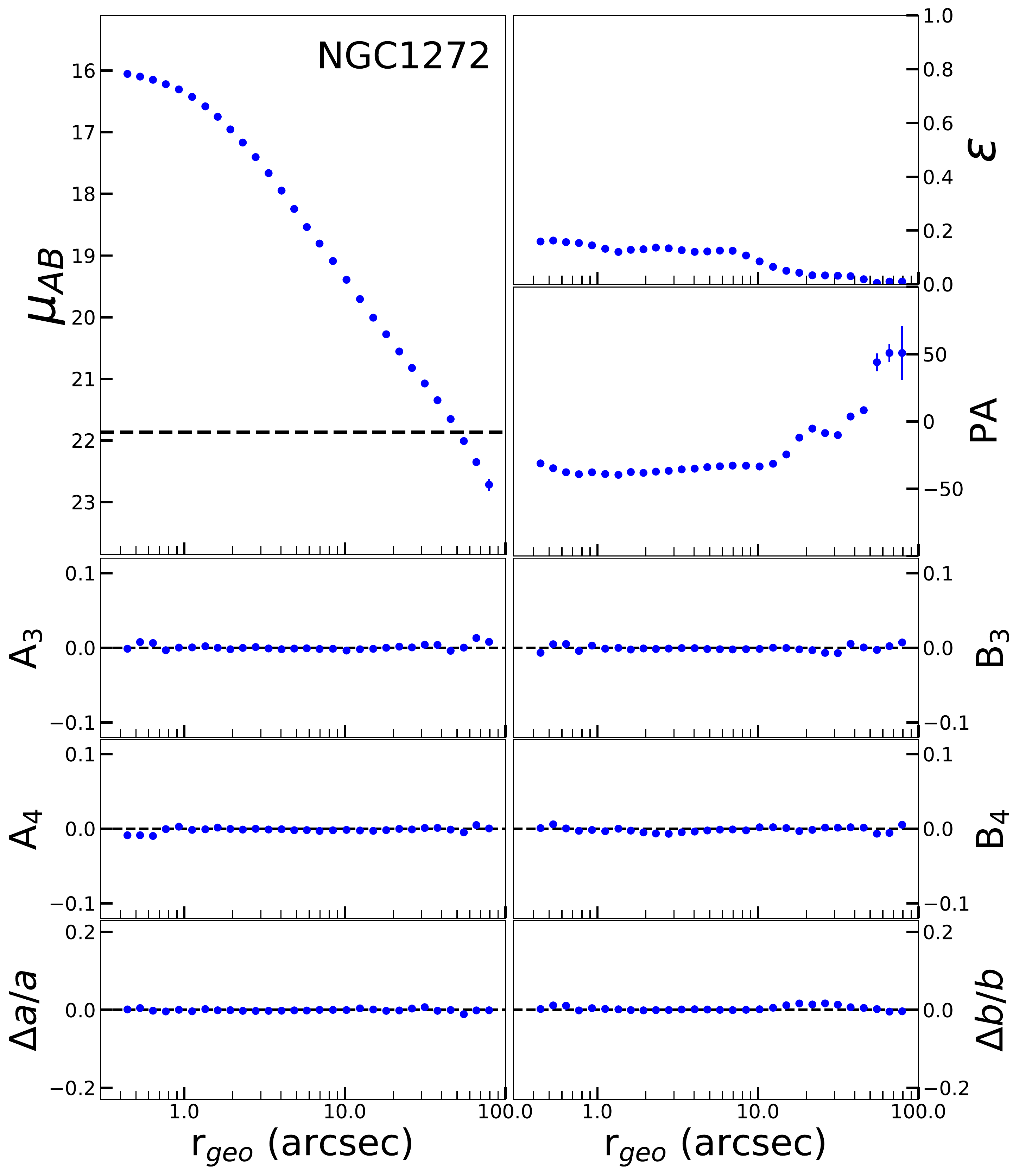}
  \end{minipage}
  \begin{minipage}[b][13.45cm][t]{0.41\textwidth}
    \includegraphics[width=\textwidth,trim=0 0 0 0]{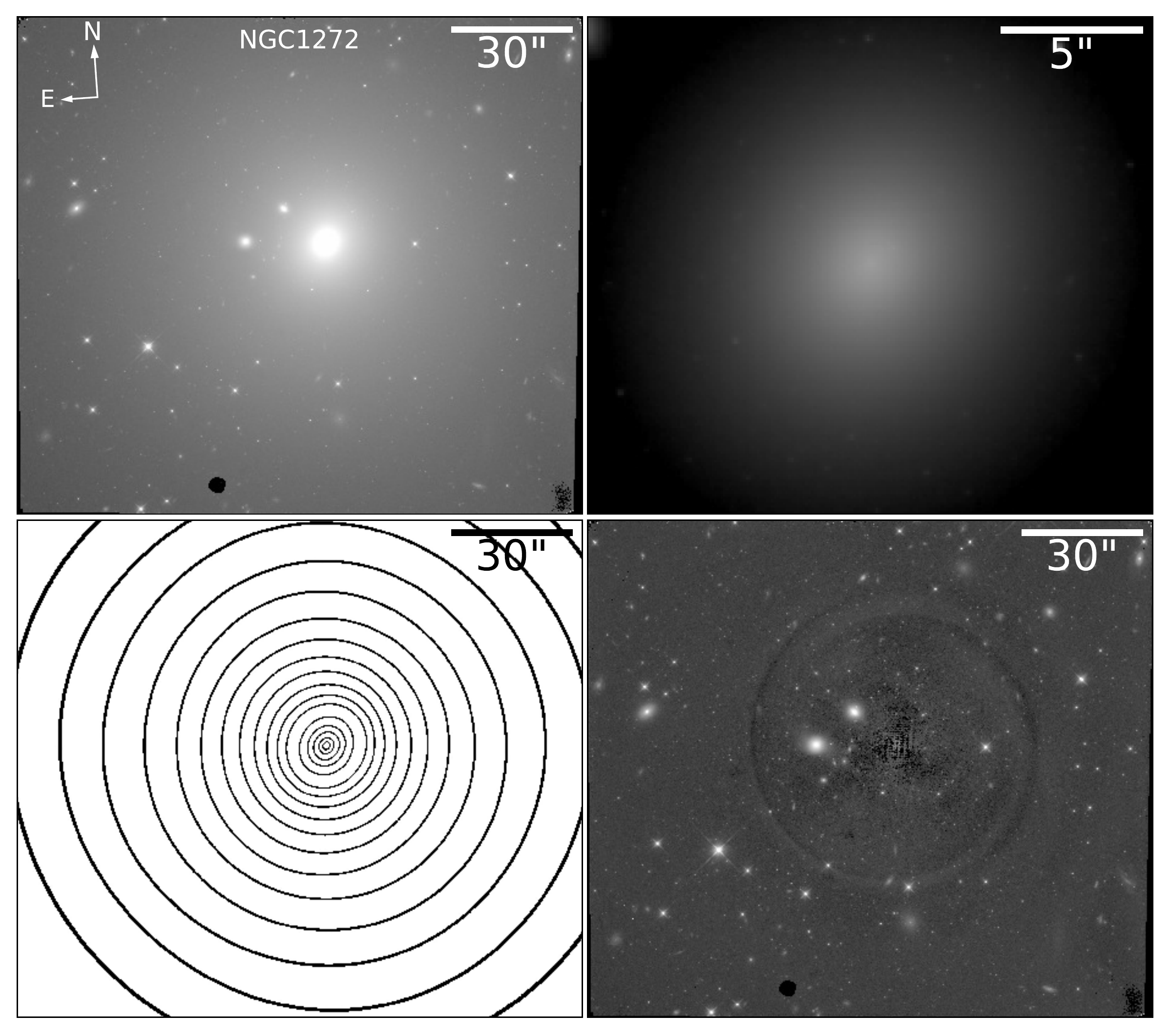}
    \caption{\small NGC~1272 features a large PA twist beyond
      ${\sim}10$ arcsec that coincides with a steep decrease in
      the ellipticity, and the outermost isophotes are almost
      perfectly round. NGC~1272 has two small
      companions $15$ and $21$ arcsec northeast of the center. \\
      Scale: $1$ arcsec = $376$ pc.  }
  \end{minipage}\\
  \vspace{-1.3cm}
  \begin{minipage}[b][13.55cm][t]{0.56\textwidth}
    \includegraphics[width=\textwidth]{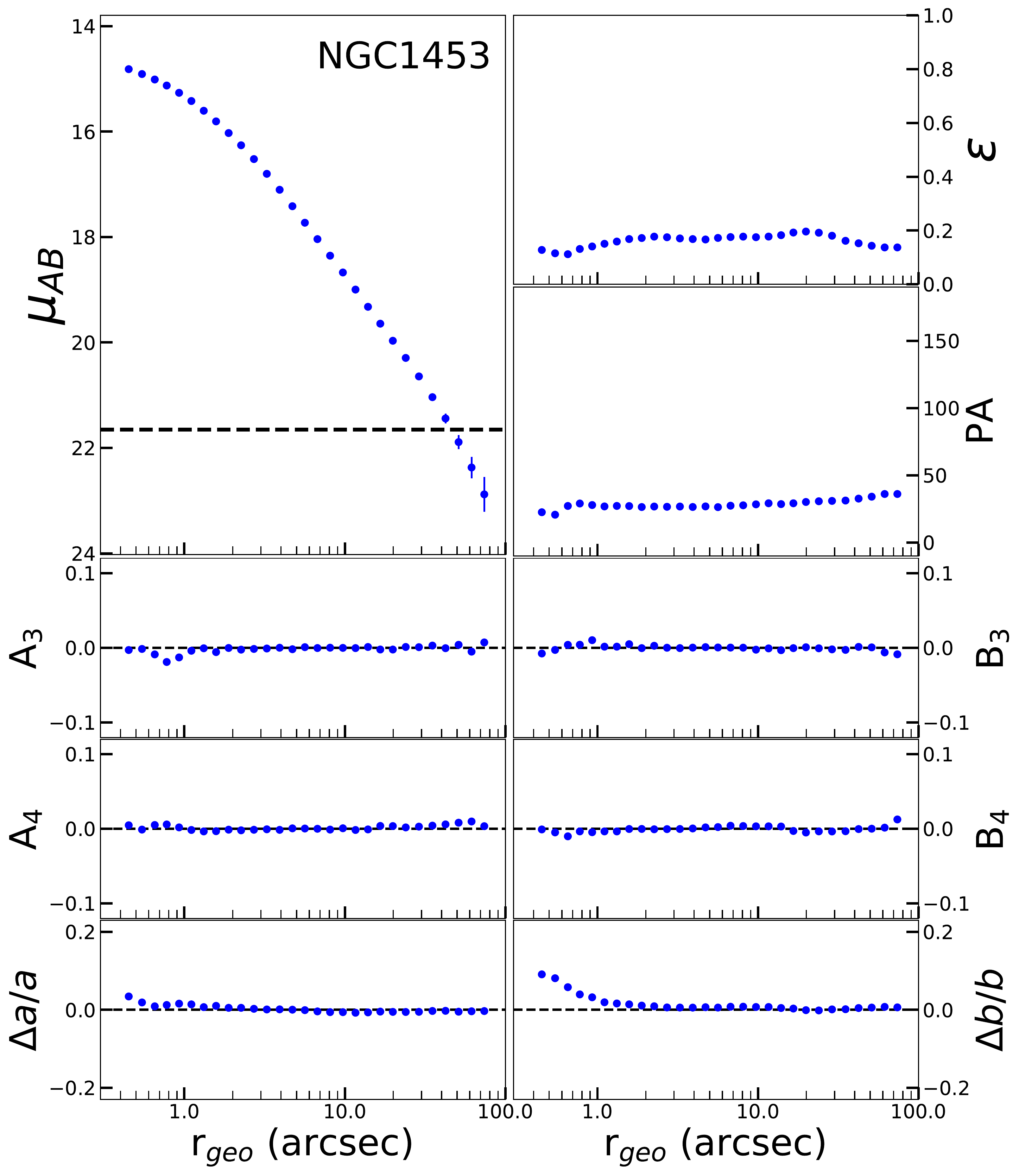}
  \end{minipage}
  \begin{minipage}[b][13.45cm][t]{0.41\textwidth}
    \includegraphics[width=\textwidth,trim=0 0 0 0]{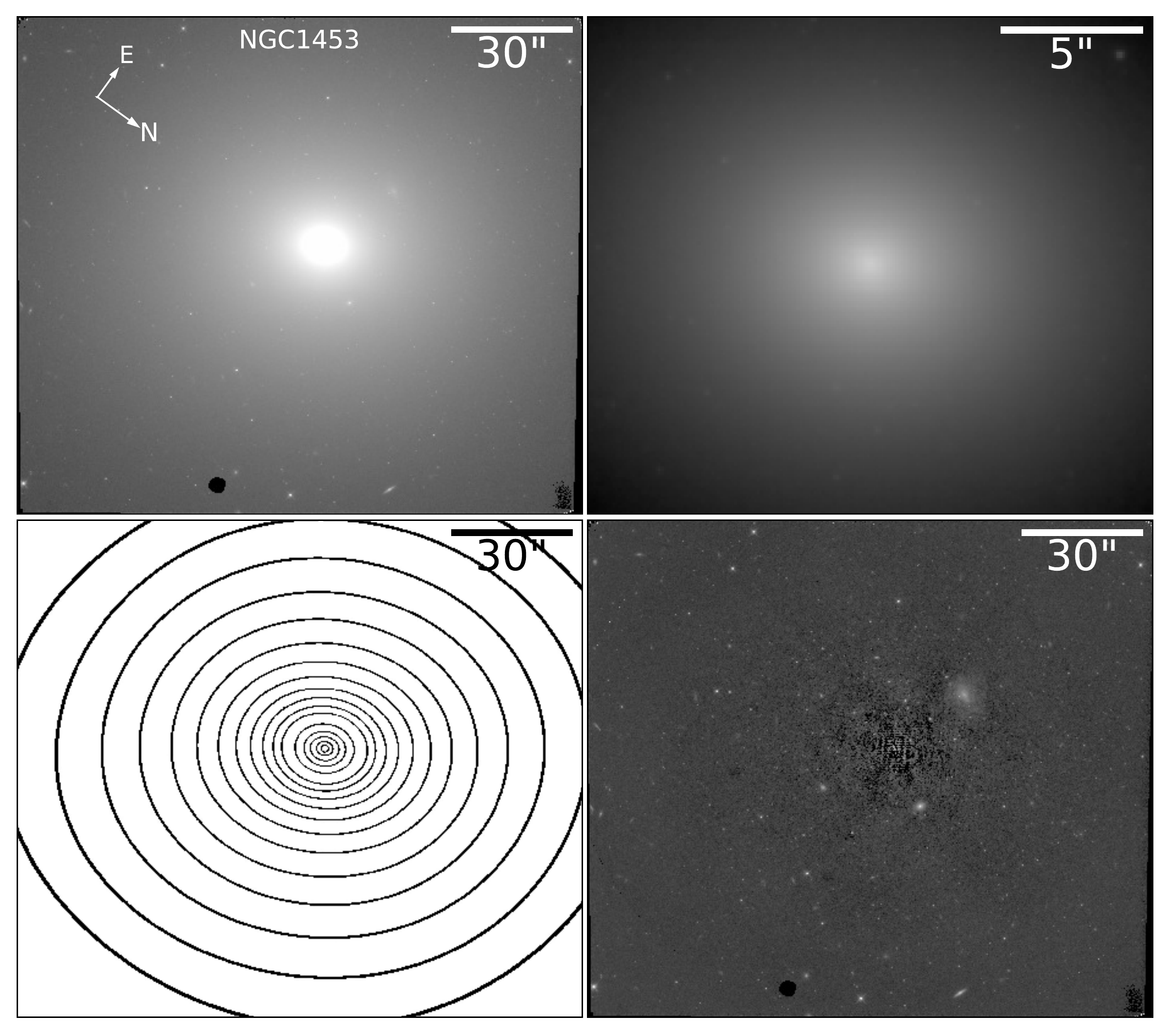}
    \caption{\small NGC~1453 has a noticeable central drift within
      ${\sim}0.6$ arcsec. The extent of the drift is difficult to
      determine, as it is close to the resolution limit. The
      ellipticity varies only slightly. \\
      Scale: $1$ arcsec = $273$ pc.}
  \end{minipage}\\
\end{figure*}

\begin{figure*}[!tbp]
  \centering\offinterlineskip
  \begin{minipage}[b][13.55cm][t]{0.56\textwidth}
    \includegraphics[width=\textwidth]{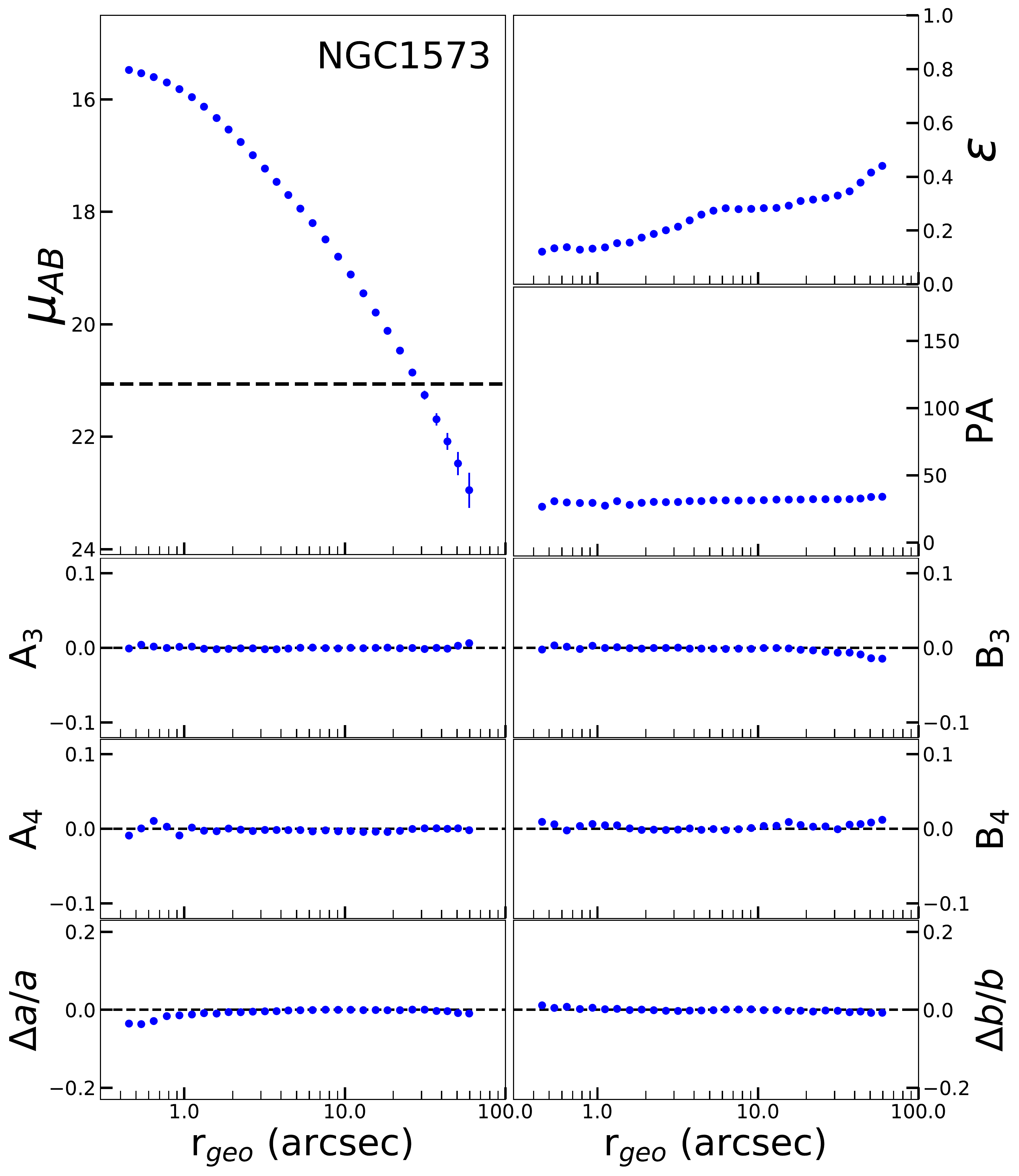}
  \end{minipage}
  \begin{minipage}[b][13.45cm][t]{0.41\textwidth}
    \includegraphics[width=\textwidth,trim=0 0 0 0]{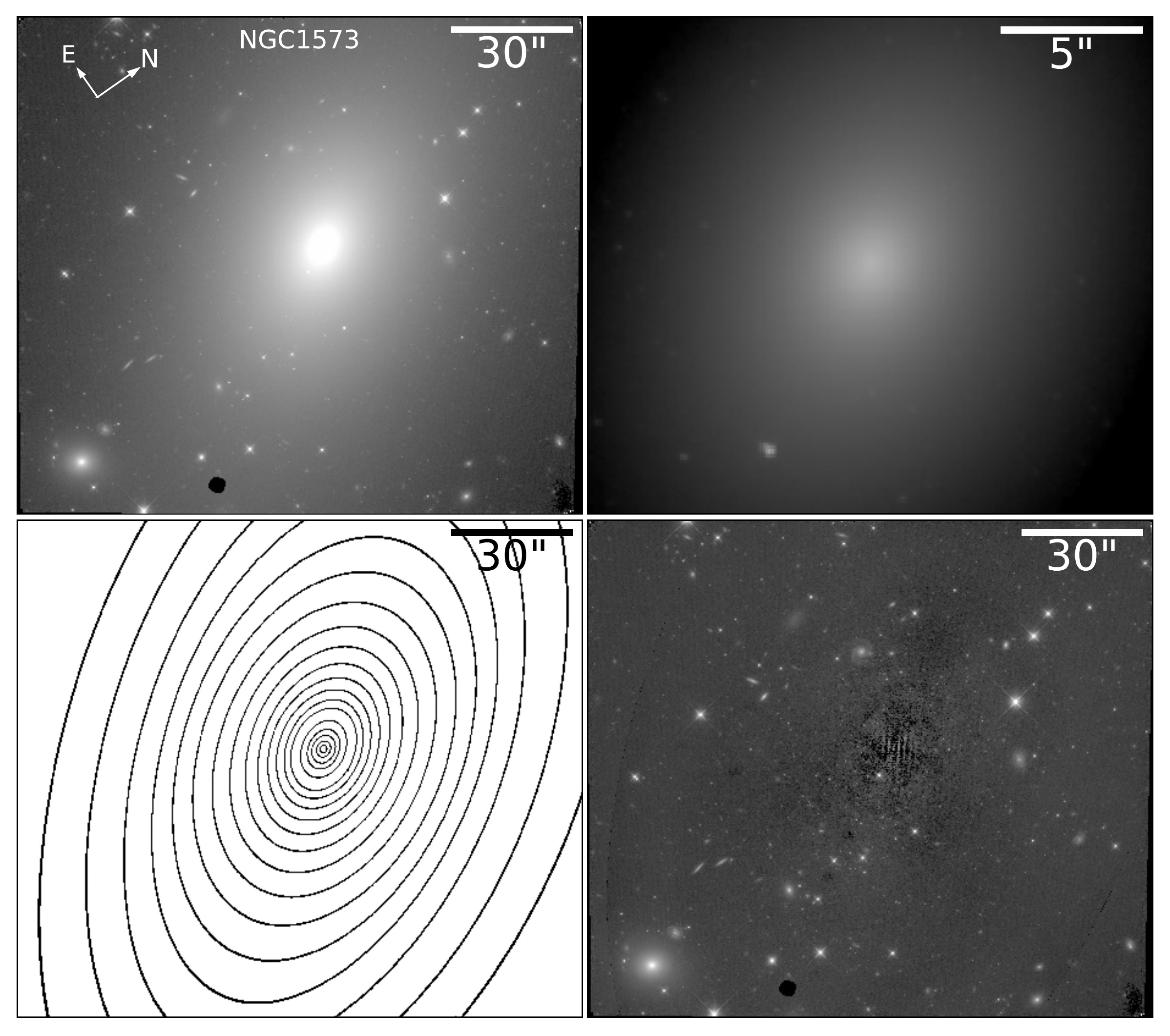}
    \caption{\small NGC~1573 has significantly increasing ellipticity
      with radius; the other parameters are fairly constant.\\
      Scale: $1$ arcsec = $315$ pc.}
  \end{minipage}
  \vspace{-1.3cm}
  \begin{minipage}[b][13.55cm][t]{0.56\textwidth}
    \includegraphics[width=\textwidth]{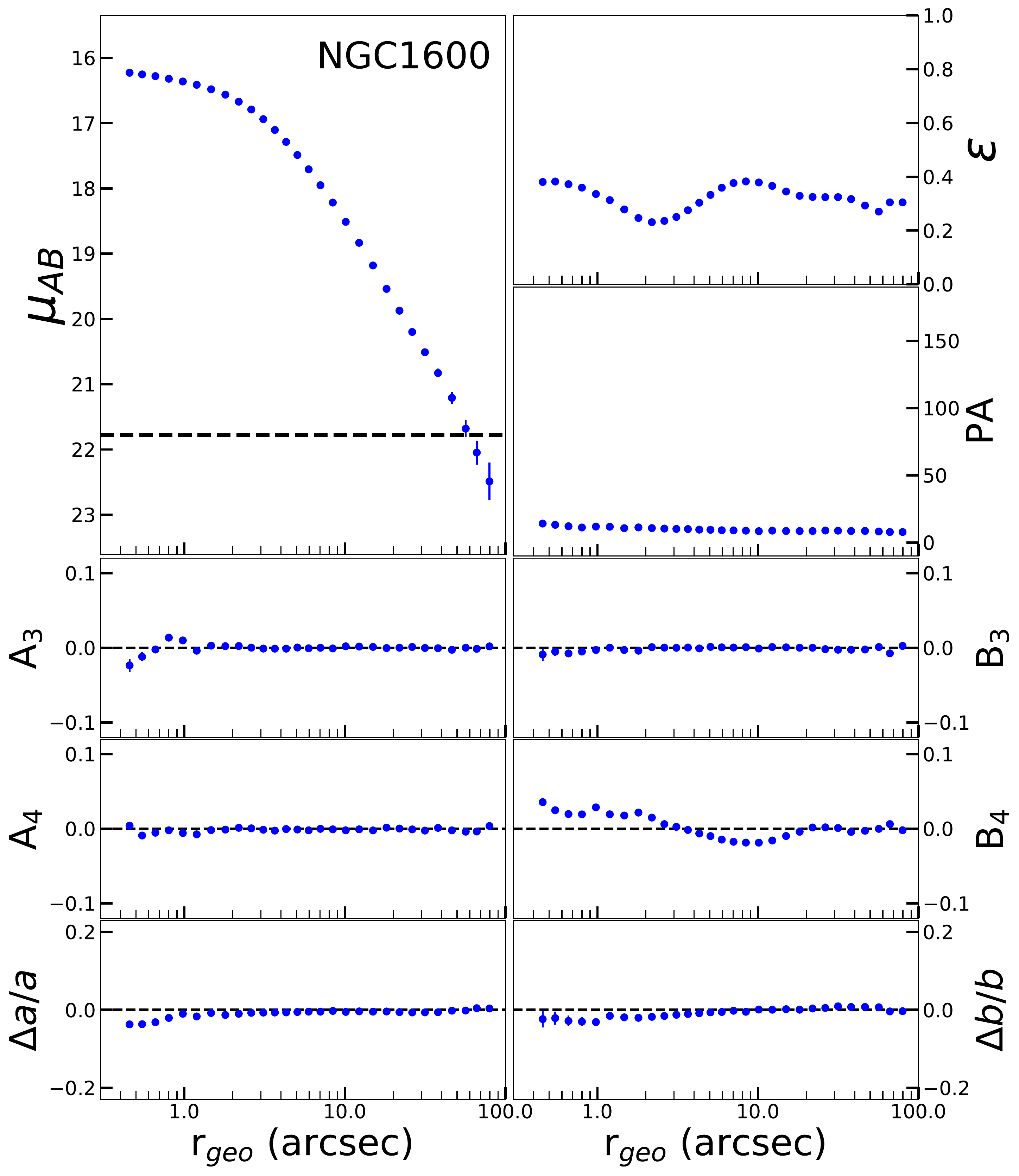}
  \end{minipage}
  \begin{minipage}[b][13.45cm][t]{0.41\textwidth}
    \includegraphics[width=\textwidth,trim=0 0 0 0]{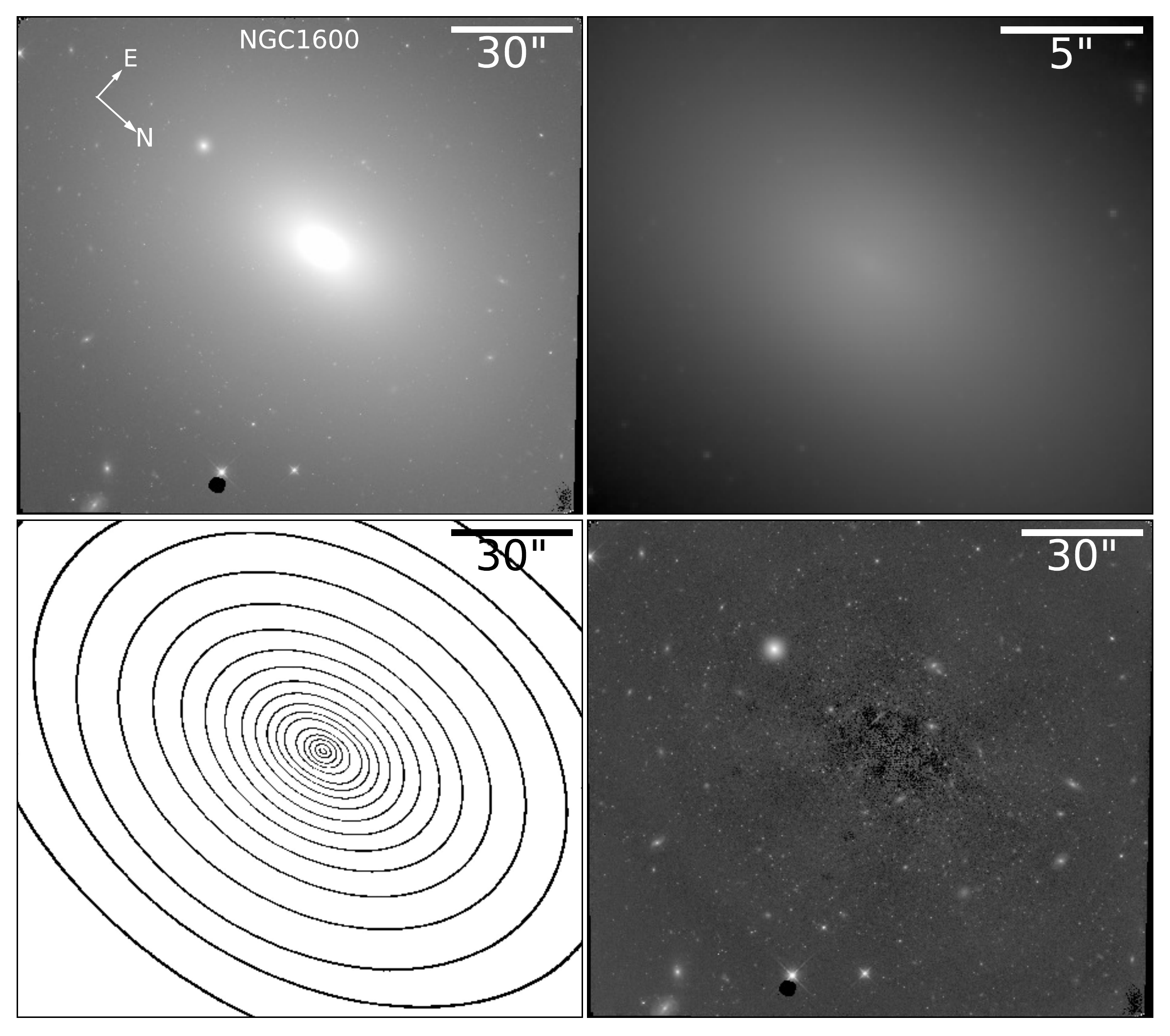}
    \caption{\small NGC~1600 has a variable ellipticity with an absolute
      minimum at ${\sim}2$ arcsec.
      The inner isophotes are significantly disky, but transition to boxy
      isophotes around ${\sim}10$ arcsec,
      beyond which they return to pure ellipticity.\\
      Scale: $1$ arcsec = $309$ pc.}
  \end{minipage}\\
\end{figure*}

\begin{figure*}[!tbp]
  \centering\offinterlineskip
  \begin{minipage}[b][13.55cm][t]{0.56\textwidth}
    \includegraphics[width=\textwidth]{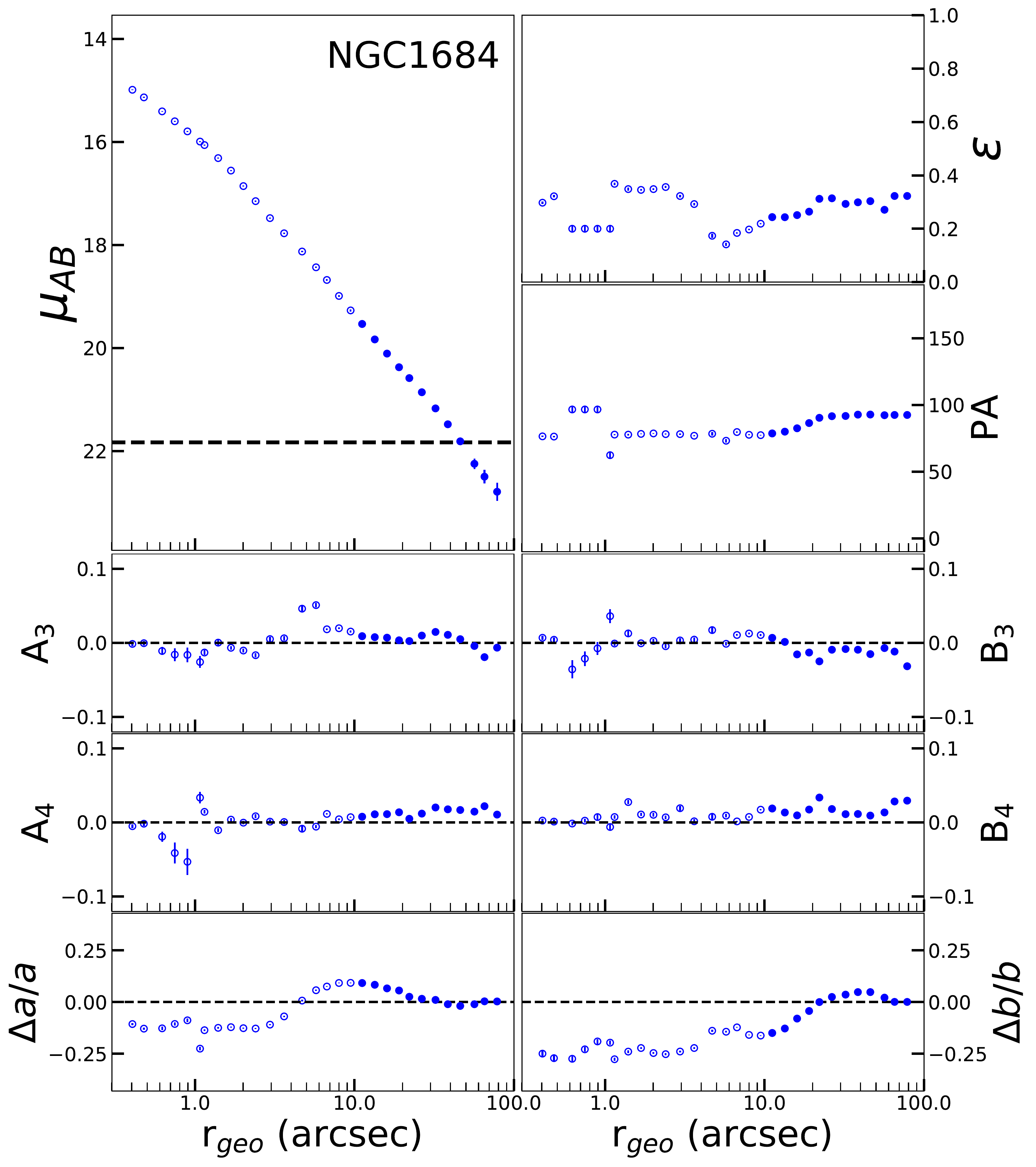}
  \end{minipage}
  \begin{minipage}[b][13.45cm][t]{0.41\textwidth}
    \includegraphics[width=\textwidth,trim=0 0 0 0]{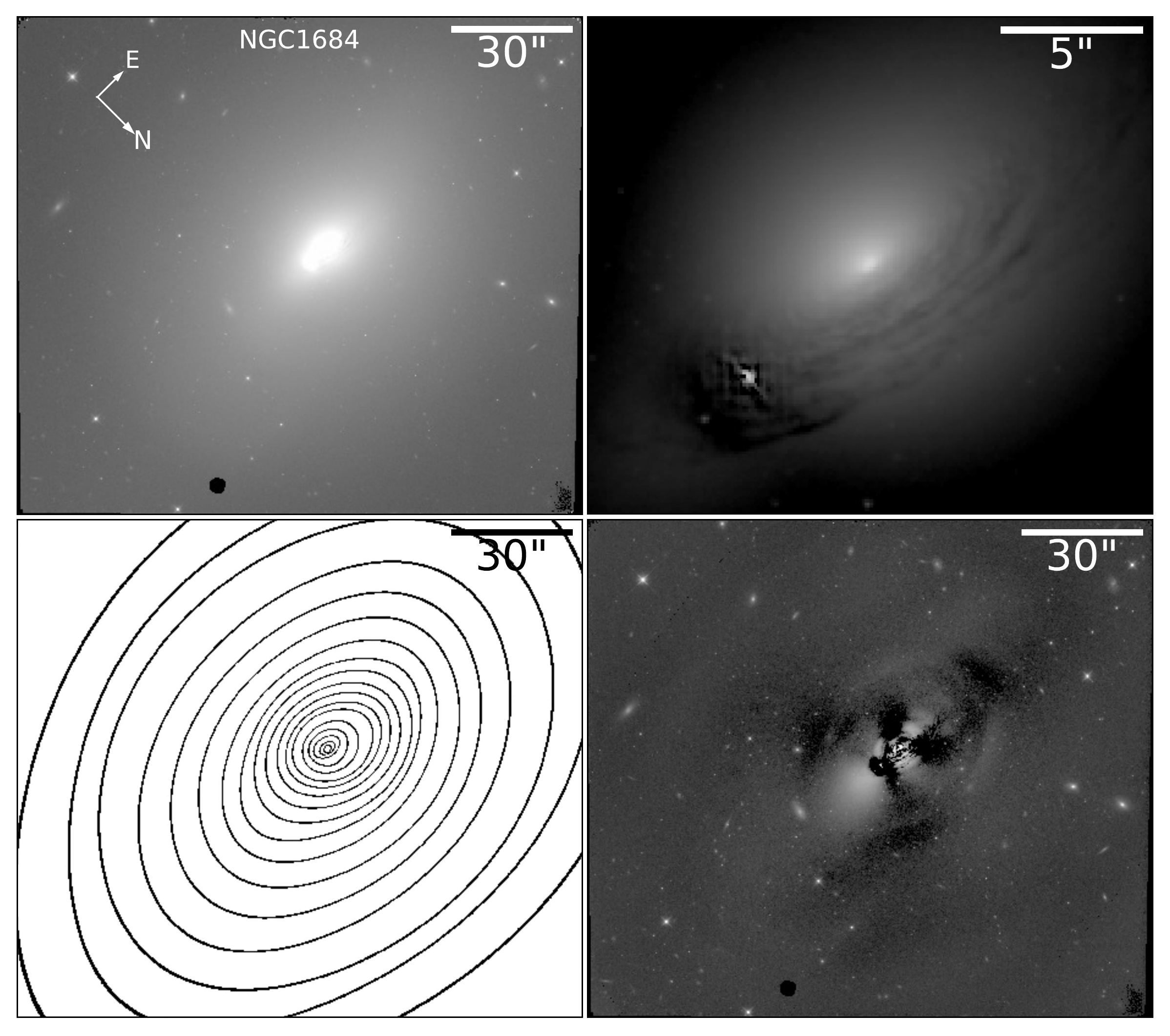}
    \caption{\small NGC~1684 has a central dust disk that distorts the
      isophotes within the central $11$ arcsec. Outside of this disk,
      the isophotes are disky. NGC~1684 has a small companion $6$ arcsec west
      of the center. \\
      Scale: $1$ arcsec = $308$ pc.}
  \end{minipage}\\
  \vspace{-1.3cm}
  \begin{minipage}[b][13.55cm][t]{0.56\textwidth}
    \includegraphics[width=\textwidth]{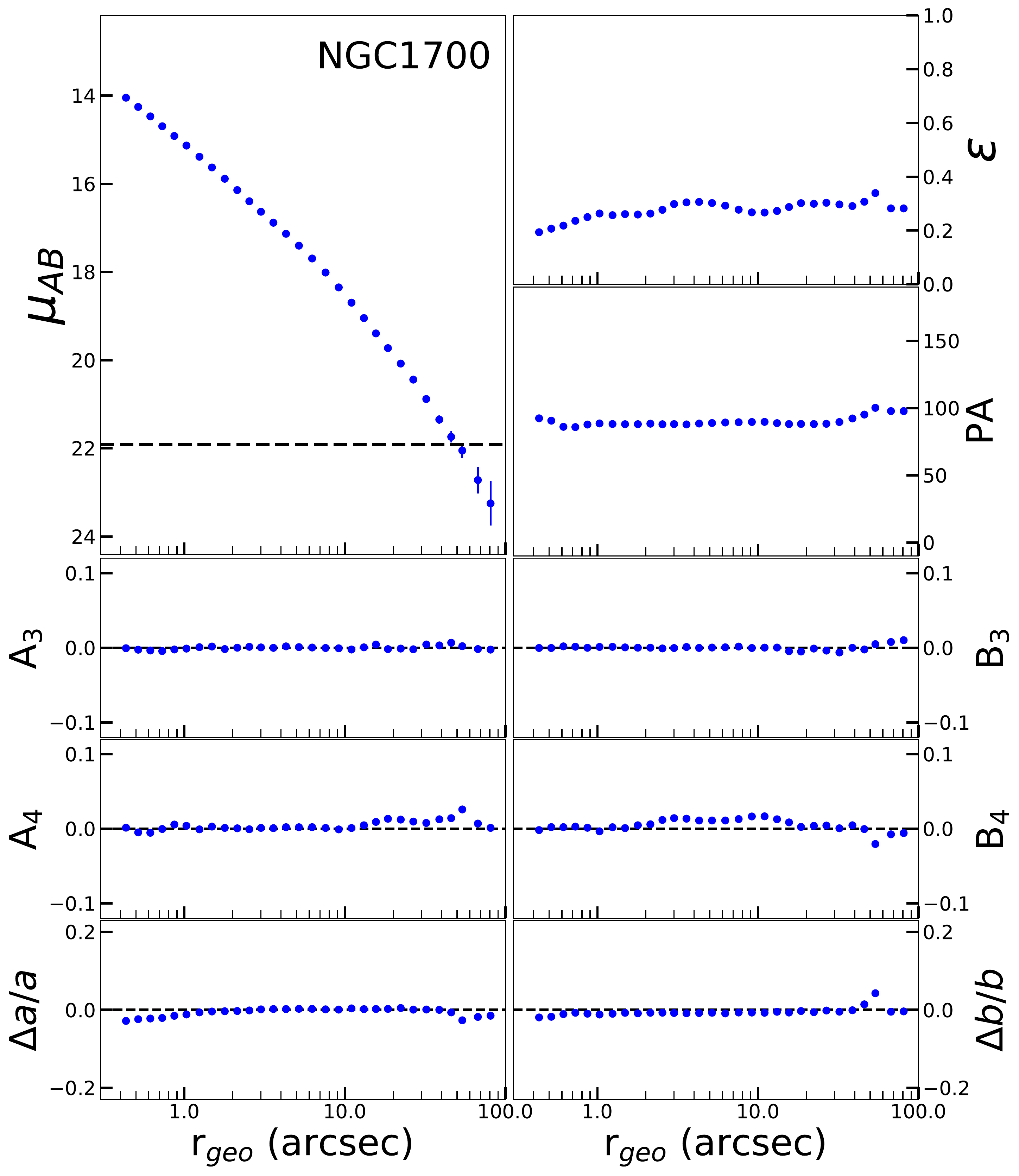}
  \end{minipage}
  \begin{minipage}[b][13.45cm][t]{0.41\textwidth}
    \includegraphics[width=\textwidth,trim=0 0 0 0]{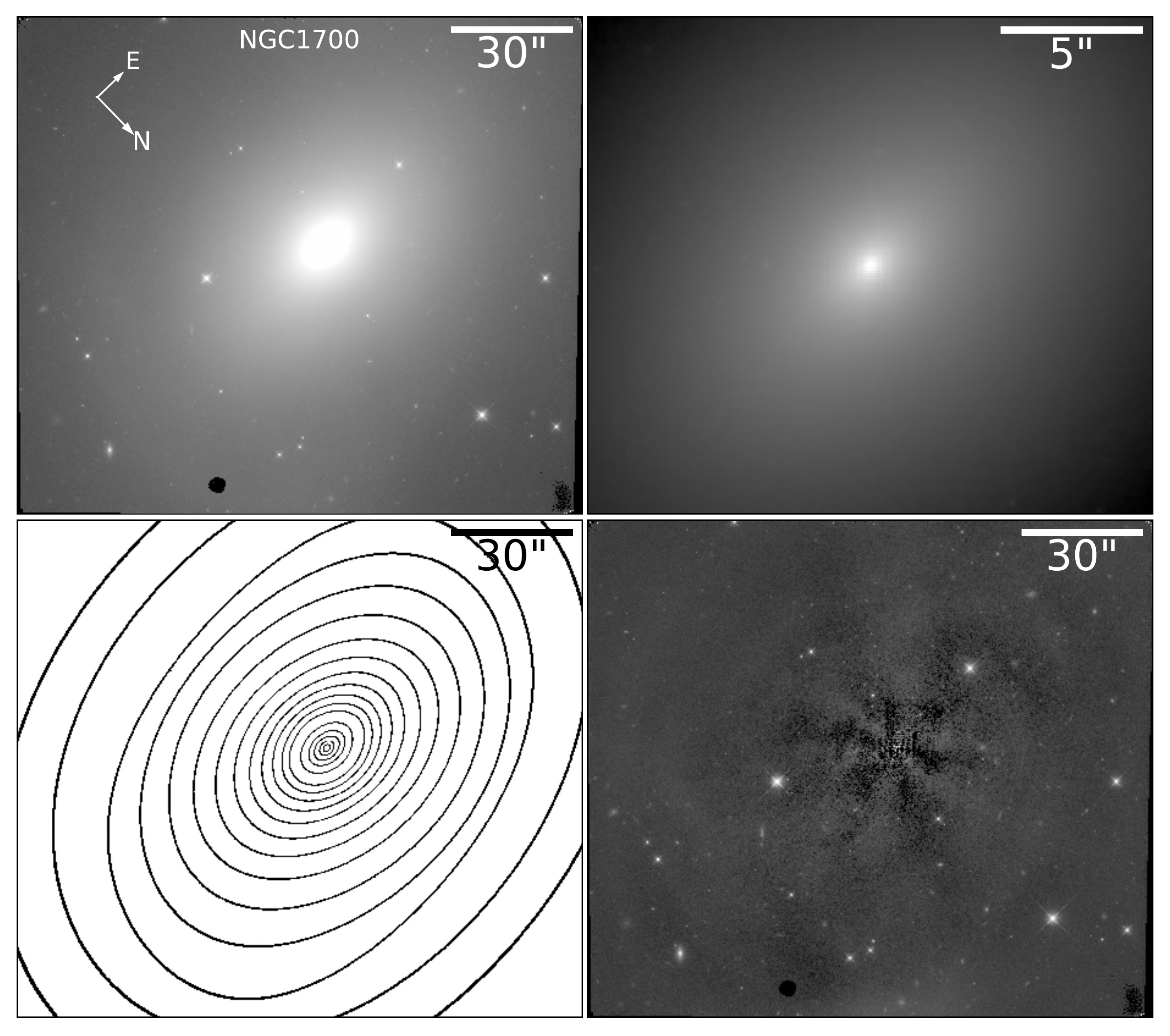}
    \caption{\small NGC~1700 has variable ellipticity that reaches a peak at
      ${\sim}4$ arcsec. Beyond this value the ellipticity has small
      oscillations around ${\sim}0.3$. Between ${\sim}2$ arcsec and
      ${\sim}12$ arcsec the
      isophotes are disky and elliptical otherwise.\\
      Scale: $1$ arcsec = $264$ pc.}
  \end{minipage}\\
\end{figure*}

\begin{figure*}[!tbp]
  \centering\offinterlineskip
  \begin{minipage}[b][13.55cm][t]{0.56\textwidth}
    \includegraphics[width=\textwidth]{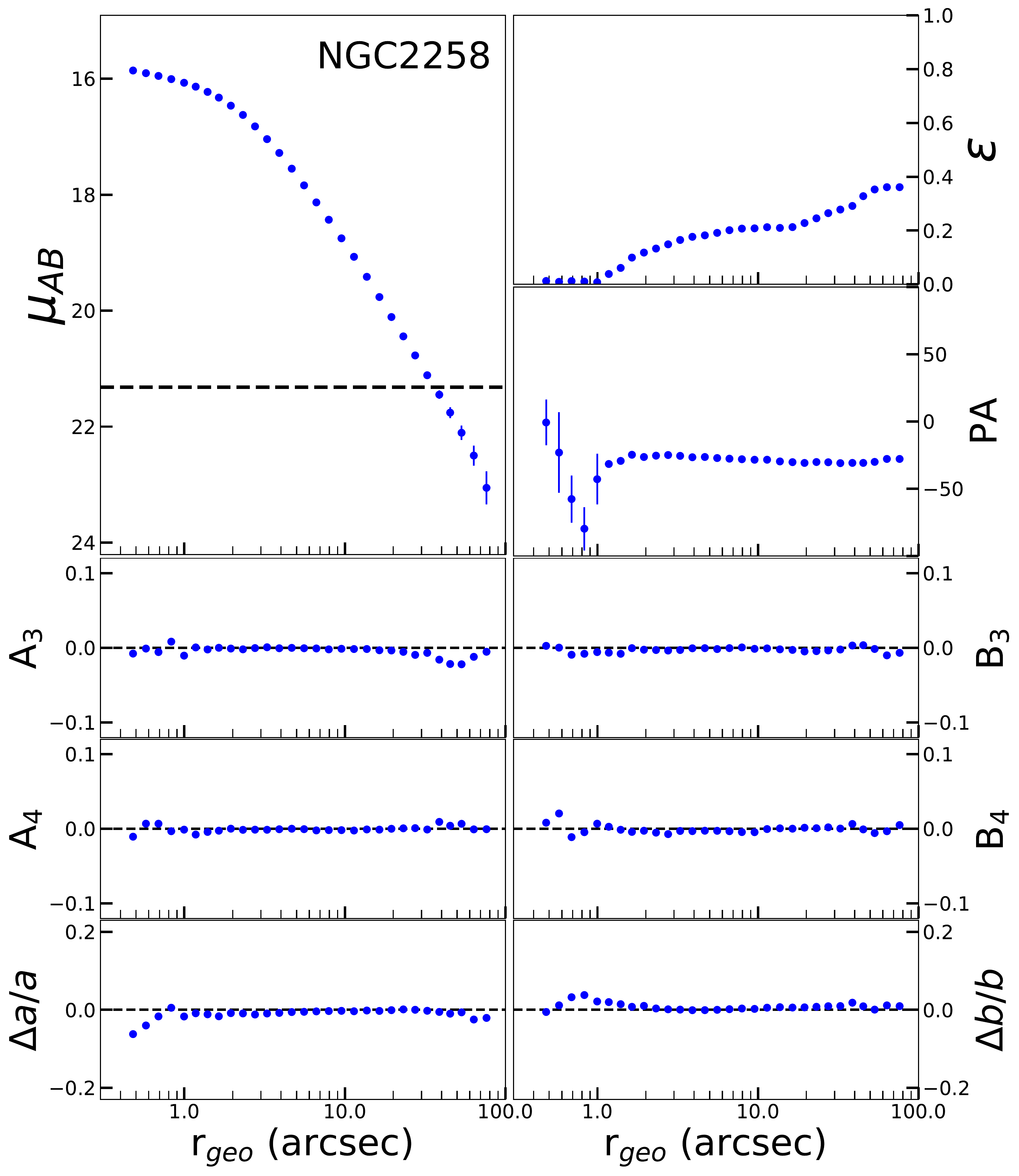}
  \end{minipage}
  \begin{minipage}[b][13.45cm][t]{0.41\textwidth}
    \includegraphics[width=\textwidth,trim=0 0 0 0]{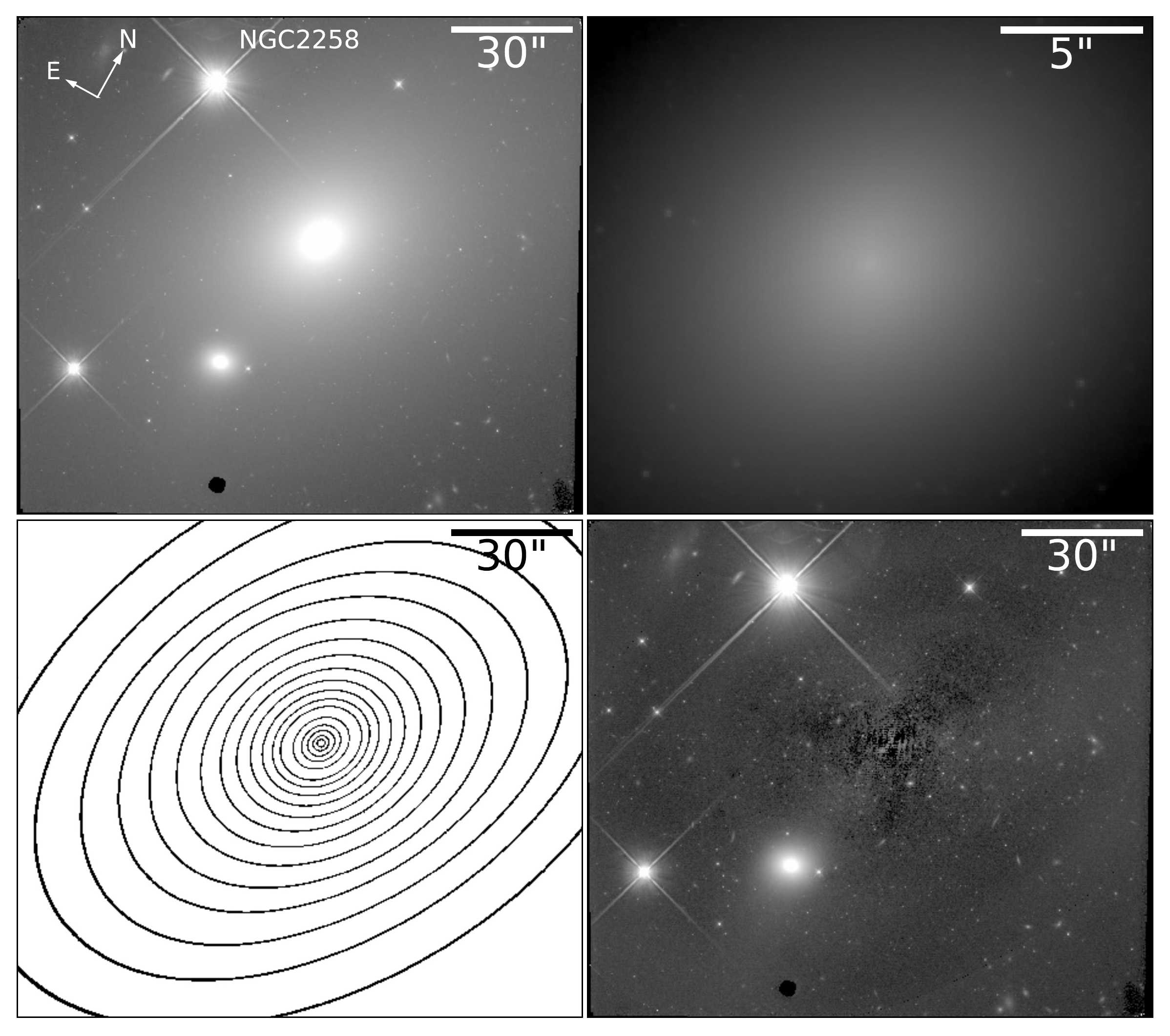}
    \caption{\small NGC~2258 has a small companion $39$ arcsec to the south.
      Its ellipticity rises strongly with radius.\\
      Scale: $1$ arcsec = $286$ pc.}
  \end{minipage}\\
  \vspace{-1.3cm}
  \begin{minipage}[b][13.55cm][t]{0.56\textwidth}
    \includegraphics[width=\textwidth]{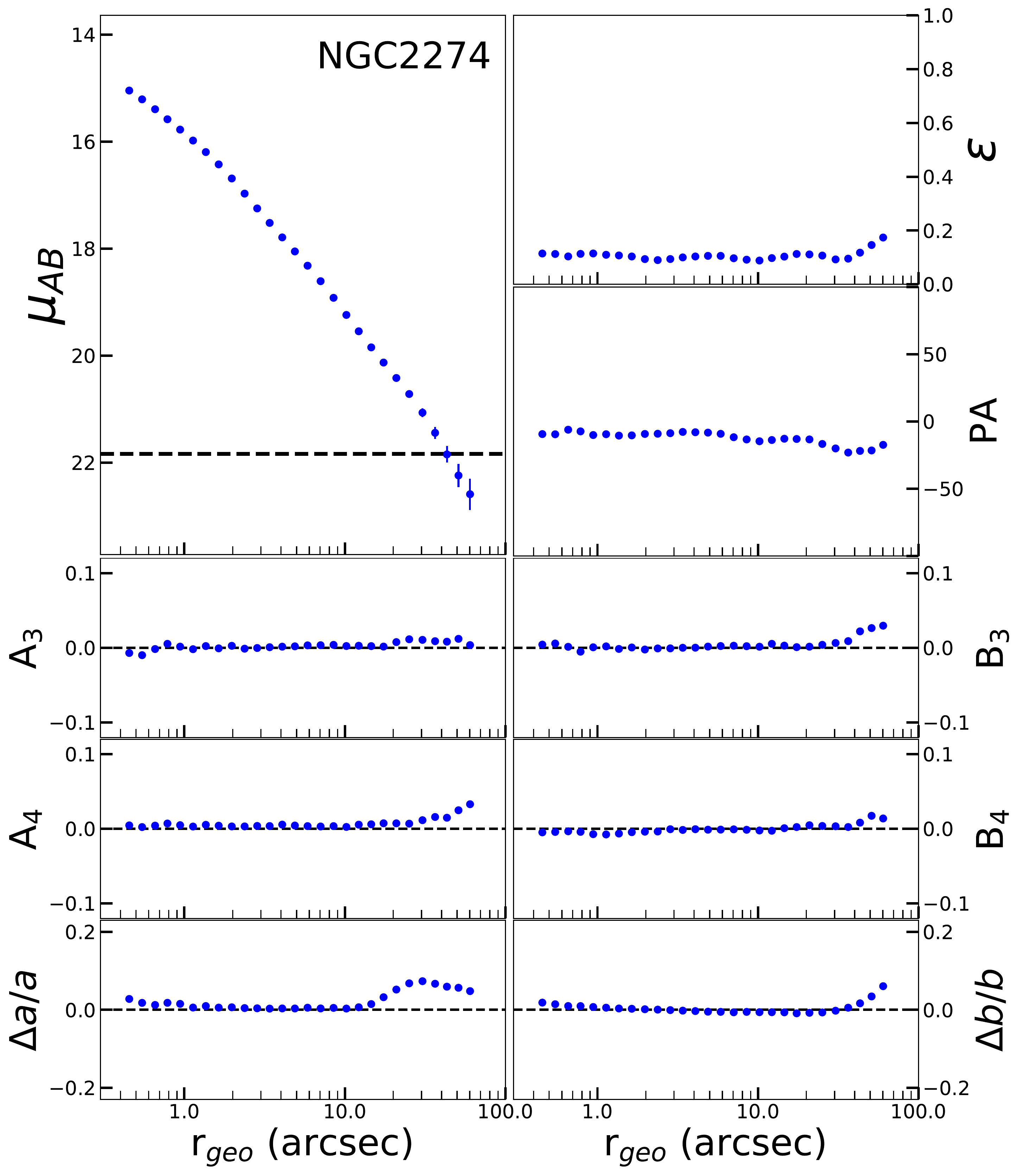}
  \end{minipage}
  \begin{minipage}[b][13.45cm][t]{0.41\textwidth}
    \includegraphics[width=\textwidth,trim=0 0 0 0]{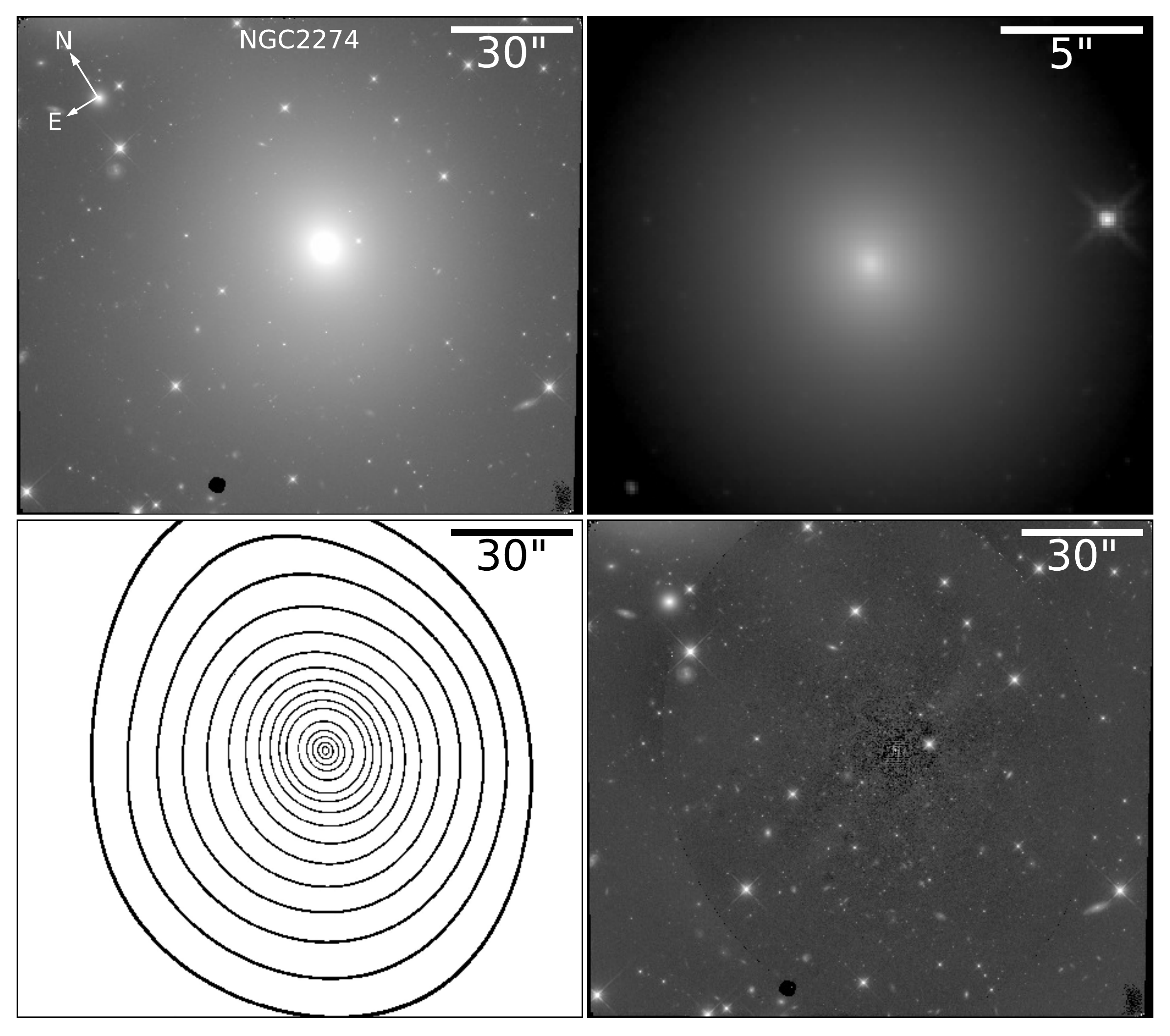}
    \caption{\small NGC~2274 has an extremely flat ellipticity, which
      remains at ${\sim}0.1$ until ${\sim}30$ arcsec, beyond which it
      rises to ${\sim}0.18$. The outermost isophotes become
      disky, corresponding to the radius where ellipticity increases. 
      Beyond ${\sim}12$ arcsec the isophote centers begins to drift,
      reaching a ${\sim}5\%$ shift along
      the semi-major and semi-minor axes by the outermost isophotes.\\
      Scale: $1$ arcsec = $358$ pc.}
  \end{minipage}\\
\end{figure*}

\begin{figure*}[!tbp]
  \centering\offinterlineskip
  \begin{minipage}[b][13.55cm][t]{0.56\textwidth}
    \includegraphics[width=\textwidth]{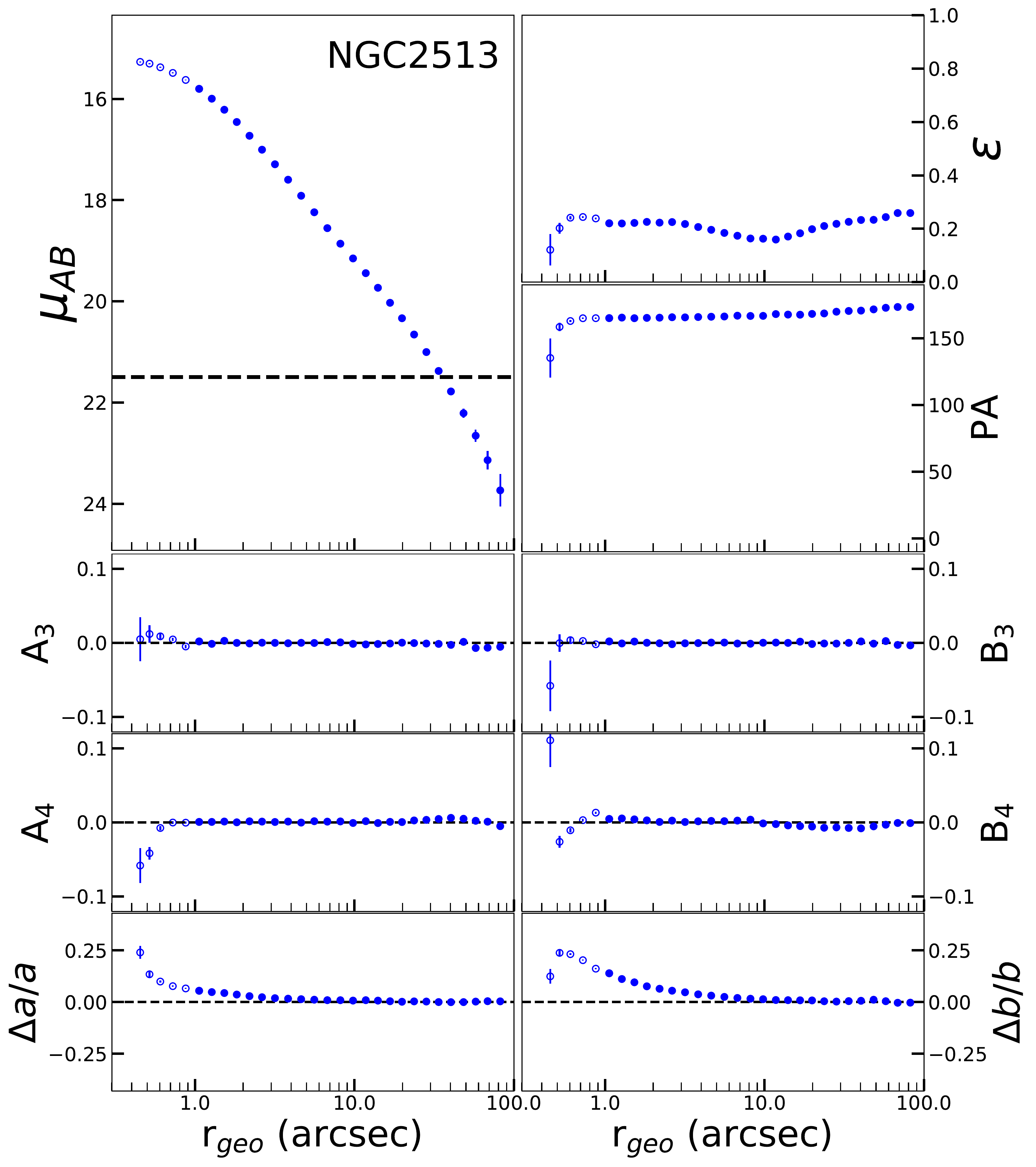}
  \end{minipage}
  \begin{minipage}[b][13.45cm][t]{0.41\textwidth}
    \includegraphics[width=\textwidth,trim=0 0 0 0]{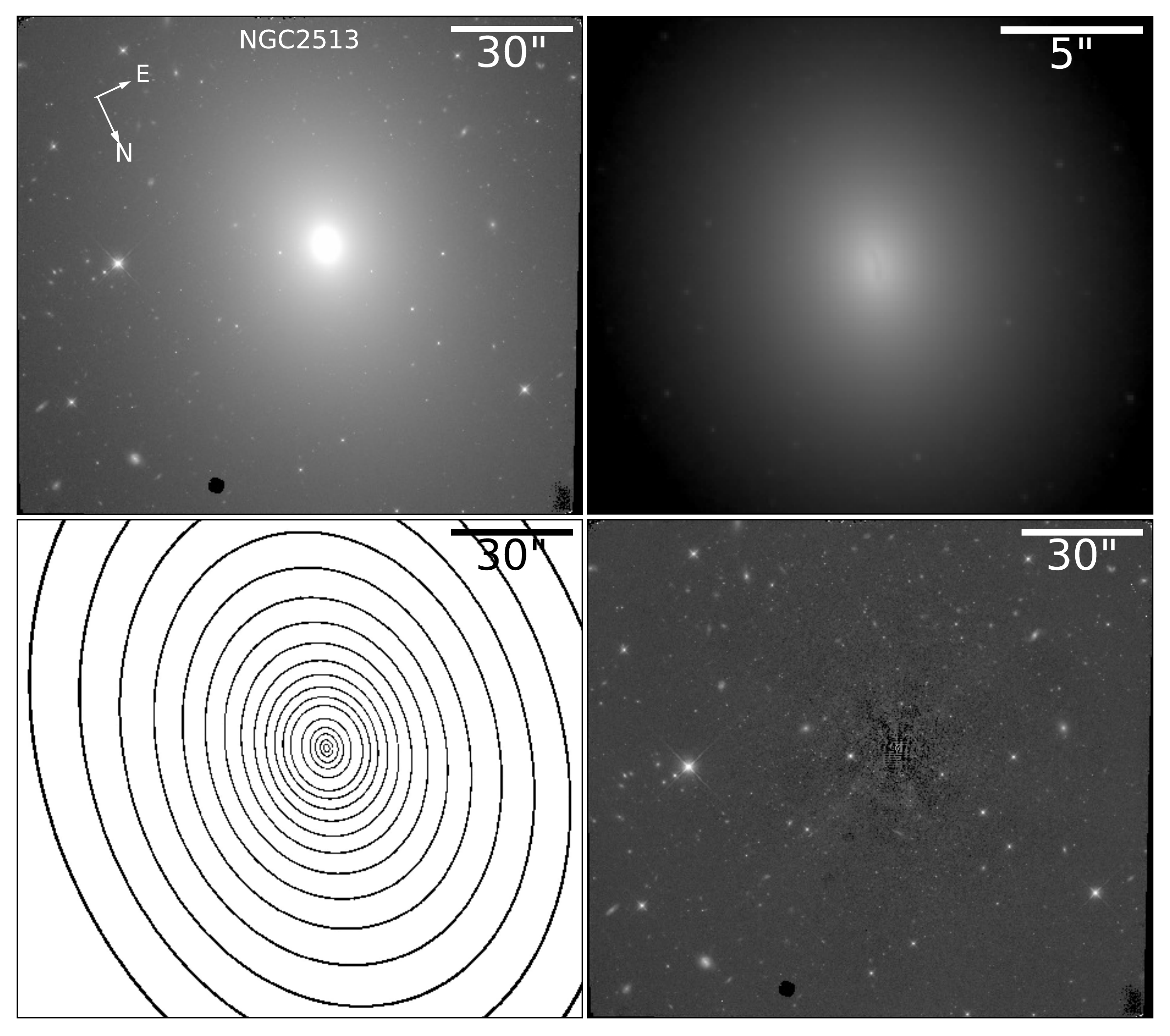}
    \caption{\small NGC~2513 has a small, compact dust disk that
      obscures isophotes within $1$ arcsec. The ellipticity
      decreases to a minimum at
      ${\sim}10$ arcsec, beyond which increases somewhat. \\
      Scale: $1$ arcsec = $343$ pc.  }
  \end{minipage}\\
  \vspace{-1.3cm}
  \begin{minipage}[b][13.55cm][t]{0.56\textwidth}
    \includegraphics[width=\textwidth]{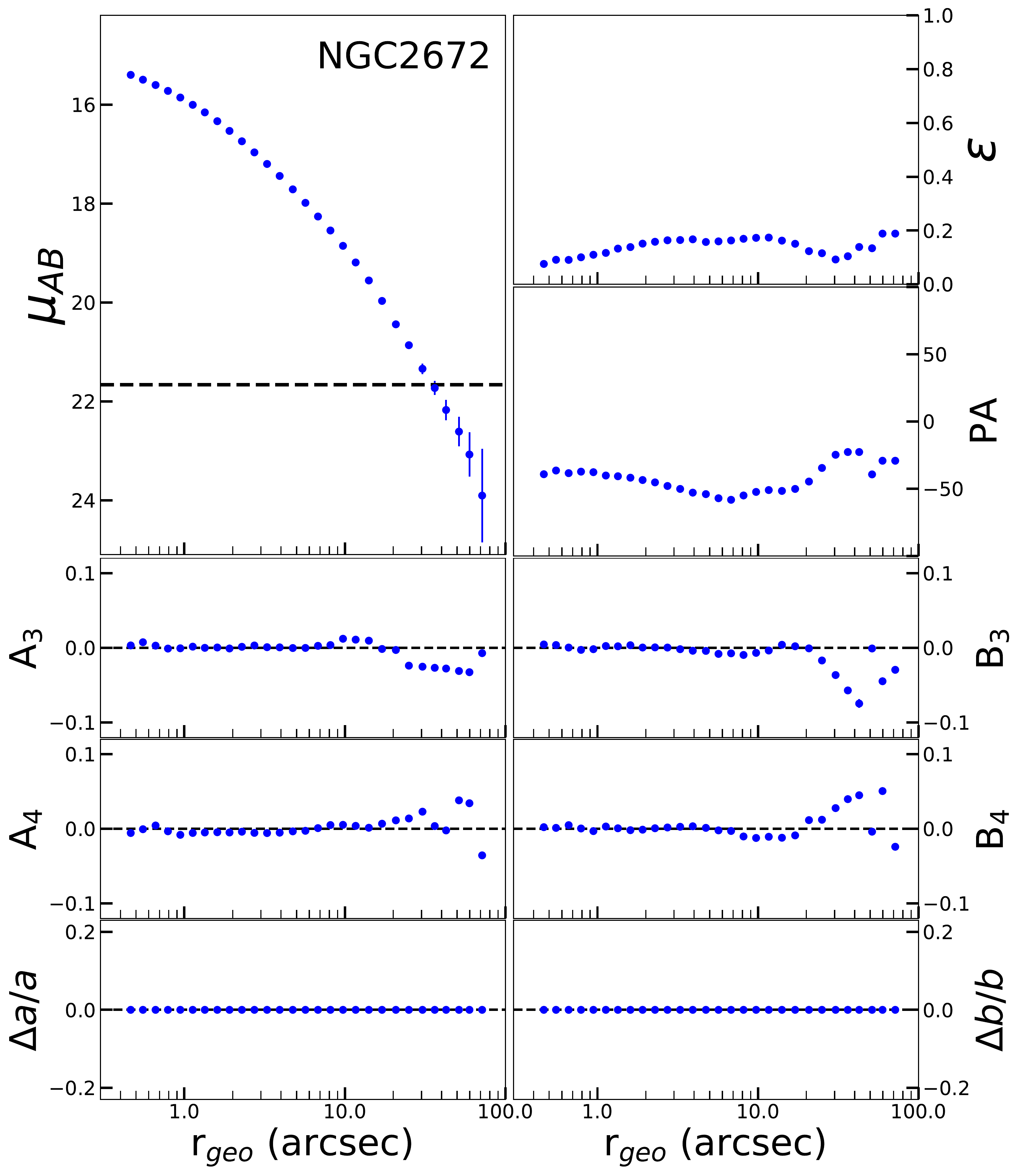}
  \end{minipage}
  \begin{minipage}[b][13.45cm][t]{0.41\textwidth}
    \includegraphics[width=\textwidth,trim=0 0 0 0]{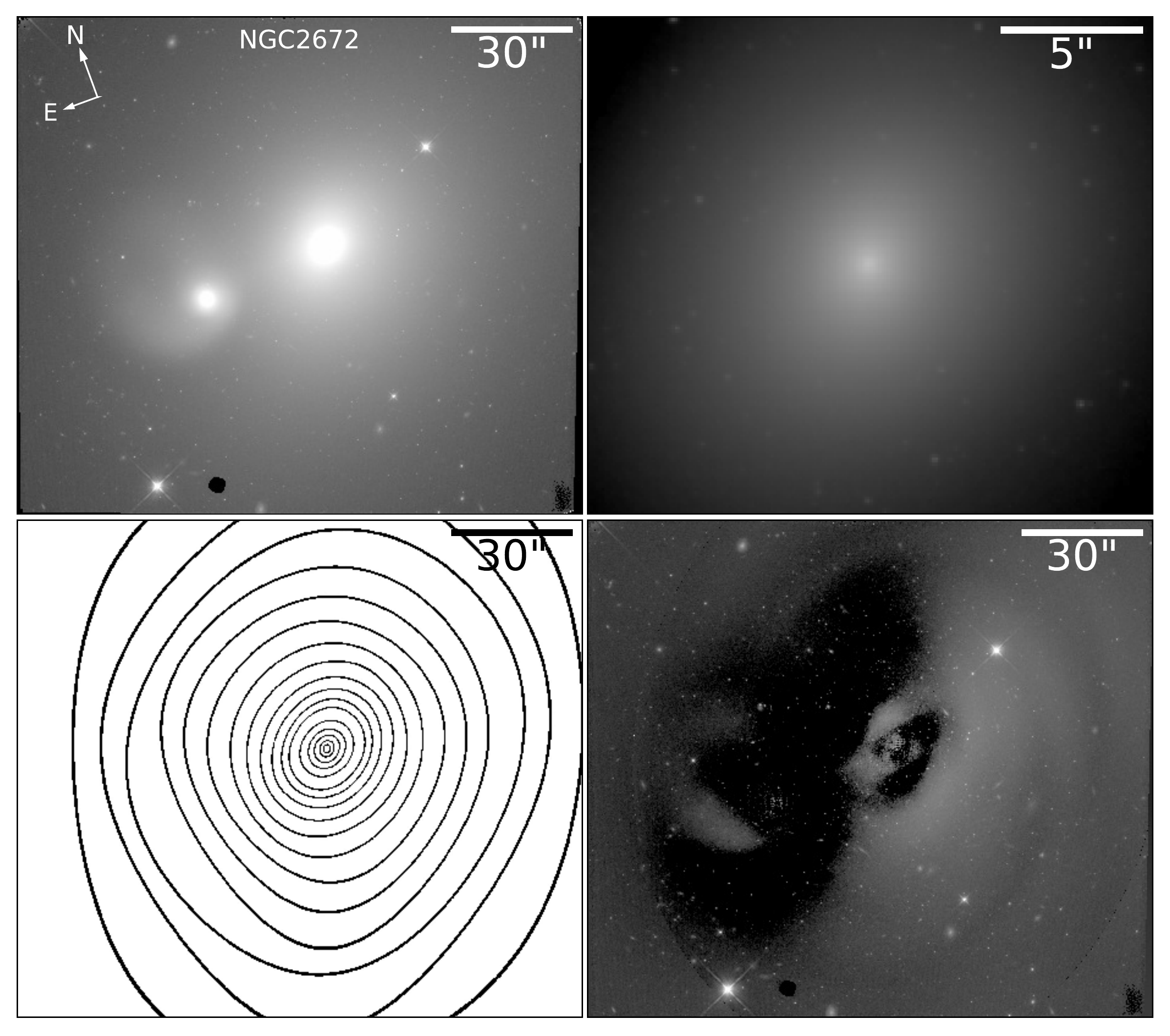}
    \caption{\small NGC~2672 has a large
      neighbor $33$ arcsec east of its center. The tidal interaction of the
      two galaxies distorts the isophotes as follows: 
      first, the ellipticity has a
      local minimum at ${\sim}33$ arcsec. The PA, which is otherwise
      stable, has a noticeable jump as it approaches $r_{geo}\,{\sim}\,33$
      arcsec. The isophotes at this radius also become very
      disky because of the presence of the companion, 
      but are boxy at other radii. \\
      Scale: $1$ arcsec = $298$ pc.}
  \end{minipage}\\
\end{figure*}

\begin{figure*}[!tbp]
  \centering\offinterlineskip
  \begin{minipage}[b][13.55cm][t]{0.56\textwidth}
    \includegraphics[width=\textwidth]{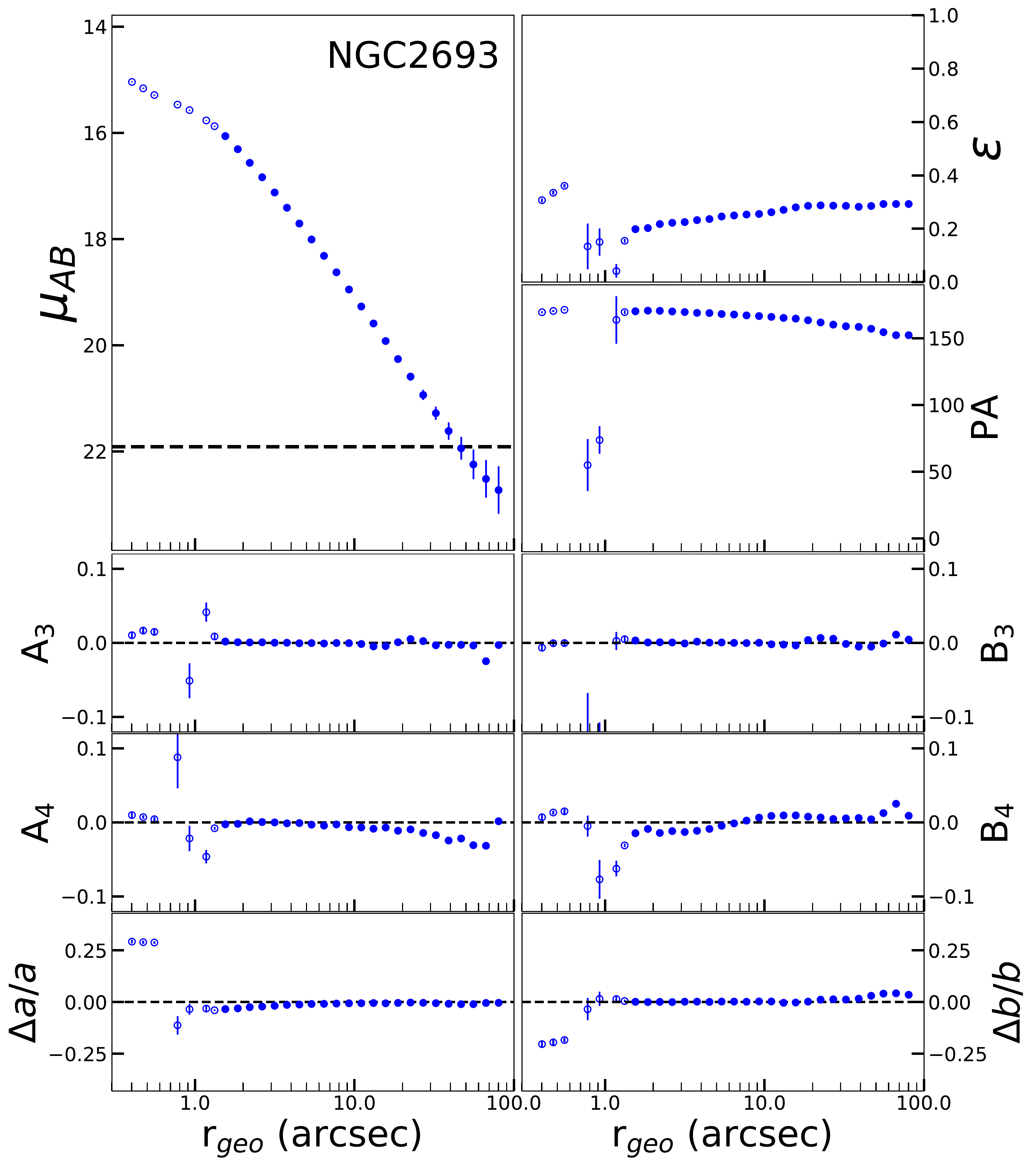}
  \end{minipage}
  \begin{minipage}[b][13.45cm][t]{0.41\textwidth}
    \includegraphics[width=\textwidth,trim=0 0 0 0]{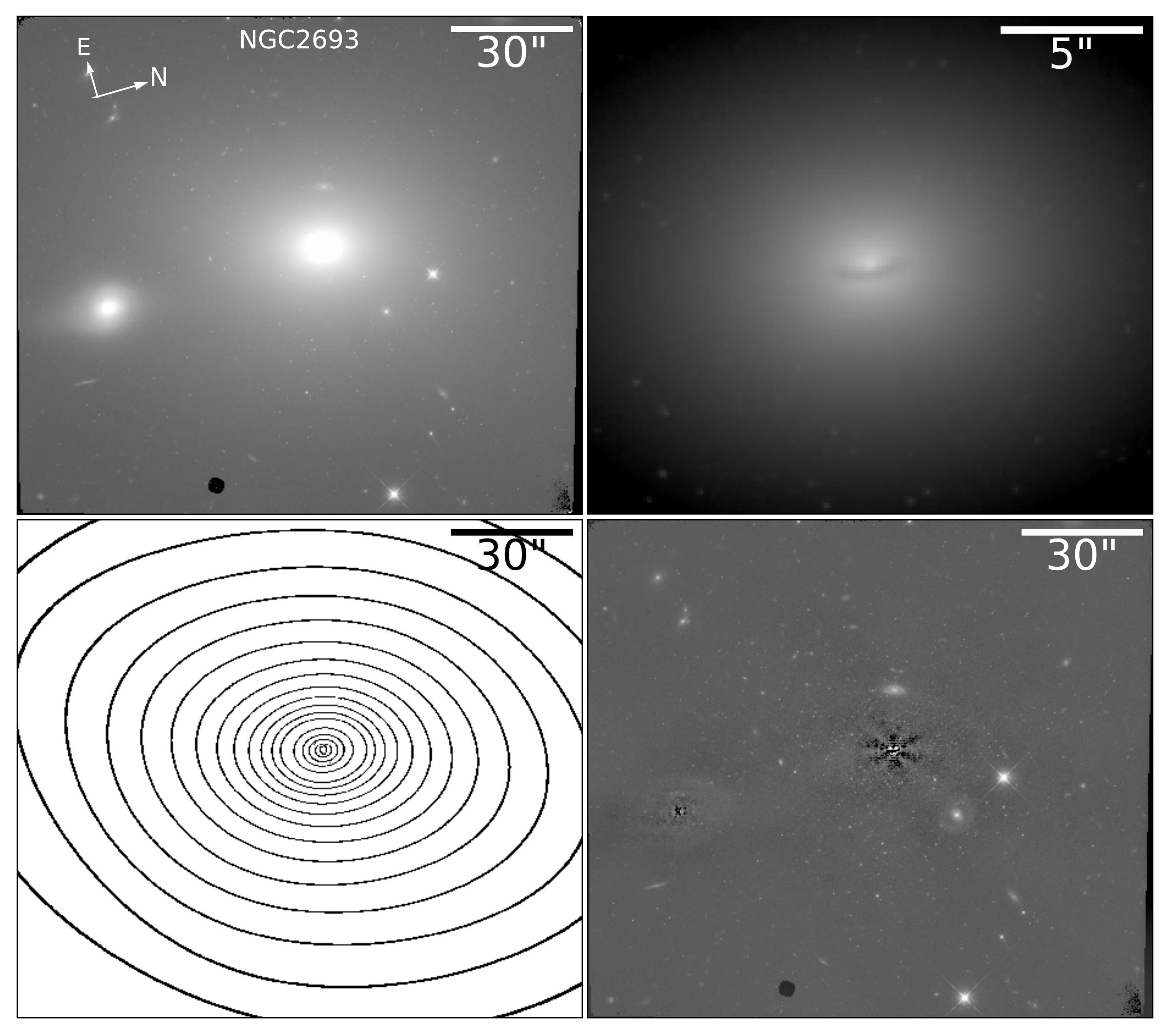}
    \caption{\small NGC~2693 has a small compact dust disk that
      extends $1.5$ arcsec (radius) from the center. The inner isophotes are
      boxy, but become slightly disky beyond ${\sim}5$
      arcsec. NGC~2693 has a nearby companion
      $55$ arcsec to the south. \\
      Scale: $1$ arcsec = $361$ pc.}
  \end{minipage}
  \vspace{-1.3cm}
  \begin{minipage}[b][13.55cm][t]{0.56\textwidth}
    \includegraphics[width=\textwidth]{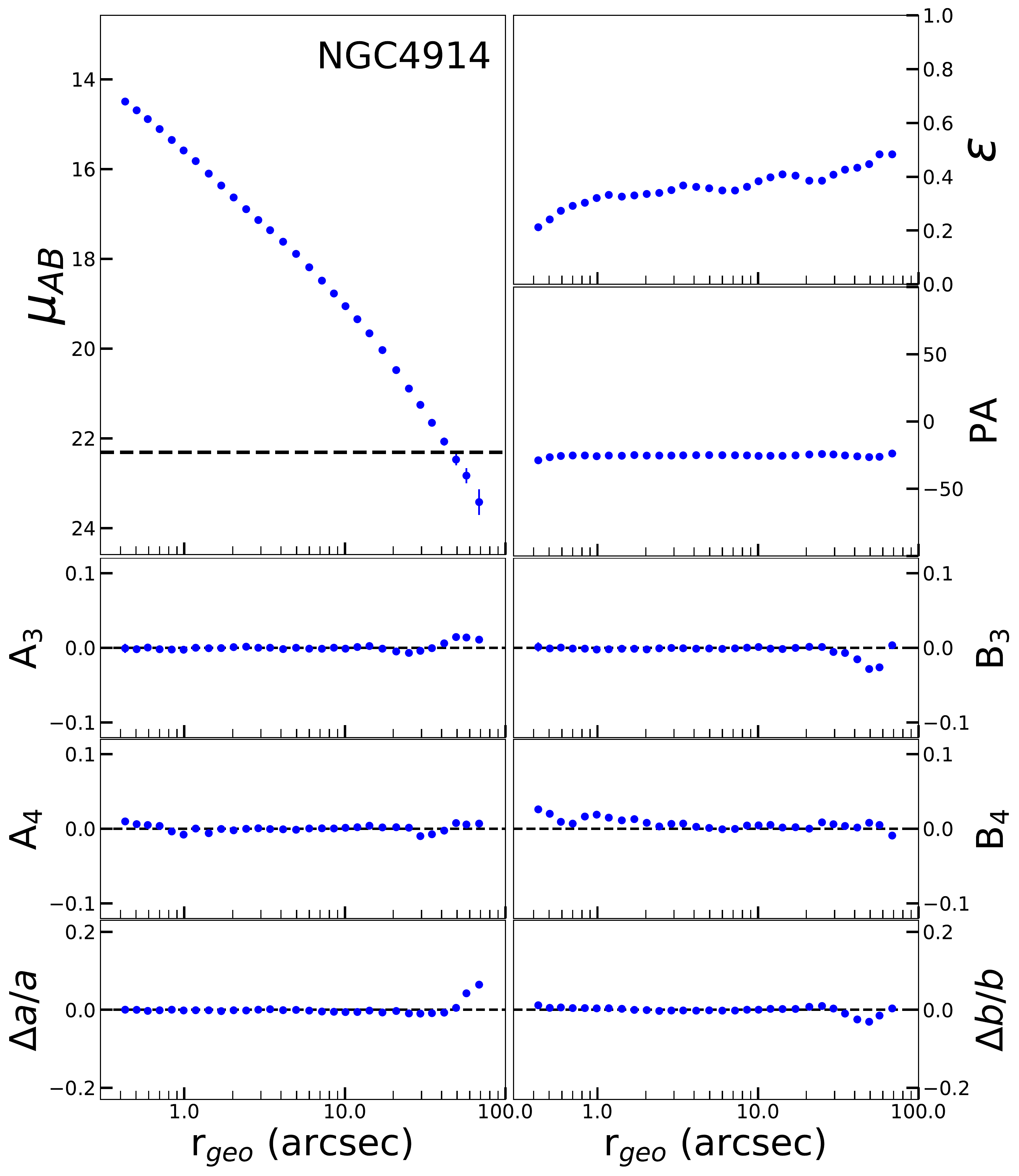}
  \end{minipage}
  \begin{minipage}[b][13.45cm][t]{0.41\textwidth}
    \includegraphics[width=\textwidth,trim=0 0 0 0]{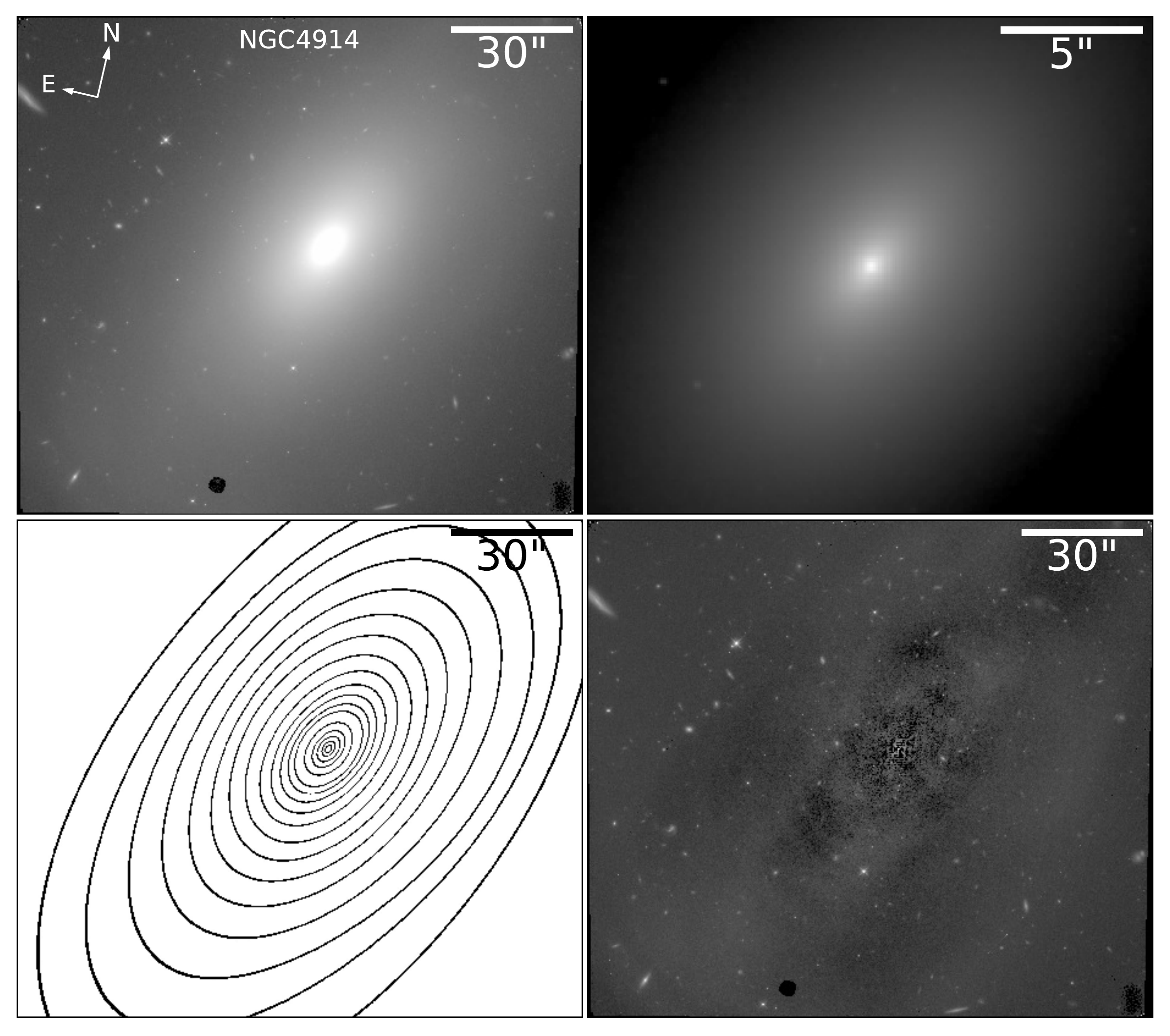}
    \caption{\small NGC~4914 has a steeply rising ellipticity profile,
      which reaches nearly $0.5$ for the outermost
      isophotes. Isophotes are disky within ${\sim}1.5$ arcsec
      and nearly elliptical farther out.\\
      Scale: $1$ arcsec = $361$ pc.}
  \end{minipage}\\
\end{figure*}

\begin{figure*}[!tbp]
  \centering\offinterlineskip
  \begin{minipage}[b][13.55cm][t]{0.56\textwidth}
    \includegraphics[width=\textwidth]{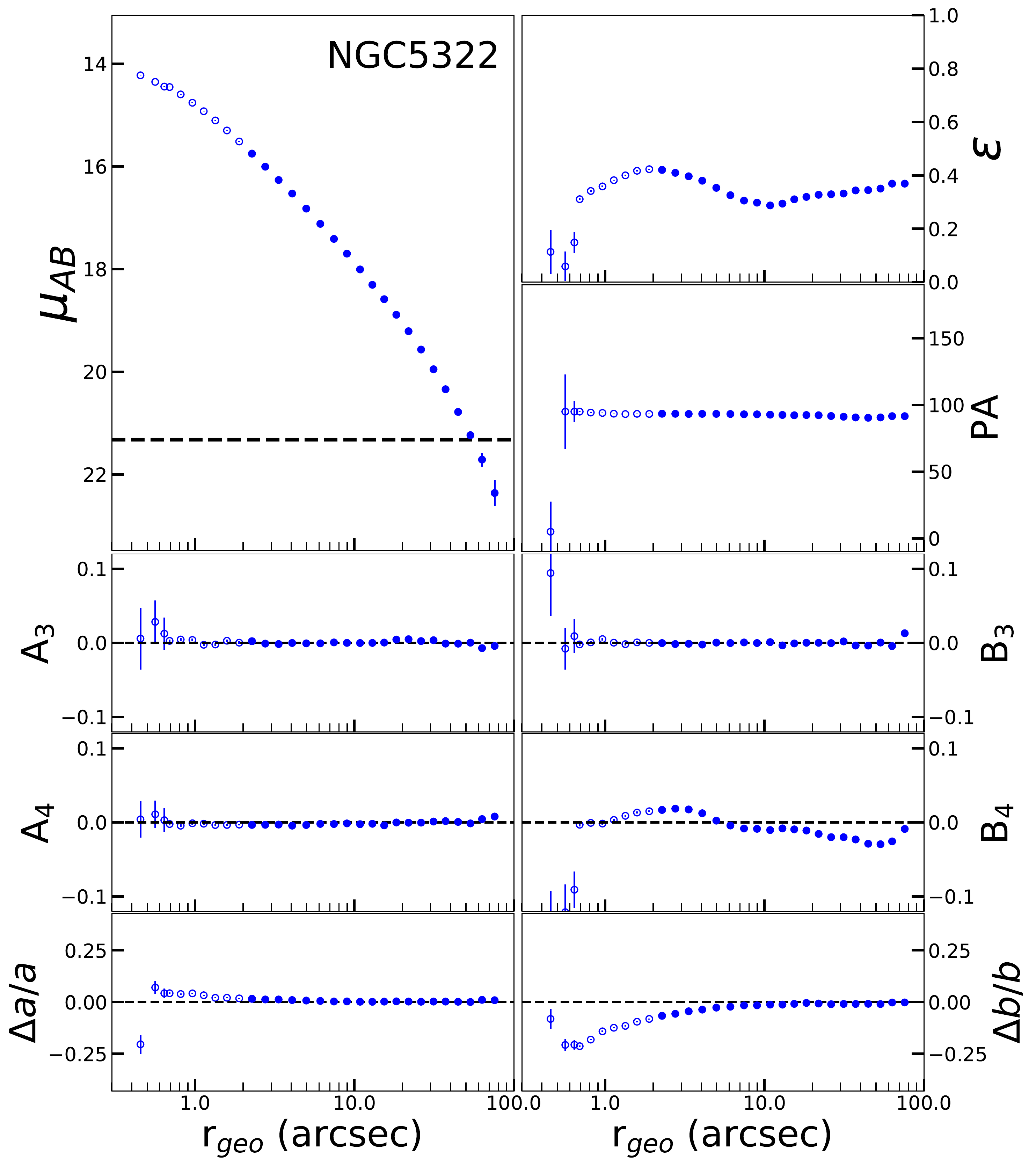}
  \end{minipage}
  \begin{minipage}[b][13.45cm][t]{0.41\textwidth}
    \includegraphics[width=\textwidth,trim=0 0 0 0]{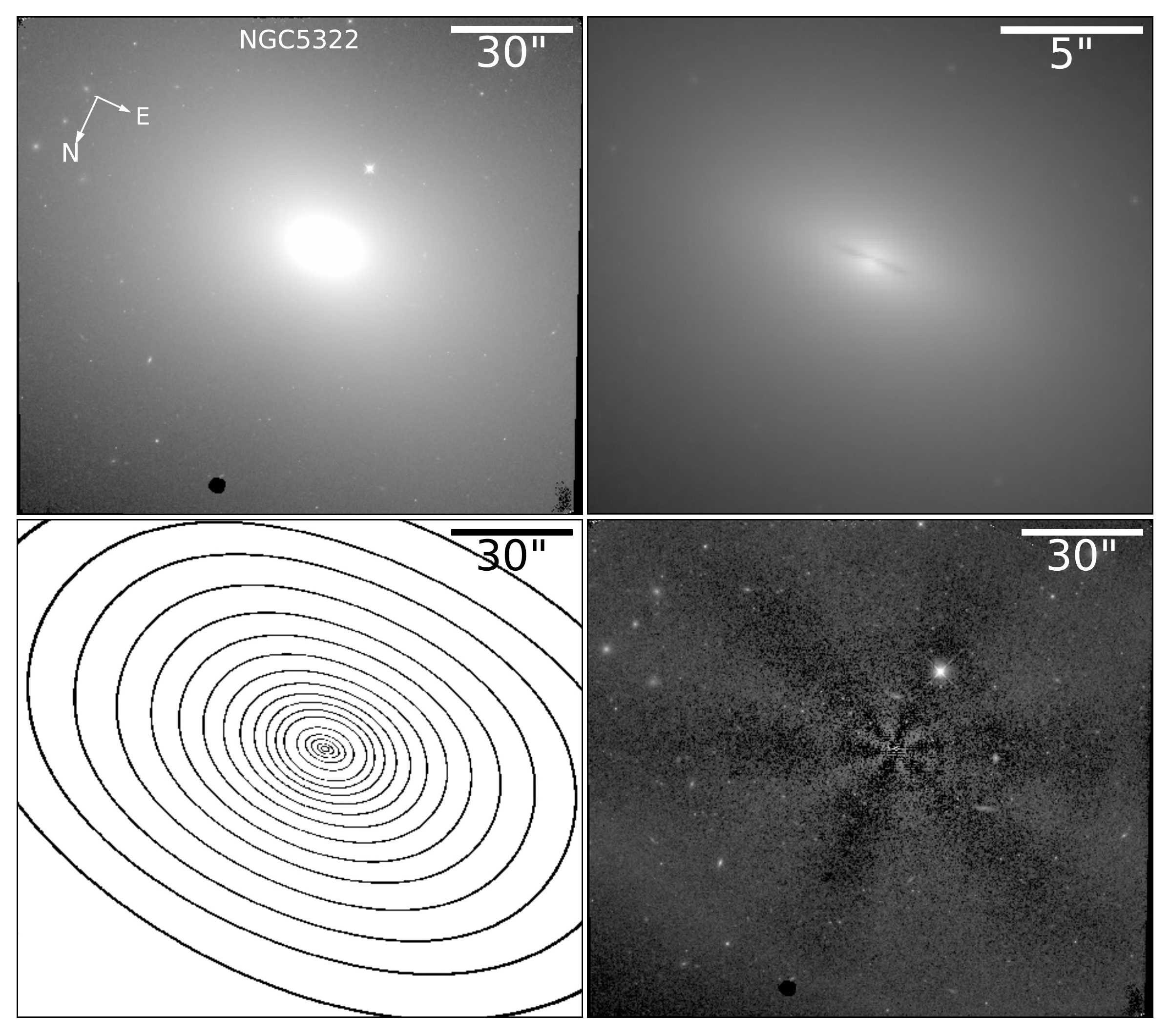}
    \caption{\small NGC~5322 has a small compact dust disk that
      extends $2$ arcsec in radius from the center. The ellipticity reaches a
      maximum at the outer edge of the dust disk, and it decreases to
      a minimum at ${\sim}10$ arcsec, beyond which it gradually
      rises. The isophotes at the edge of the dust disk are disky, but
      become more boxy with increasing radius. Isophotes
      beyond ${\sim}7$ arcsec are significantly boxy.\\
      Scale: $1$ arcsec = $166$ pc.}
  \end{minipage}\\
  \vspace{-1.3cm}
  \begin{minipage}[b][13.55cm][t]{0.56\textwidth}
    \includegraphics[width=\textwidth]{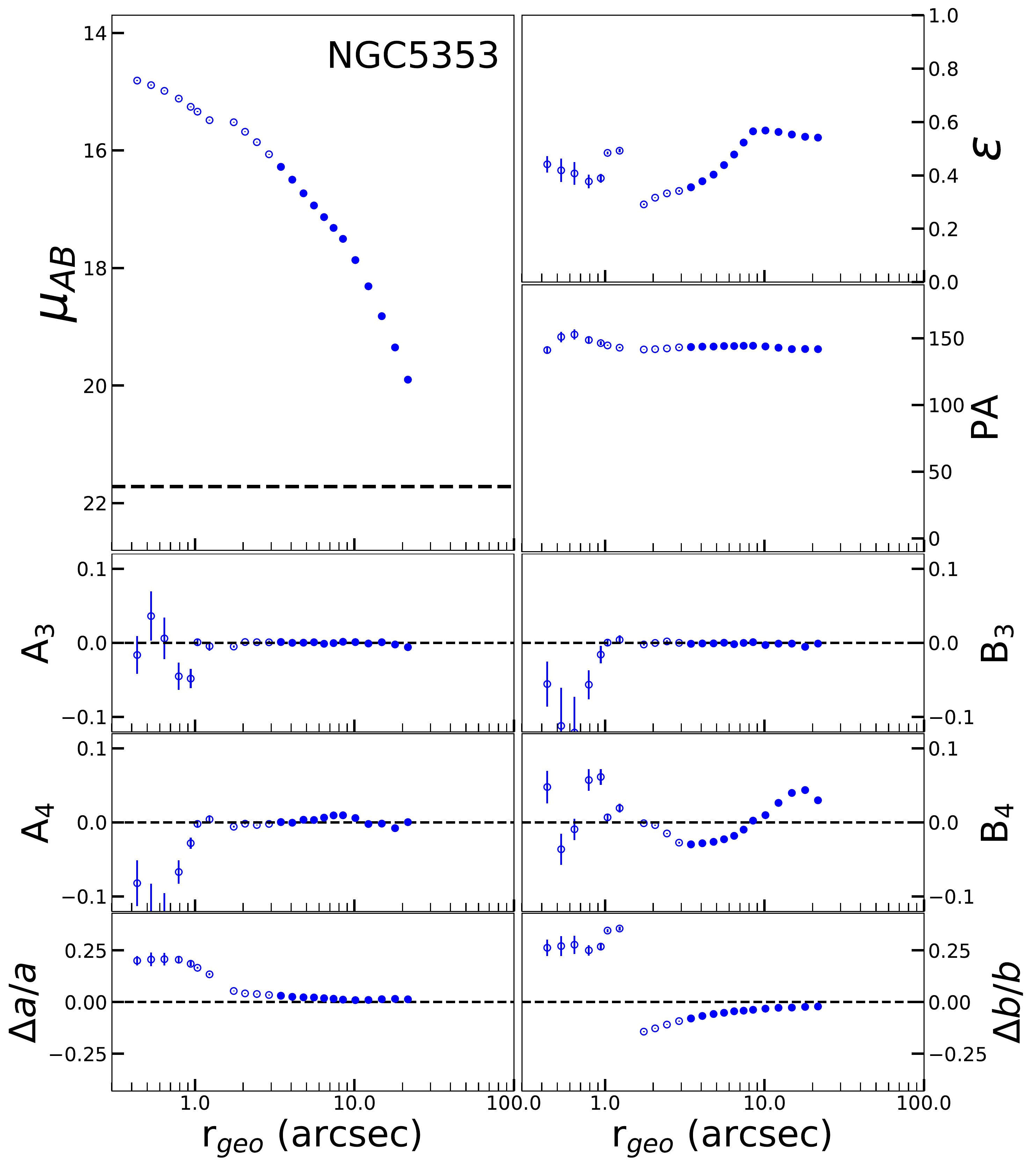}
  \end{minipage}
  \begin{minipage}[b][13.45cm][t]{0.41\textwidth}
    \includegraphics[width=\textwidth,trim=0 0 0 0]{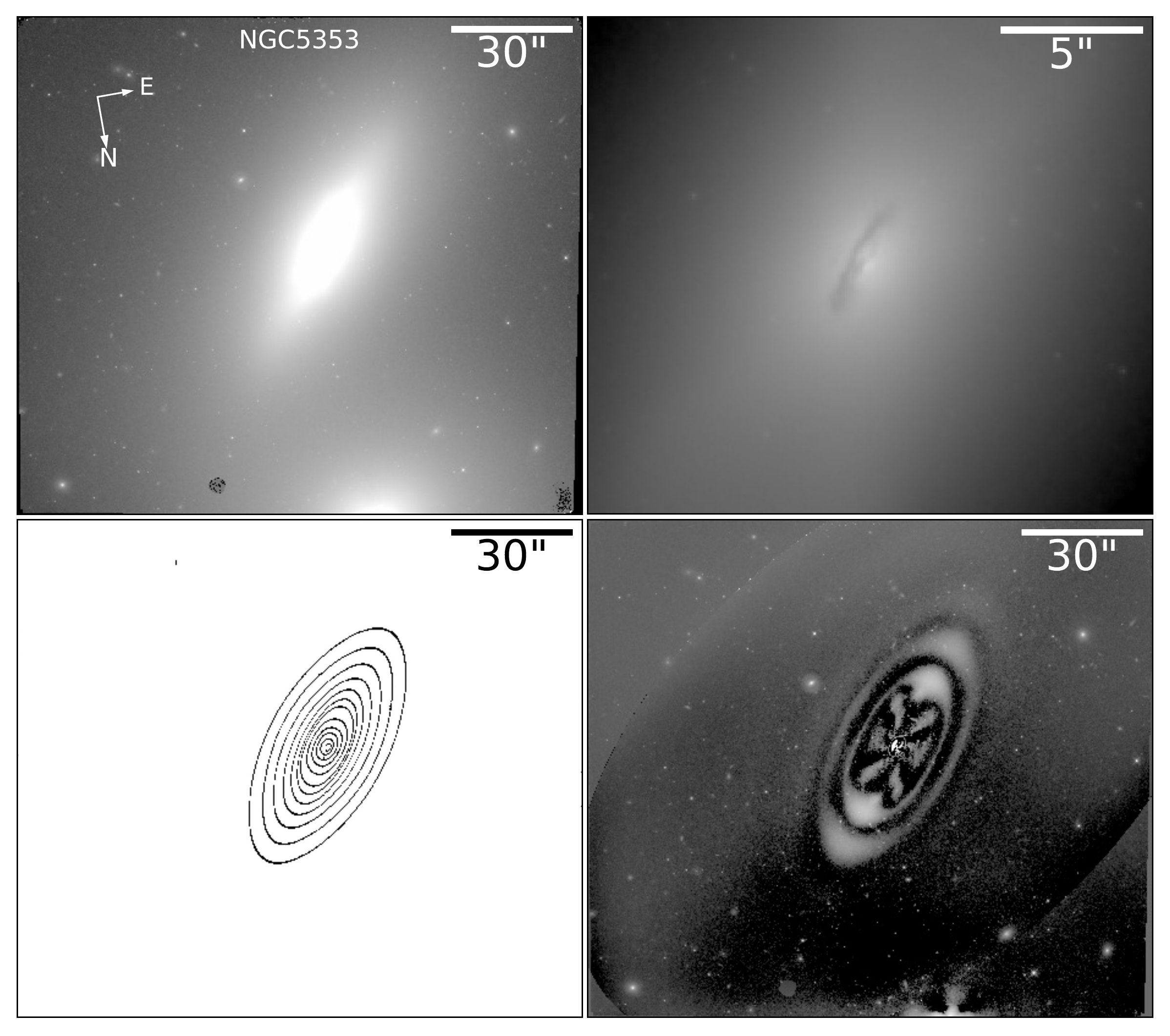}
    \caption{\small NGC~5353 has a dust disk that extends $3$ arcsec
      from its center. It also has a large nearby neighbor to the north that is
      mostly off-frame. The outer isophotes are difficult to
      disentangle from the companion galaxy, and so isophotes beyond
      $25$ arcsec are truncated. The ellipticity reaches a peak value of
      ${\sim}0.6$ at a radius of ${\sim}10$ arcsec, after which the
      ellipticity very gradually decreases. The isophotes between
      ${\sim}3$-$8$ arcsec are very boxy, but at ${\sim}10$ arcsec
      this quickly shifts and the isophotes beyond ${\sim}10$ arcsec
      are very disky. Throughout the dramatic shifts in isophotal
      shapes, the PA remains constant.\\
      Scale: $1$ arcsec = $199$ pc.}
  \end{minipage}\\
\end{figure*}

\begin{figure*}[!tbp]
  \centering\offinterlineskip
  \begin{minipage}[b][13.55cm][t]{0.56\textwidth}
    \includegraphics[width=\textwidth]{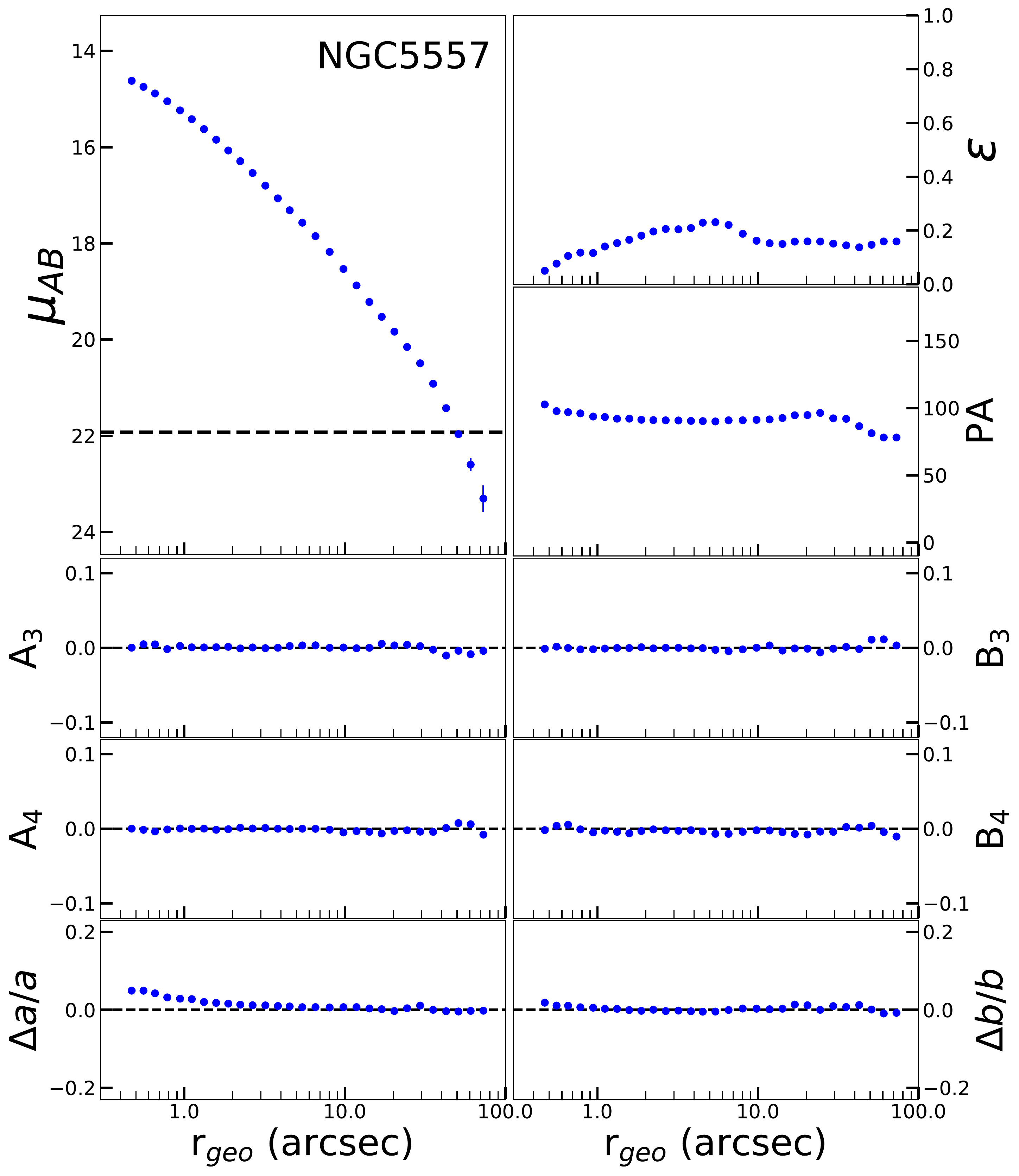}
  \end{minipage}
  \begin{minipage}[b][13.45cm][t]{0.41\textwidth}
    \includegraphics[width=\textwidth,trim=0 0 0 0]{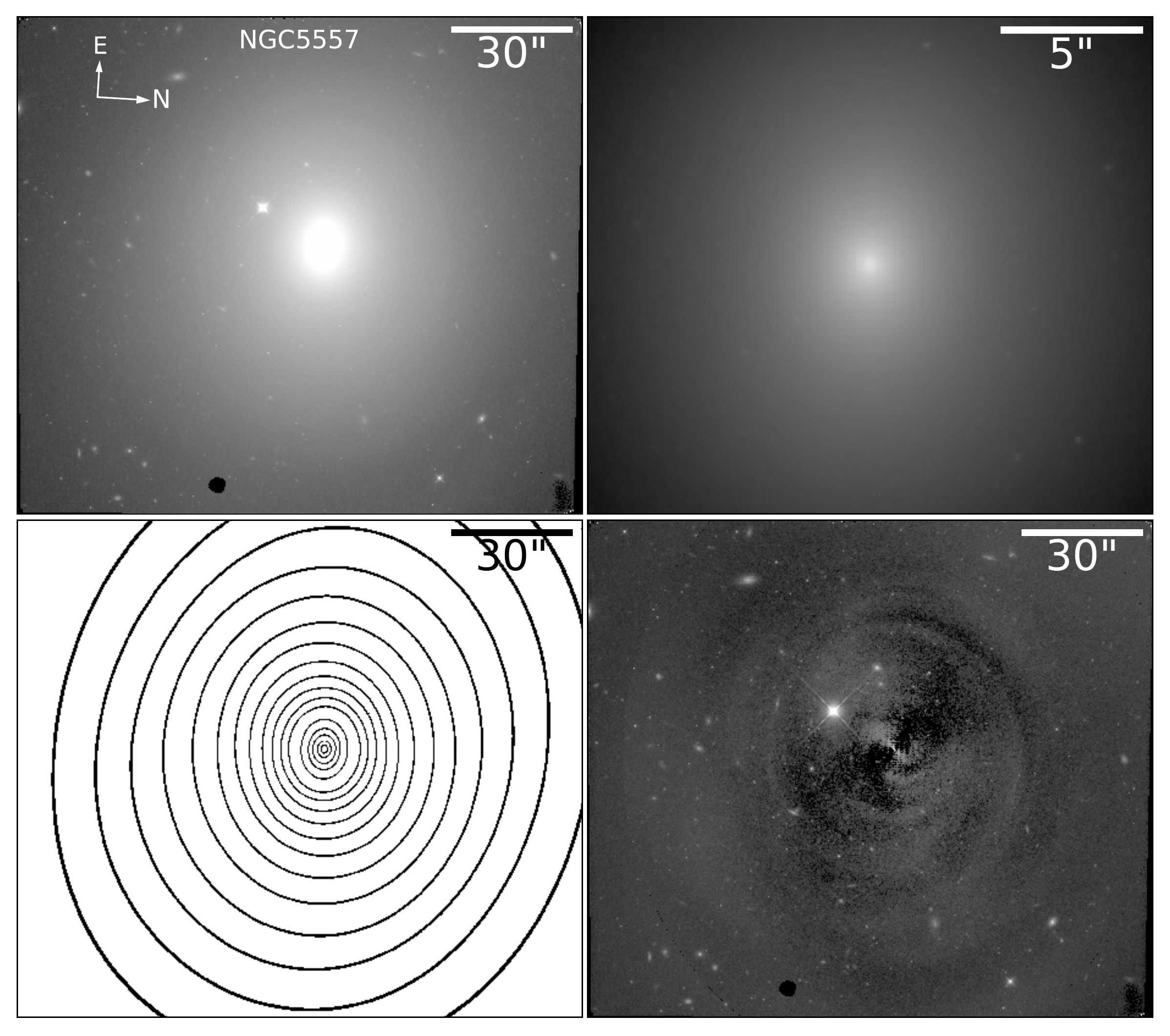}
    \caption{\small NGC~5557 has a small but varying ellipticity and
      modest PA shift at larger radii. It has a small companion to the 
      southeast.\\
      Scale: $1$ arcsec = $247$ pc. }
  \end{minipage}\\
  \vspace{-1.3cm}
  \begin{minipage}[b][13.55cm][t]{0.56\textwidth}
    \includegraphics[width=\textwidth]{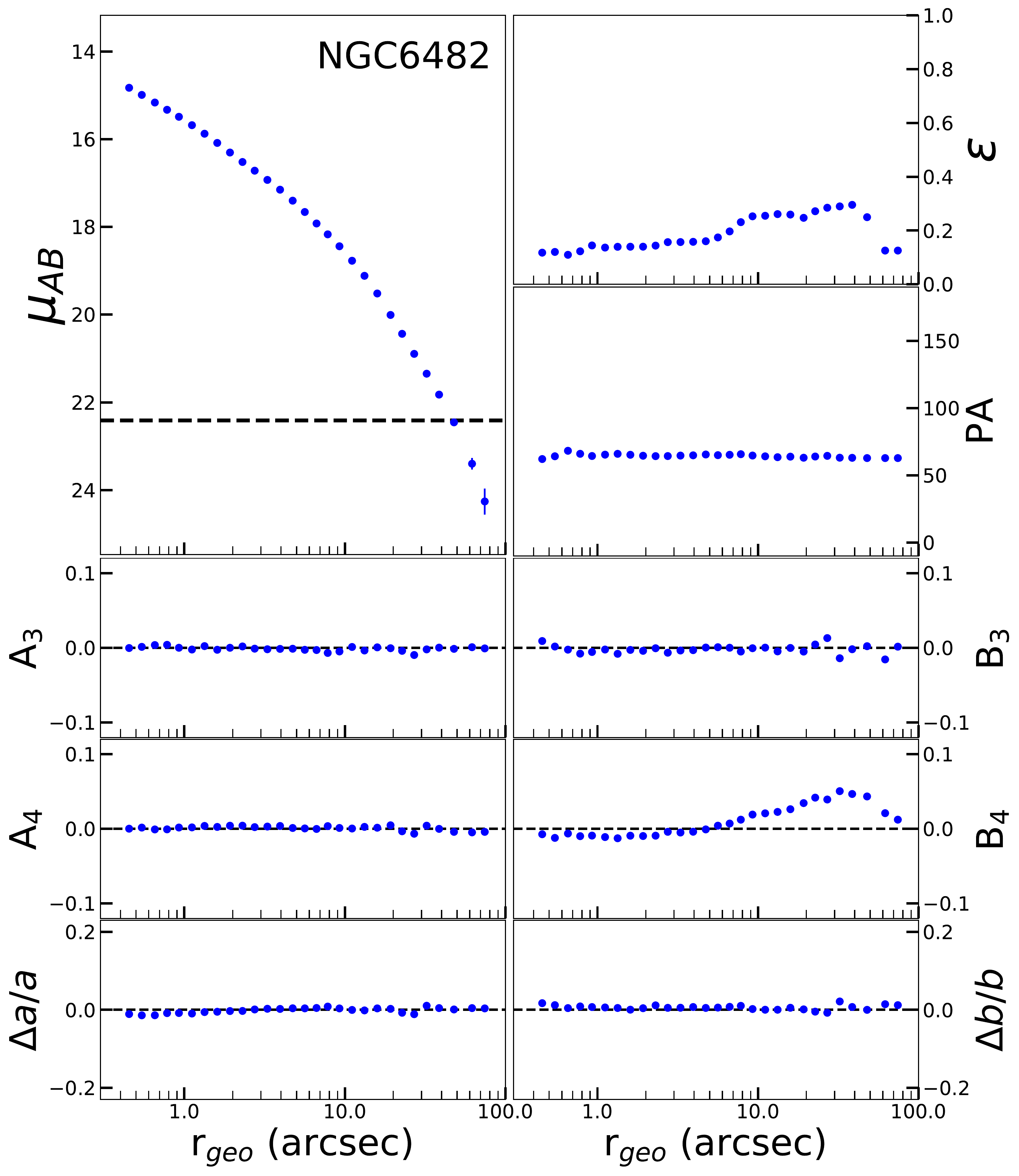}
  \end{minipage}
  \begin{minipage}[b][13.45cm][t]{0.41\textwidth}
    \includegraphics[width=\textwidth,trim=0 0 0 0]{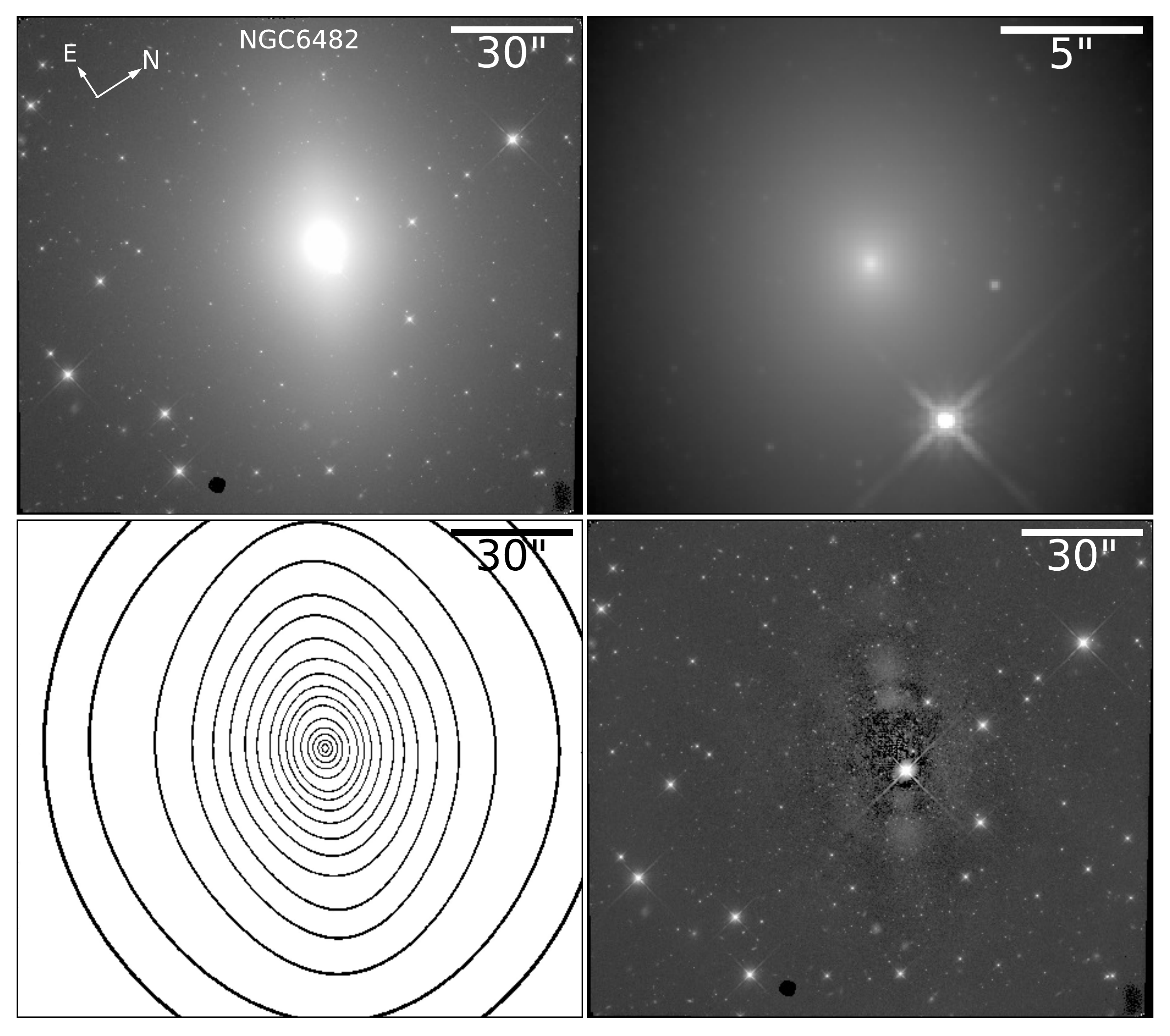}
    \caption{\small NGC~6482 has slightly boxy isophotes between
      ${\sim}0.5$ and $2.5$ arcsec. The isophotes become very disky
      beyond ${\sim}8$ arcsec, but eventually become nearly 
      elliptical (and round) in the outermost isophotes.\\
      Scale: $1$ arcsec = $298$ pc.}
  \end{minipage}\\
\end{figure*}

\begin{figure*}[!tbp]
  \centering\offinterlineskip
  \begin{minipage}[b][13.55cm][t]{0.56\textwidth}
    \includegraphics[width=\textwidth]{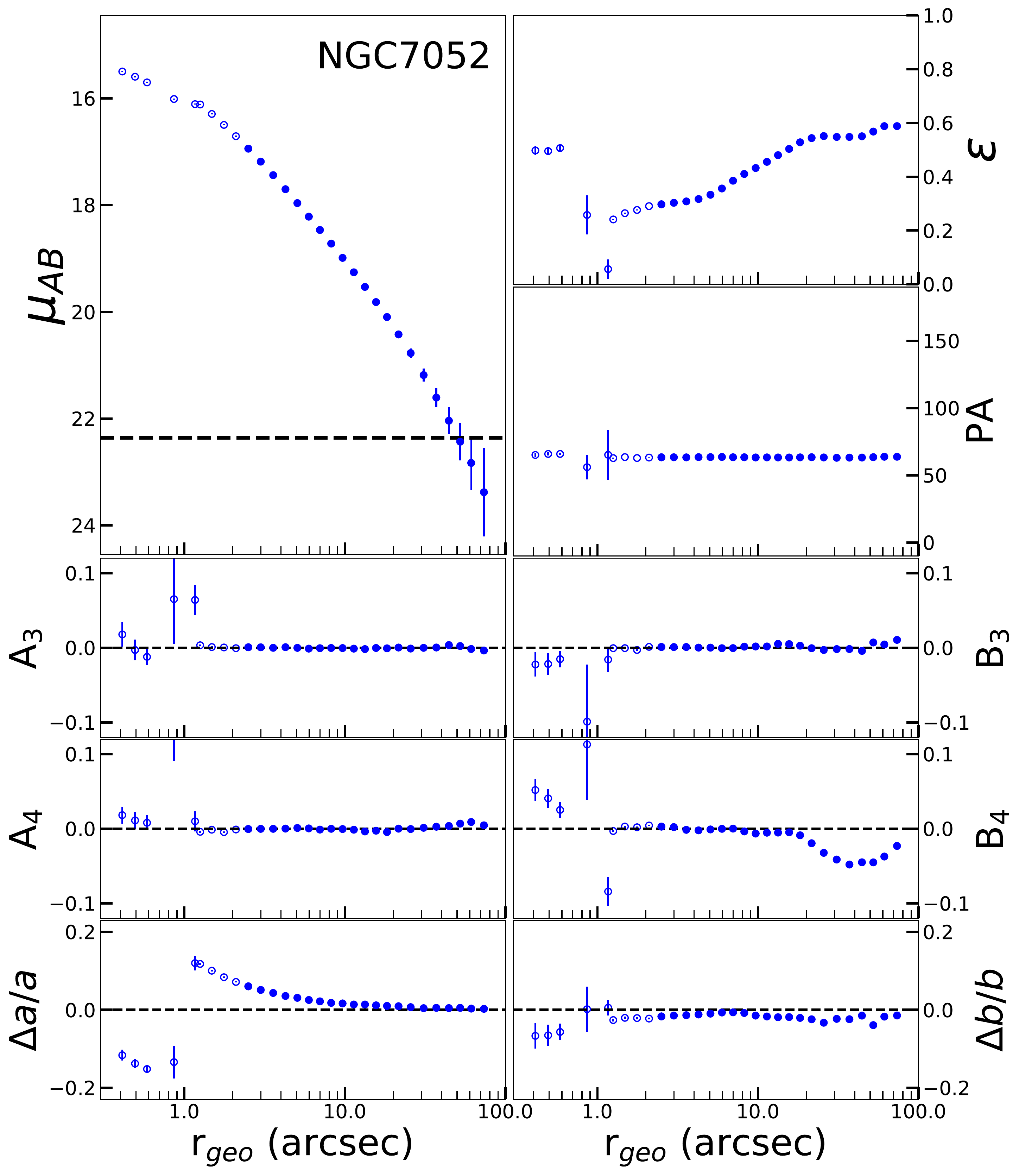}
  \end{minipage}
  \begin{minipage}[b][13.45cm][t]{0.41\textwidth}
    \includegraphics[width=\textwidth,trim=0 0 0 0]{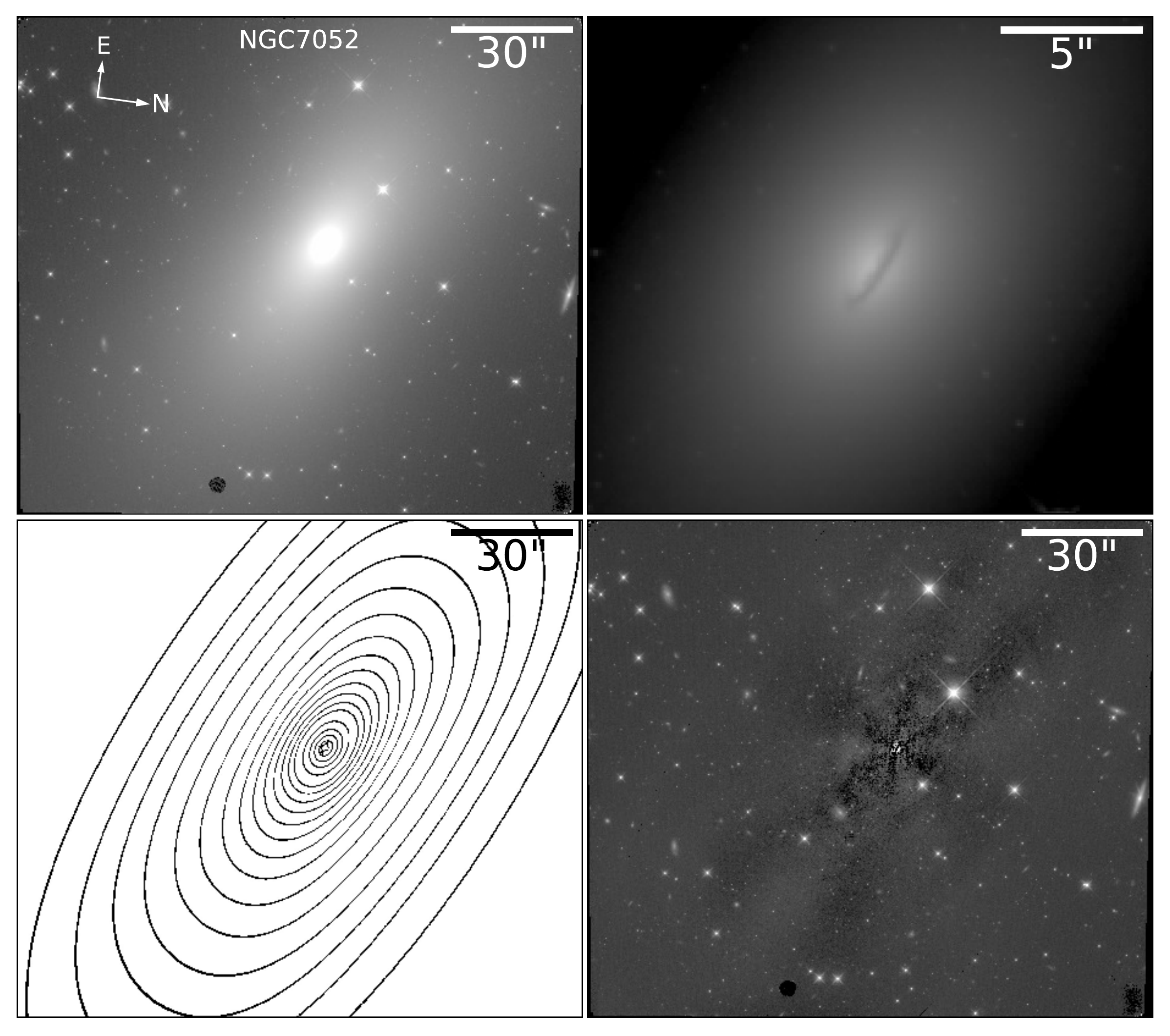}
    \caption{\small NGC~7052 has a compact dust disk that extends to a radius 
      of $2$ arcsec from its center. The ellipticity rises dramatically
      outside the dust disk, and reaches a peak of nearly $0.6$ in the
      outermost isophotes. The
      isophotes beyond ${\sim}10$ arcsec are very boxy.\\
      Scale: $1$ arcsec = $336$ pc.}
  \end{minipage}\\
  \vspace{-1.3cm}
  \begin{minipage}[b][13.55cm][t]{0.56\textwidth}
    \includegraphics[width=\textwidth]{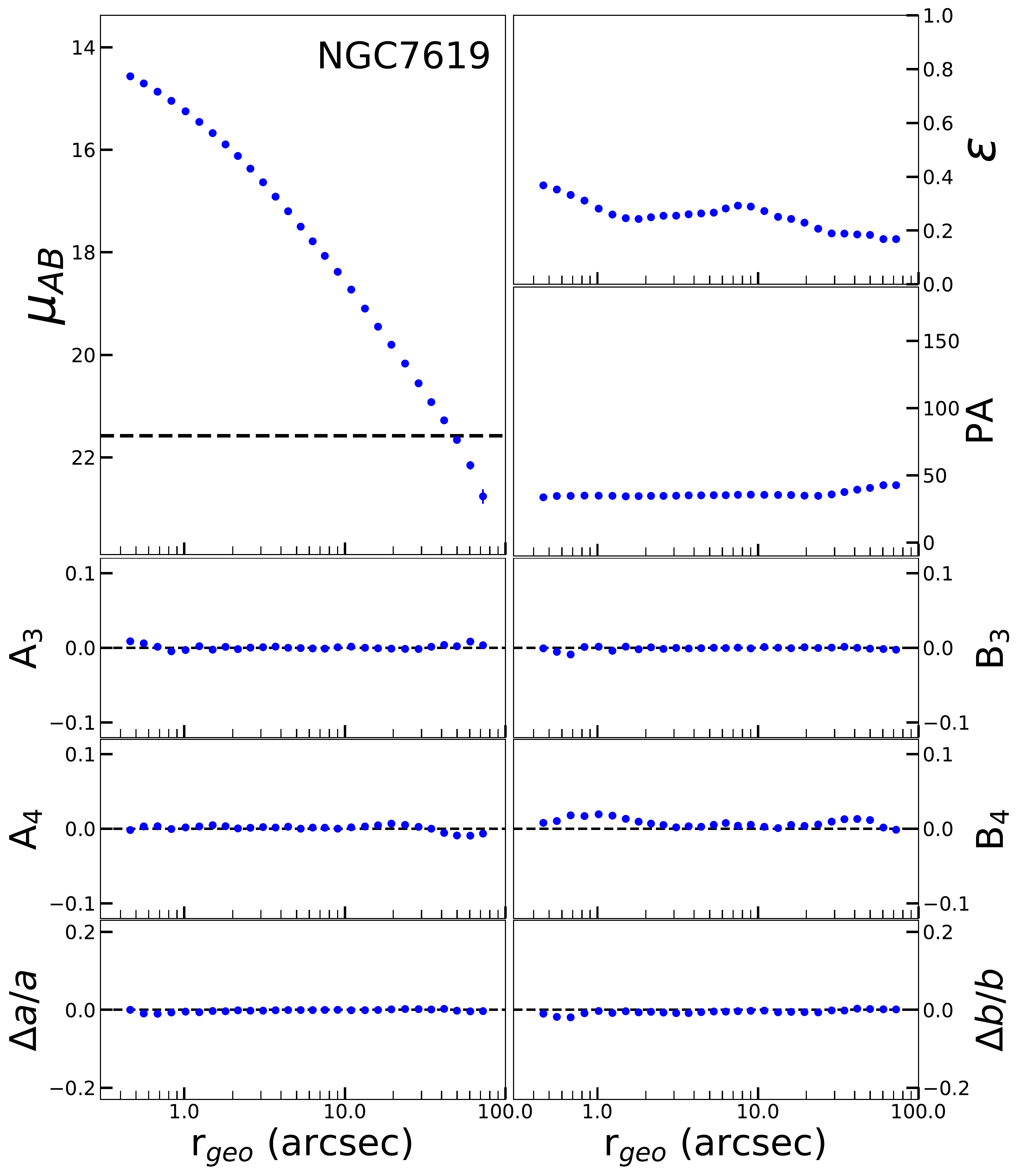}
  \end{minipage}
  \begin{minipage}[b][13.45cm][t]{0.41\textwidth}
    \includegraphics[width=\textwidth,trim=0 0 0 0]{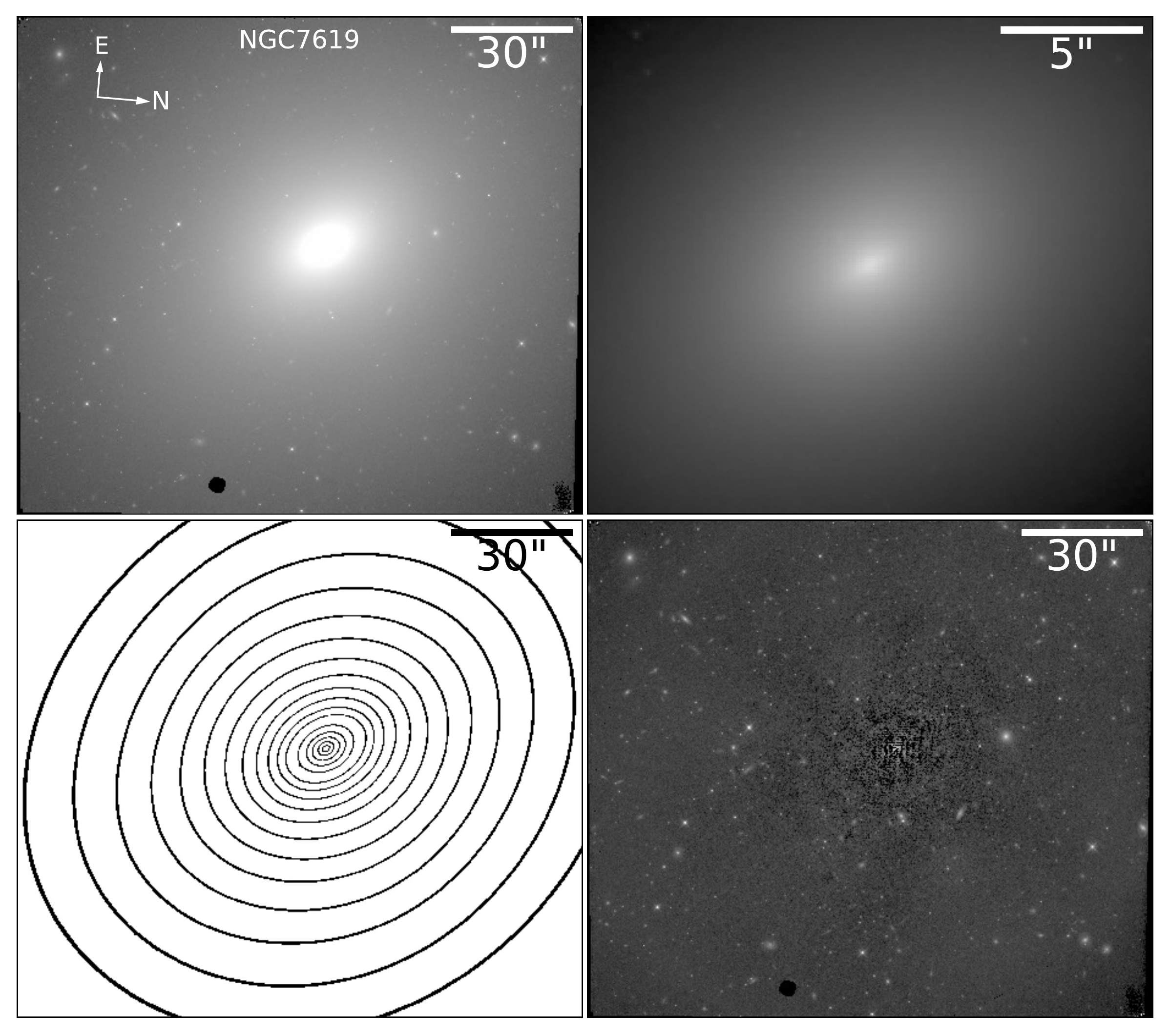}
    \caption{\small NGC~7619 has an ellipticity profile that decreases
      with increasing radius. The isophotes between ${\sim}0.6$ to $2$
      arcsec and ${\sim}30$ to $60$ arcsec are slightly disky.\\
      Scale: $1$ arcsec = $262$ pc.}
  \end{minipage}
\end{figure*}

\end{document}